\documentclass[11pt,a4paper]{article}

\usepackage{amsmath,amsfonts,amssymb,mathrsfs,mathabx}
\usepackage[dvips]{graphicx}
\usepackage{tikz}
\usepackage{comment}

\usepackage{booktabs}
\usepackage[footnotesize,bf]{caption}
\usepackage{graphicx}
\usepackage{txfonts}

\def \dl{\Delta}

\newcommand \bra[1]{\langle {#1} |}
\newcommand \ket[1]{|{#1} \rangle}

.tfm scaled 1000 
.tfm scaled 1000 
.tfm scaled 1000 

\usepackage{hyperref}

\hypersetup{
    unicode=true,          
    pdftoolbar=false,        
    pdfmenubar=false,        
    pdffitwindow=false,     
    pdfstartview={FitH},    
    pdftitle={Classical irregular block, N=2 pure gauge theory},    
    pdfauthor={M. Piatek, A. R. Pietrykowski},     
    pdfsubject={Integrable systems},   
    pdfcreator={M. Piatek, A. R. Pietrykowski},   
    pdfproducer={M. Piatek, A. R. Pietrykowski}, 
    pdfkeywords={Integrable systems} {Supersymetry}
                {Seiberg-Witten theory} {AGT correspondence} {Mathieu equation}, 
    pdfnewwindow=true,      
    colorlinks=true,       
    linkcolor=black,          
    citecolor=blue,        
    filecolor=blue,      
    urlcolor=blue           
}

\def \d {\delta}
\def \e {\epsilon}

\def \f {\phi}

\def \r {\rho}

\def \th{\theta}

\def \z {\zeta}

\def \La{\Lambda}
\def \pt{\partial}

\newcommand \cbr[1]{\left({#1}\right)}
\newcommand \sbr[1]{\left[{#1}\right]}
\newcommand \pbr[1]{\left\{{#1}\right\}}
\newcommand \ad[2]{\mbox{ad}_{#1}\!\left( #2 \right)}

\newlength{\intwidth}

\numberwithin{equation}{section}

\newcommand{\preprintsize}{
      \headheight=5pt                              
     \topmargin= 0 cm \headsep=0.1cm
     \oddsidemargin= 0cm
      \evensidemargin= 0cm  
      \textheight = 23truecm \textwidth=16truecm      %
}

\preprintsize

\begin{document}

\begin{center}
{\LARGE\bf Classical limit of irregular blocks and Mathieu functions}
\end{center}

\begin{center}
{\large{Marcin Piatek}$^{\,a,\,c,\;}$}\footnote{\href{mailto:piatek@fermi.fiz.univ.szczecin.pl}{e-mail: piatek@fermi.fiz.univ.szczecin.pl}}
\hskip 1.0cm
{\large{Artur R. Pietrykowski}$^{\,b,\,c,\;}$}\footnote{\href{mailto:pietrie@theor.jinr.ru}{e-mail: pietrie@theor.jinr.ru}}

\vskip 4mm
${}^{a}$
Institute of Physics and CASA*, University of Szczecin\\
ul. Wielkopolska 15, 70-451 Szczecin, Poland

\vskip 4mm
${}^{b}$
Institute of Theoretical Physics\\
University of Wroc{\l}aw\\
pl. M. Borna 9, 50-204 Wroc{\l}aw, Poland

\vskip 4mm
${}^{c}$
Bogoliubov Laboratory of Theoretical Physics,\\
Joint Institute for Nuclear Research, 141980 Dubna, Russia
\end{center}

\vskip .5cm

\begin{abstract}\noindent
The Nekrasov--Shatashvili limit of the ${\cal N}\!=\!2$ SU(2) 
pure gauge  ($\Omega$-deformed) super Yang--Mills theory encodes the information 
about the spectrum of the Mathieu operator. 
On the other hand, the Mathieu equation emerges entirely 
within the frame of two-dimensional
conformal field theory ($2d$ CFT) as the classical limit of the 
null vector decoupling equation for some degenerate irregular block.
Therefore, it seems to be possible to investigate the spectrum of the
Mathieu operator employing the techniques of $2d$ CFT. 
To exploit this strategy, a full correspondence between the Mathieu equation and its 
realization within $2d$ CFT has to be established. 
In our previous paper \cite{Piatek:2014lma},
we have found that the expression of the Mathieu eigenvalue given in terms of the classical irregular
block exactly coincides with the well known weak coupling expansion of this eigenvalue 
in the case in which the auxiliary parameter is the noninteger Floquet 
exponent. In the present work we verify that the formula for the corresponding
eigenfunction obtained from the irregular block reproduces the so-called Mathieu exponent 
from which the noninteger order elliptic cosine and sine functions may be constructed. 
The derivation of the Mathieu equation within the formalism of $2d$ CFT is based on
conjectures concerning the asymptotic behaviour of irregular blocks in the classical limit. 
A proof of these hypotheses is sketched.
Finally, we speculate on how it could be possible to use the methods of $2d$ CFT in order 
to get from the irregular block the eigenvalues of the
Mathieu operator in other regions of the coupling constant.
\end{abstract}

\newpage
{\small \hrule \tableofcontents \vskip .5cm\hrule}

\section{Introduction}
In a last few years much attention was paid to the study of the connections among
two-dimensional conformal field theory ($2d$ CFT), ${\cal N}=2$ supersymmetric gauge 
theories and integrable systems, cf.~e.g.~\cite{Nekrasov:2010ka,Teschner:2010je,Kozlowski:2010tv,
Meneghelli:2013tia,Vartanov:2013ima,Belavin:2011js,Fateev:2011hq,Alba:2010qc,Tai:2010ps,Muneyuki:2011qu,
Itoyama:2015xia,Poghossian:2010pn,Fucito:2011pn,Maruyoshi:2010,Bonelli:2009zp,Bonelli:2011na,
Piatek:2011tp,Ferrari:2012gc,Ferrari:2012qj,Ferrari:2014nba,Tan:2013tq}.\footnote{See also the volume
\cite{Teschner:2014oja} edited by 
J.~Teschner and refs.~therein.}
This kind of research was inspired by the discovery of certain dualities, 
in particular, the AGT \cite{Alday:2009aq} and Bethe/gauge \cite{NS:2009,Nekrasov:2009uh,Nekrasov:2009ui}
correspondences.\footnote{See also \cite{Nekrasov:2011bc,Nekrasov:2012xe,Nekrasov:2013xda}.}

The AGT correspondence states that the Liouville field theory (LFT) 
correlators on the Riemann surface $C_{g,n}$ with genus $g$ 
and $n$ punctures can be identified with the
partition functions of a class $T_{g,n}$ of 
four-dimensional ${\cal N}=2$ supersymmetric SU(2) quiver gauge
theories:
\begin{equation}\label{agt}
\left\langle\prod\limits_{i=1}^{n}{\sf V}_{\Delta_i}\right\rangle^{\!\!\rm LFT}_{\!\!C_{g,n}}
= Z^{(\sigma)}_{T_{g,n}}.
\end{equation}

Let us recall that for a given pant 
decomposition $\sigma$ of the Riemann surface $C_{g,n}$, both sides of 
the equation above have an integral representation. Indeed, 
LFT correlators can be factorized according to the pattern 
given by the pant decomposition of $C_{g,n}$ and written as 
an integral over a continuous spectrum of the Liouville theory 
in which, for each pant decomposition $\sigma$, 
the integrand is built out of the holomorphic and the anti-holomorphic 
Virasoro conformal blocks 
${\cal F}_{c,\Delta_p}^{(\sigma)}[\Delta_i](\sf Z)$ 
and
$\bar{\cal F}_{c,\Delta_p}^{(\sigma)}[\Delta_i](\bar{\sf Z})$ 
multiplied by the DOZZ 3-point functions \cite{Dorn:1994xn,Zamolodchikov:1995aa}.  
The Virasoro conformal block ${\cal F}_{c,\Delta_p}^{(\sigma)}[\Delta_i](\sf Z)$  
on $C_{g,n}$ depends on the following quantities:
the cross ratios of the vertex operators  locations denoted symbolically by ${\sf Z}$,
the external conformal weights $\lbrace\Delta_i\rbrace_{i=1,\ldots,n}$,
the intermediate conformal weights $\lbrace\Delta_p\rbrace_{p=1,\ldots,3g-3+n}$ 
and the central charge $c$.

On the other hand, the partition function 
$Z^{(\sigma)}_{T_{g,n}}$ can be written as the integral over the holomorphic 
times the anti-holomorphic Nekrasov partition functions \cite{Nekrasov:2002qd,Nekrasov:Okounkov:2003}:
$$
Z^{(\sigma)}_{T_{g,n}} =
\int [da]\,{\cal Z}_{\rm Nekrasov}^{(\sigma)}\,\bar{\cal Z}_{\rm Nekrasov}^{(\sigma)},
$$
where $[da]$ is some appropriate measure. The Nekrasov partition function can be written as a product
of three factors
${\cal Z}_{\rm Nekrasov}\!=\!{\cal Z}_{\rm class}{\cal Z}_{\rm 1-loop}{\cal Z}_{\rm inst}$.
The first two factors 
${\cal Z}_{\rm class}{\cal Z}_{\rm 1-loop}=:{\cal Z}_{\rm pert}$ describe the contribution coming 
from perturbative calculations. Supersymmetry implies that there are contributions to 
${\cal Z}_{\rm pert}$ only at the
tree- (${\cal Z}_{\rm class}$) and 1-loop (${\cal Z}_{\rm 1-loop}$) levels. 
${\cal Z}_{\rm inst}$ is the instanton contribution. The Nekrasov partition function 
${\cal Z}_{\rm Nekrasov}(\tilde q, \tilde a, \tilde m,\epsilon_1,\epsilon_2)$
depends on the set of parameters: $\tilde q$, $\tilde a$, $\tilde m$, $\epsilon_1$, $\epsilon_2$.
The components of $\tilde q\!=\!\lbrace\exp 2\pi\tau_1, \ldots, \exp 2\pi\tau_{3g-3+n}\rbrace$ are
the gluing parameters associated with the pant decomposition of $C_{g,n}$, where the
$\tau_p \!=\! \frac{\theta_p}{2\pi} + \frac{4\pi i}{g_{p}^{2}}$ are the complexified gauge couplings. 
The multiplet $\tilde m\!=\!\lbrace m_{1},\ldots, m_{n}\rbrace$ contains the mass parameters.
Moreover, $\tilde a\!=\!\lbrace a_{1},\ldots,a_{3g-3+n}\rbrace$, where the $a$'s are the vacuum 
expectation values of the scalar fields in the vector multiplets. 
Finally, $\epsilon_1$, $\epsilon_2$ represent the complex $\Omega$-background parameters.

Comparing the integral representations of both sides of eq.~(\ref{agt}) 
it is possible, thanks to AGT hypothesis, to identify separately  in the holomorphic 
and anti-holomorphic sectors the Virasoro conformal blocks 
${\cal F}_{c,\Delta_p}[\Delta_i](\sf Z)$ on $C_{g,n}$  and the instanton sectors
${\cal Z}_{\rm inst}$ of the Nekrasov partition functions for
the super Yang--Mills theories $T_{g,n}$.

Soon after its discovery, the AGT conjecture has been extended to the $2d$ conformal 
Toda/$4d$ SU(N) gauge theories correspondence \cite{Wyllard:2009hg,Mironov:2009by},
and to the so-called `nonconformal' cases \cite{Gaiotto:2009,Marshakov:2009,Alba:2009fp} 
(see also \cite{Tan:2013tq,Hadasz:2010xp,Gaiotto:2012sf,Maulik:2012wi}), 
which will be of main interest in the present work.

The AGT correspondence works at the level of the quantum Liouville
field theory. It is intriguing to ask, however, what happens
if we proceed to the semiclassical limit of the Liouville correlation functions. 
This is the limit in which the central charge $c$,
the external $\Delta_i$ and intermediate $\Delta_p$ conformal weights tend to infinity in such a way
that their ratios are fixed $\Delta_p/c=\Delta_i/c={\rm const.}$, cf.~\cite{Zamolodchikov:1995aa}. 
For the standard parametrization of the central charge $c=1+6{\sf Q}^2$, 
where ${\sf Q}=b+\frac{1}{b}$ and for {\it heavy} weights 
$(\Delta_p,\Delta_i)=\frac{1}{b^2}(\delta_p,\delta_i)$ with $\delta_p,\delta_i={\cal O}(b^0)$,
the classical limit corresponds to $b\to 0$.
It is commonly believed that in the classical limit the conformal 
blocks behave exponentially with respect to ${\sf Z}$:
$$
{\cal F}\;\stackrel{b\to 0}{\sim}\;{\rm e}^{\frac{1}{b^2}f}.
$$
The function $f$ is known as the {\it classical conformal block.}

The AGT correspondence dictionary says that $b=\!\!\sqrt{\epsilon_2/\epsilon_1}$.
Therefore, the semiclassical limit $b\to 0$ of the conformal blocks
corresponds to the so-called {\it Nekrasov--Shatashvili limit}
$\epsilon_2\to 0$ ($\epsilon_1$ being kept finite) 
of the Nekrasov partition functions. 
In \cite{NS:2009} it was observed that in the limit $\epsilon_2\to 0$ 
the Nekrasov partition functions have the following asymptotical behavior:
\begin{equation}\label{NSlimit}
{\cal Z}_{{\rm Nekrasov}}(\,\cdot\,,\epsilon_1, \epsilon_2)
\stackrel{\epsilon_2\to 0}{\sim}
\exp\left\lbrace\frac{1}{\epsilon_2}\,
W(\,\cdot\,,\epsilon_1)\right\rbrace,
\end{equation}
where $W(\,\cdot\,,\epsilon_1)=W_{{\rm pert}}(\,\cdot\,,\epsilon_1)
+W_{{\rm inst}}(\,\cdot\,,\epsilon_1)$ is
the {\it effective twisted superpotential}
of the corresponding two-dimensional gauge theories
restricted to the two-dimensional $\Omega$-background.

\begin{figure}[t] \label{fig1}
\centering
\includegraphics[width=400pt]{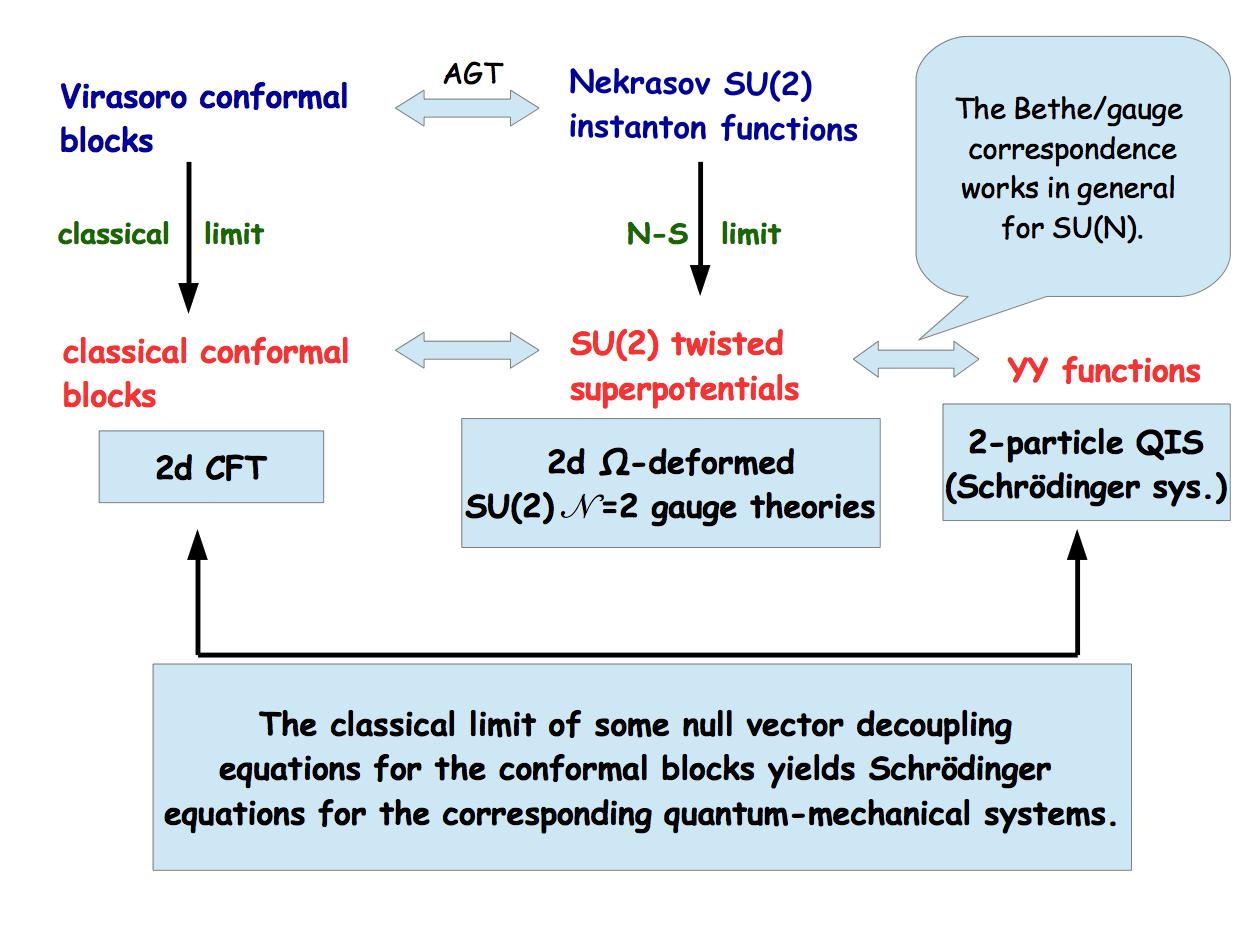}

\caption{The triple correspondence in the case of the Virasoro classical conformal blocks
links the latter to the SU(2) instanton twisted superpotentials which describe the spectra of some
quantum--mechanical systems. The Bethe/gauge correspondence on the r.h.s.~connects
the SU(N) ${\cal N}=2$ SYM theories with the N--particle quantum integrable systems. 
An extension of the above triple relation to the case ${\rm N}>2$ needs to consider on the l.h.s.~the 
classical limit of the $W_{\rm N}$ symmetry conformal blocks according to the known extension 
\cite{Wyllard:2009hg} of the AGT conjecture.}
\end{figure}

The twisted superpotentials play a pivotal role in the already mentioned 
{\it Bethe/gauge correspondence} \cite{NS:2009,Nekrasov:2009uh,Nekrasov:2009ui} 
which maps supersymmetric vacua of the ${\cal N}\!=\!2$ theories 
to Bethe states of quantum integrable systems (QIS's). 
A result of that duality is that the twisted superpotentials are identified with
the Yang--Yang (YY) functions \cite{YY} which describe the spectra of some QIS's.
Therefore, combining both the classical/Nekrasov--Shatashvili limit of 
the AGT duality and the Bethe/gauge correspondence one thus 
gets a triple correspondence which connects the classical blocks 
with the twisted superpotentials and then with the Yang--Yang functions (cf.~Fig.1).

For example, the twisted superpotentials
for the ${\cal N}\!=\!2$ SU(N) $N_f=0$ (pure gauge) and the ${\cal N}\!=\!2^*$ 
SU(N) SYM theories determine respectively the spectra 
of the N--particle periodic Toda (pToda) and the 
elliptic Calogero--Moser (eCM) models \cite{NS:2009}. In the case of the SU(2) gauge group these 
QIS's are simply quantum--mechanical systems whose dynamics is described by some 
Schr\"{o}dinger equations. Concretely, for the 2--particle pToda
and eCM models these Schr\"{o}dinger equations correspond to
the celebrated Mathieu and Lam\'{e} equations 
with energy eigenvalues expressed in terms of the twisted superpotentials. 
This correspondence can be used to investigate 
nonperturbative effects in the Mathieu and Lam\'{e} quantum--mechanical systems, cf.~\cite{Basar:2015xna}.
On the other hand, the Mathieu and Lam\'{e} equations emerge entirely within the framework of
$2d$ CFT as the classical limit of the null vector decoupling (NVD) equations for
the 3--point degenerate irregular block and for the 2--point block (projected 2--point function)
on the torus with one degenerate {\it light} operator \cite{Piatek:2014lma,Maruyoshi:2010,Piatek:2013ifa}.
It turns out that the classical irregular block $f_{\rm irr}$ and the classical
1--point block on the torus $f_{\rm torus}$ determine the spectra of the Mathieu and Lam\'{e}
operators in the same way as their gauge theory counterparts, 
i.e.:~$W_{\rm inst}^{{\rm SU(2)}, N_f=0}$ and $W_{\rm inst}^{{\rm SU(2)},{\cal N}=2^*}$.
Therefore, it seems that there is a way to study the spectrum of the Mathieu and Lam\'{e} operators  
using two-dimensional conformal field theory methods.\footnote{For interesting questions which can be
studied in this way, see the conclusions of the present work.}
However, in order to exploit this possibility it is necessary to establish a full correspondence
between the Mathieu and Lam\'{e} equations and their realizations within $2d$ CFT.
The missing element is to understand how the solutions of the equations
obtained in the classical limit from the NVD equations 
are connected to the eigenfunctions of the Mathieu and Lam\'{e} operators.
It is also important to know what kind of solutions are possible to be obtained.
An answer to these questions in the case of the Mathieu equation is our 
main goal in the present paper.

The organization of the paper is as follows. In section \ref{sec2} the
necessary tools of $2d$ CFT are introduced. In section \ref{sec3}
the simplest irregular blocks are defined and some of their properties are described.
In particular, an exponentiation of the pure gauge irregular block
within the classical limit is proved at the leading order.
After that, the NVD equations for certain degenerate irregular 
blocks are derived. Section \ref{sec4} is devoted to the derivation of
the Mathieu equation within the formalism of $2d$ CFT. The calculation presented there provides 
formulas for the Mathieu eigenvalue and the related eigenfunction 
in terms of the classical limit of irregular blocks.
It is shown that these formulas reproduce the well known noninteger order weak coupling 
expansion of the Mathieu eigenvalue and the corresponding Mathieu function.
In subsection \ref{subsect} a factorization property of 
the degenerate irregular block with the light operator and its representation in the classical limit
as a product of light and heavy parts is proved at the leading order. 
This factorization property is crucial for deriving
the Mathieu equation. Section \ref{sec5} contains our conclusions. 
In particular, the problems that are still open and 
the possible extensions of the present work are discussed.

\section{Conformal blocks in the operator formalism}
\label{sec2}
\subsection{Chiral vertex operators}
Starting from the Belavin--Polyakov--Zamolodchikov axioms \cite{Belavin:1984vu}, Moore and Seiberg 
\cite{Moore:1988qv}
have constructed formalism of the so-called {\it rational conformal field theories} 
(RCFT's),\footnote{A very similar formalism can be found in \cite{FFK}.} where
\begin{itemize}
\item[---] the operator algebra of local fields contains purely holomorphic subalgebra
${\cal A}$ called {\it chiral} or {\it vertex algebra};
\item[---] the Hilbert space of states of the theory is a direct sum of irreducible 
representations of the algebra ${\cal A}\oplus{\cal A}$:
\begin{equation}\label{Hilbert}
{\cal H}\;=\;\bigoplus\limits_{i=1}^{N}\;{\cal U}_i \otimes {\cal U}_i\;.
\end{equation}
\end{itemize}
In RCFT's the summation  in (\ref{Hilbert}) is over a
discrete finite set. However, one can generalize and successfully 
apply the Moore--Seiberg formalism to the case of 
two-dimensional conformal field theories with continuous spectrum, 
cf.~e.g.~\cite{Teschner:2001rv,Teschner:2003en}. 
In such a case the direct sum in eq.~(\ref{Hilbert}) becomes a direct integral.

In any 2d CFT there exist at least two chiral fields, i.e., the identity operator and its 
descendant --- the holomorphic component of the energy-momentum tensor 
$T(z)=\sum_{n\in\mathbb{Z}}z^{-n-2}L_{n}$.
Therefore, each chiral algebra ${\cal A}$ contains as a subalgebra 
the Virasoro algebra
$\textrm{Vir}=\bigoplus_{n\in\mathbb{Z}}\mathbb{C}\,L_{n}\,\bigoplus \,\mathbb{C}\,c$,
\begin{equation}
\label{Virasoro}
[L_n,L_m ] \;=\; (n-m) L_{n+m} +{c\over 12 }(n^3-n)\delta_{n+m,0} \ .
\end{equation}

In the Moore--Seiberg formalism the `physical' fields of \cite{Belavin:1984vu} are built 
out of more fundamental objects --- the so-called {\it chiral vertex operators} (CVO's). 
These are intertwining operators acting between representations of the vertex algebra.
In the present paper we confine ourselves to the simplest case when ${\cal A}=\!{\rm Vir}$
and define CVO's as operators acting between Verma modules.

Let ${\cal V}_{c, \Delta}^{\,n}$ be the free vector space generated by  all vectors of the form
\begin{equation}
\label{Basis}
|\,\nu^n_{\Delta,I}\,\rangle=L_{-I}|\,\nu_{\Delta}\,\rangle = 
L_{-k_{1}}\ldots L_{-k_{j-1}}L_{-k_{j}} |\,\nu_{\Delta}\,\rangle
\end{equation}
where
$I=(k_{1},\ldots, k_{j-1}, k_{j})$ is an ordered ($k_{1}\geq \ldots\geq
k_{j}\geq 1$) sequence of positive integers of the length $|I|\equiv k_{1}+\ldots+k_{j}=n$,
and $|\,\nu_{\Delta}\,\rangle$ is the highest weight vector:
\begin{equation}
\label{HW}
L_{0}|\,\nu_{\Delta}\,\rangle=
\Delta|\,\nu_{\Delta}\,\rangle,\;\;\;\;\;\; 
L_{n}|\,\nu_{\Delta}\,\rangle= 0\;\;\;\;\;\forall\;n>0\;.
\end{equation}
The $\mathbb Z$-graded representation of the Virasoro algebra
determined on the space:
$$
{\cal V}_{c,\Delta}\;=\;\bigoplus_{n=0}^{\infty}{\cal V}_{c,\Delta}^{\,n}
$$
by the relations (\ref{Virasoro}) and (\ref{HW}) is called
the Verma module of the central charge $c$ and the highest weight $\Delta$.
The dimension of the subspace ${\cal V}_{c,\Delta}^{\,n}$ of all homogeneous elements of degree $n$
is given by the number  $p(n)$ of partitions of $n$ (with the convention $p(0)=1$).
It is an eigenspace of $L_{0}$ with the eigenvalue $\Delta+n$.

On ${\cal V}_{c,\Delta}^{n}$ there exists the symmetric
bilinear form $\langle\,\cdot\,|\,\cdot\,\rangle$
uniquely defined by the relations
$$
\langle\,\nu_{\Delta}\,|\,\nu_{\Delta}\,\rangle\;=\;1\;\;\;{\rm and}\;\;\;
(L_n)^{\dagger}\;=\;L_{-n}.
$$
The Gram matrix $G_{c,\Delta}$ of the form $\langle\,\cdot\,|\,\cdot\,\rangle$
is block-diagonal in the basis 
$\left\lbrace|\,\nu_{\Delta, I}\,\rangle\right\rbrace$ with blocks
$$
\Big[G_{c,\Delta}^{n}\Big]_{IJ}\;=\;
\langle\, \nu_{\Delta, I}^n\,|\,\nu_{\Delta, J}^n\,\rangle
\;=\;\langle\, \nu_{\Delta}\,|(L_{-I})^{\dagger}L_{-J}|\,\nu_{\Delta}\,\rangle.
$$
In particular, one finds
\begin{itemize}
\item[---]
$n=1$: $\lbrace L_{-1}|\,\nu_\Delta\,\rangle \rbrace$,
\begin{eqnarray*}
G^{n=1}_{c,\Delta} = \langle L_{-1}\nu_\Delta\,|\,L_{-1}\nu_\Delta\rangle
=\langle\nu_\Delta\,|\,L_{1}L_{-1}\nu_\Delta\rangle = 2\Delta,
\end{eqnarray*}
\item[---]
$n=2$: $\lbrace L_{-2}|\,\nu_\Delta\,\rangle,\;L_{-1} L_{-1}|\,\nu_\Delta\,\rangle\rbrace$,
\begin{equation*}
G^{n=2}_{c,\Delta} =
\begin{pmatrix}
\langle L_{-2}\nu_\Delta\,|\,L_{-2}\nu_\Delta\rangle  &  \langle L_{-1}^2\nu_\Delta\,|\,L_{-2}\nu_\Delta\rangle \\
\langle L_{-2}\nu_\Delta\,|\,L_{-1}^{2}\nu_\Delta\rangle & \langle L_{-1}^{2}\nu_\Delta\,|\,L_{-1}^{2}\nu_\Delta\rangle  \\
\end{pmatrix}
=
\begin{pmatrix}
    \frac{c}{2}+4\Delta  & 6\Delta \\
    6\Delta & 4\Delta(2\Delta+1) \\
\end{pmatrix},
\end{equation*}
\item[---]
$n=3$: $\lbrace L_{-3}|\,\nu_\Delta\,\rangle,\;
L_{-2}L_{-1}|\,\nu_\Delta\,\rangle,\; L_{-1}L_{-1}L_{-1}|\,\nu_\Delta\,\rangle\rbrace$,
\begin{equation*}
G^{n=3}_{c,\Delta}=
  \begin{pmatrix}
    2c+6\Delta  & 10\Delta & 24\Delta \\
    10\Delta & \Delta(c+8\Delta+8) & 12\Delta(3\Delta+1) \\
    24\Delta & 12\Delta(3\Delta+1) & 24\Delta(\Delta+1)(2\Delta+1) \\
  \end{pmatrix}.
\end{equation*}
\end{itemize}

The Verma module ${\cal V}_{c,\Delta}$ is irreducible if and only if the form 
$\langle\,\cdot\,|\,\cdot\,\rangle$ is non-degenerate.
The criterion for irreducibility is vanishing of the determinant $\det G_{c,\Delta}^{n}$
of the Gram matrix, known as the Kac determinant, given by the formula 
\cite{K1,FF1,FF2,Th,KW,KR}:
\begin{equation}
\label{det} \textrm{det}\; G^{\,n}_{c,\Delta} \;=\; C_n
\prod_{\substack{r,s\in\,\mathbb{N},\\ s\leq r \\ 1\leq rs\leq n  } } \Phi_{rs}(c,\Delta)^{p(n-rs)}.
\end{equation}
In the equation above $C_n$ is a constant and
$$
\label{phi} \Phi_{rs}(c,\Delta)\;=\; \textstyle \left( \Delta
+{r^2-1\over 24}(c-13) + {rs-1\over 2}\right) \left( \Delta
+{s^2-1\over 24}(c-13) + {rs-1\over 2}\right) +{(r^2-s^2)^2\over
16}.
$$
The Kac determinant vanishes for
\begin{eqnarray*}
\Delta_{rs}(c) &=& \frac{(13 - c)(r^2 + s^2)+ \sqrt{(c-25)(c-1)}
(r^2
- s^2)- 24 rs - 2 + 2c}{48},\nonumber\\[8pt]
&& r, s\in\mathbb{Z},\;\;\;r\geq 1,\;s\geq 1,\;\;\;1\leq rs\leq n
\end{eqnarray*}
or
\begin{eqnarray*}
\label{lad}
c_{rs}(\Delta) &=& 13 - 6 \left(T_{rs}(\Delta)+{1\over T_{rs}(\Delta)}\right),\\[5pt]
T_{rs}(\Delta) &=& \frac{rs -1 + 2\,\Delta  +
    {\sqrt{{\left( r - s \right) }^2 +
        4\,\left(r\,s -1 \right) \,\Delta  +
        4\,{\Delta }^2}}}{  r^2-1},\nonumber\\[8pt]
&& r, s\in\mathbb{Z},\;\;\;r\geq 2,\;s\geq 1,\;\;\;1\leq rs\leq
n.\nonumber
\end{eqnarray*}
For these values of $\Delta$ and $c$ the representations ${\cal V}_{c,\Delta_{rs}(c)}$ or ${\cal
V}_{c_{rs}(\Delta),\Delta}$ are reducible.

The set $\left\lbrace \Delta_{rs}(c) \right\rbrace$ of the degenerate conformal weights 
can be parametrized as follows
\begin{equation}\label{degenerate}
\Delta_{rs}(c) \;=\; \Delta_0 + \frac{\beta^2_{rs}}{4},
\;\;\;\;\;\;\;\;\;\;\;\;\;
\beta_{rs} \;=\; r\beta_{+} + s\beta_{-},
\end{equation}
where
$$
\beta_{\pm}(c) =
\frac{\sqrt{1-c}\pm\sqrt{25-c}}{2\sqrt{6}},
\;\;\;\;\;\;\;\;\;\;\;\;
\Delta_0 = -\frac{1}{4}(\beta_{+} + \beta_{-})^{2}=\frac{c-1}{24}.
$$
Sometimes, it is also convenient to use the alternative parametrization:\footnote{
Here $\beta_{+}\!\left(1+6\left(b+\frac{1}{b}\right)^{\!2}\right)=ib$ and 
$\beta_{-}\!\left(1+6\left(b+\frac{1}{b}\right)^{\!2}\right)\;=\;\frac{i}{b}$.}
\begin{equation}
\label{wagadegeneratb} \Delta_{rs}(c) \;=\; \frac{{\sf Q}^2}{4}-
\frac{1}{4}\left(rb + \frac{s}{b} \right)^{2}
\end{equation}
for which the central charge is given by $c =1+6{\sf Q}^2$ with ${\sf Q}=b + b^{-1}$.

The non-zero element $|\,\chi_{rs}\,\rangle\in{\cal V}_{c,\Delta_{rs}(c)}$ of degree 
$n=rs$ is called a null vector if
$\label{zerowydef} L_0\,|\,\chi_{rs}\,\rangle =(\Delta_{rs} +
rs)\,|\,\chi_{rs}\,\rangle$,
and
$L_{k}\,|\,\chi_{rs}\,\rangle = 0$, $\forall\,k>0$.
Hence, $|\,\chi_{rs}\,\rangle$ is the highest weight state which generates its own Verma
module ${\cal V}_{c,\Delta_{rs}(c)+rs}$,
which is a submodule of ${\cal V}_{c,\Delta_{rs}(c)}$. One can prove that
each submodule of the Verma module ${\cal V}_{c,\Delta_{rs}(c)}$ is generated by
a null vector. Then,  the module ${\cal V}_{c,\Delta_{rs}(c)}$ is irreducible if and only if
it does not contain null vectors with positive degree.

For non-degenerate values of $\Delta$, i.e.~for $\Delta\neq\Delta_{rs}(c)$, there
exists in ${\cal V}_{c,\Delta}^{n}$ the
`dual' basis $\lbrace|\,\nu^{t,n}_{\Delta, I}\,\rangle\rbrace$
whose elements are defined by the relation
$
\langle\,\nu^{t,n}_{\Delta, I}\,|\,\nu^{n}_{\Delta, J}\,\rangle=
\delta_{IJ}
$
for all
$|\,\nu^{n}_{\Delta, J}\,\rangle
\in\lbrace|\,\nu_{\Delta, J}^{n}\,\rangle\rbrace$.
The dual basis vectors $|\,\nu^{t,n}_{\Delta,I}\,\rangle$ have
the following representation in the standard basis
$$
|\,\nu^{t,n}_{\Delta, I}\,\rangle=\sum_{J,|J|=n}
\Big[ G_{c,\Delta}^n\Big]^{IJ} |\,\nu^{n}_{\Delta, J}\,\rangle,
$$
where $\Big[ G_{c,\Delta}^n\Big]^{IJ}$ is the inverse of the Gram matrix $\Big[
G_{c,\Delta}^{n}\Big]_{IJ}$.

Let ${\cal V}_{\Delta}$
be the Verma module with the highest weight state $|\,\nu_\Delta\,\rangle$.
The chiral vertex operator is the linear map
$$
V{^{\Delta_3}_\infty}{^{\Delta_2}_{\:z}}{^{\Delta_1}_{\;0}} :
{\cal V}_{\Delta_2} \otimes {\cal V}_{\Delta_1}
\to
{\cal V}_{\Delta_3}
$$
such that for all $|\,\xi_2\,\rangle \in {\cal V}_{\Delta_2}$ the operator
$$
V(\xi_2 | z) \equiv
V{^{\Delta_3}_\infty}{^{\Delta_2}_{\:z}}{^{\Delta_1}_{\;0}}
(|\,\xi_2\,\rangle\otimes\,\cdot\,):{\cal V}_{\Delta_1}
\to
{\cal V}_{\Delta_3}
$$
satisfies the following conditions
\begin{eqnarray}
\label{CVO}
\left[L_n , V\!\left(\nu_{2}|z\right)\right] &=& z^{n}\left(z
\frac{\partial}{\partial z} + (n+1)\Delta_2
\right) V\!\!\left(\nu_{2}|z\right)\,,\;\;\;\;\;\;\;\;n\in\mathbb{Z}
\\
\label{CVO2}
V\!\!\left(L_{-1}\xi_{2}|z\right) &=& \frac{\partial}{\partial z}V\!\!\left(\xi_{2}|z\right),
\\
\label{CVO3}
V\!\!\left(L_{n}\xi_{2}|z\right) &=& \sum\limits_{k=0}^{n+1}
\left(\,_{\;\;k}^{n+1}\right)
(-z)^{k}\left[L_{n-k}, V\!\!\left(\xi_{2}|z\right)\right]\,,
\;\;\;\;\;\;\;\;n>-1,
\\
V\!\!\left(L_{-n}\xi_{2}|z\right) &=&\sum\limits_{k=0}^{\infty}
\left(\,_{\;\;n-2}^{n-2+k}\right)
z^k\,L_{-n-k}\,V\!\!\left(\xi_{2}|z\right)
\nonumber
\\
\label{CVO5}
&+& (-1)^{n}\sum\limits_{k=0}^{\infty}
\left(\,_{\;\;n-2}^{n-2+k}\right)
z^{-n+1-k}\,\,V\!\!\left(\xi_{2}|z\right)\,L_{k-1},
\;\;\;\;\;n>1
\end{eqnarray}
and
$$
\left\langle\,\nu_{\Delta_3}\,|V\!\left(\nu_{\Delta_2}
|z\right)|\,\nu_{\Delta_1}\,\right\rangle
\;=\;z^{\Delta_3 -\Delta_{2}-\Delta_1}.
$$
The commutation relation (\ref{CVO}) defines the primary vertex operator
corresponding to the highest weight state 
$|\,\nu_2\,\rangle\in{\cal V}_{\Delta_2}$.
Eqs.~(\ref{CVO2})--(\ref{CVO5}) characterize the decendant CVO's.

\subsection{The 3-point block}
For a given triple $\Delta_1,\Delta_2, \Delta_3$ of conformal weights
we define the trilinear map
$$
\rho{^{\Delta_3}_\infty}{^{\Delta_2}_{\:z}}{^{\Delta_1}_{\;0}} :
{\cal V}_{\Delta_3}\otimes {\cal V}_{\Delta_2} \otimes {\cal V}_{\Delta_1}
\to
\mathbb{C}
$$
induced by the matrix element of a single chiral vertex operator
$$
\rho{^{\Delta_3}_\infty}{^{\Delta_2}_{\:z}}{^{\Delta_1}_{\;0}}(\xi_3,\xi_2,\xi_1)
\;=\;
\left\langle\,\xi_3\,|V\!\!\left(\xi_2
|z\right)|\,\xi_1\,\right\rangle,
\;\;\;\;\;\;\;\;
\forall\;|\,\xi_i\,\rangle \in {\cal V}_{\Delta_i},
\;\;\;\;
i=1,2,3.
$$
The form $\displaystyle \rho{^{\Delta_3}_\infty}{^{\Delta_2}_{\:z}}{^{\Delta_1}_{\;0}}$
is uniquely determined by the conditions (\ref{CVO})-(\ref{CVO5}). In particular, 
\begin{enumerate}
\item
for $L_0$-eingenstates\footnote{Note that for the
basis vectors $\lbrace |\,\nu_{i, I}\,\rangle \rbrace$
one has $\Delta_i(\nu_{i, I}) = \Delta_i + |I|$.}
$
L_0|\,\xi_i\,\rangle  = \Delta_i (\xi_i)|\,\xi_i\,\rangle\;,\;i=1,2,3
$
one gets
\begin{equation}
\label{z_dep}
\rho{^{\Delta_3}_\infty}{^{\Delta_2}_{\:z}}{^{\Delta_1}_{\;0}}
(\xi_3,\xi_2,\xi_1)
=
z^{\Delta_3(\xi_3)- \Delta_2(\xi_2)- \Delta_1(\xi_1) }
\rho{^{\Delta_3}_\infty}{^{\Delta_2}_{\:1}}{^{\Delta_1}_{\;0}}
(\xi_3,\xi_2,\xi_1)\;;
\end{equation}
\item
for basis vectors $\nu_{i, I}\equiv|\,\nu_{\Delta_i, I}\,\rangle\in{\cal V}_{\Delta_i}$,  
$i=1,2,3$ one finds
\begin{eqnarray}
\nonumber
\rho{^{\Delta_3}_\infty}{^{\Delta_2}_{\:1}}{^{\Delta_1}_{\;0}}
(\nu_{3,I},\nu_2,\nu_1)
&=& \gamma_{ \Delta_3}\!\left[\,^{\Delta_2}_{\Delta_1}\right]_I\;,
\\[8pt]
\label{ggg}
\rho{^{\Delta_3}_\infty}{^{\Delta_2}_{\:1}}{^{\Delta_1}_{\;0}}
(\nu_3,\nu_2,\nu_{1,I})
&=& \gamma_{ \Delta_1}\!\left[\,^{\Delta_2}_{\Delta_3}\right]_I\;,
\\[8pt]
\nonumber
\rho{^{\Delta_3}_\infty}{^{\Delta_2}_{\:1}}{^{\Delta_1}_{\;0}}
(\nu_3,\nu_{2,I},\nu_1)
&=& (-1)^{|I|}\gamma_{ \Delta_2}\!\left[\,^{\Delta_1}_{\Delta_3}\right]_I\;,
\end{eqnarray}
where for a given partition $I= (k_1,\dots,k_{\ell(I)}),\ k_i\geq k_j\geq 1,\ i<j$,
 \begin{equation}
\label{gamma}
\gamma_{ \Delta}\!\left[\,^{\Delta_2}_{\Delta_1}\right]_I\;\equiv \;
\prod\limits_{i=1}^{\ell(I)}\left(\Delta +k_i\Delta_2 -\Delta_1 +\sum_{i<j}^{\ell (I)}k_j \right).
\end{equation}
\end{enumerate}
In terms of the trilinear form $\rho$ (3-point block) one can spell out 
an important result known as the 
{\bf null vector decoupling theorem} (Feigin--Fuchs \cite{FF3}):\footnote{Here 
we closely follow \cite{Teschner:2001rv}.}

\medskip\noindent
{\it
Let $i,j,k\in\lbrace 1,2,3\rbrace$ be chosen such that $j\neq i$, $k\neq i$, $j\neq k$.
Let us assume that
\begin{itemize}
\item[($i$)] 
$\Delta_i=\Delta_{rs}(c)\equiv\frac{1}{24}(c-1)+\frac{1}{4}\beta_{rs}^{2}$,
$r, s\in\mathbb{Z}_{>0}$ (cf.~parametrization (\ref{degenerate})) and
\item[($ii$)]
the vector $|\,\xi_i\,\rangle$ lies in the singular submodule generated by 
the null vector $|\,\chi_{rs}\,\rangle$, i.e.:
$
|\,\xi_i\,\rangle\in{\cal V}_{c,\Delta_{rs}(c)+rs}\subset{\cal V}_{c,\Delta_{rs}(c)}.
$
\end{itemize}
Then, 
\fbox{$
\rho{^{\Delta_3}_{z_3}}{^{\Delta_2}_{\:z_{2}}}{^{\Delta_1}_{\;z_{1}}}
(\xi_3,\xi_2,\xi_1)=0
$}
if and only if 
$$
\Delta_j=\Delta_{\beta_j}\equiv\frac{1}{24}(c-1)+\frac{1}{4}\beta_{j}^{2}
\;\;\;\;\;\;and\;\;\;\;\;\;
\Delta_k=\Delta_{\beta_k}\equiv\frac{1}{24}(c-1)+\frac{1}{4}\beta_{k}^{2}
$$
satisfy the fusion rules 
$\beta_j - \beta_k = \beta_{pq}$,
where
$p\in\lbrace 1-r, 3-r,\ldots, r-1\rbrace$ and 
$q\in\lbrace 1-s, 3-s,\ldots, s-1\rbrace$.
}

\section{Quantum and classical zero flavor irregular blocks}
\label{sec3}
\subsection{Definition and basic properties}
To begin with, let us consider the following (coherent) vector in the
Verma module ${\cal V}_{c,\Delta}$ discovered by D.~Gaiotto in  
\cite{Gaiotto:2009} and constructed by A.~Marshakov, A.~Mironov and A.~Morozov 
in \cite{Marshakov:2009}:\footnote{With some abuse of nomenclature, 
we will call `zero flavor' both the Gaiotto state
and the irregular block. The reason for that is that the irregular block corresponds 
to the Nekrasov instanton function of  the ${\cal N}=2$ ($\Omega$-deformed) 
pure gauge (zero flavor $N_f=0$) 
super Yang--Mills theory, in accordance with the `non-conformal' extension of the AGT conjecture, 
see below.}
\begin{eqnarray}
\label{GaiottoPure}
|\,\Delta,\Lambda^2\,\rangle &=&
\sum\limits_{I}\Lambda^{2|I|}
\left[G^{|I|}_{c,\Delta}\right]^{(1^{|I|}) I}L_{-I}|\,\nu_{\Delta}\,\rangle\nonumber
\\
&=&
\sum\limits_{n=0}^{\infty}\Lambda^{2n}\sum\limits_{I,|I|=n}
\Big[G^{n}_{c,\Delta}\Big]^{(1^{n}) I}|\,\nu^n_{\Delta,I}\,\rangle\;.
\end{eqnarray}
The summation in eq.~(\ref{GaiottoPure}) runs over all partitions
or equivalently over their pictorial representations --- Young diagrams. The symbol $(1^{|I|})$
in eq.~(\ref{GaiottoPure}) denotes a single--row Young diagram, where the total number of boxes $|I|=n$
equals the number of columns $\ell(I)$, i.e. $\ell(I)=|I|=n$.

In \cite{Marshakov:2009} it was shown that the vector (\ref{GaiottoPure}) 
obeys the Gaiotto defining conditions: 
\begin{equation}
L_{0}|\Delta,\Lambda^2\rangle =
\left(\Delta+\frac{\Lambda}{2}\frac{\partial}{\partial\Lambda}\right)|\Delta,\Lambda^2\rangle,
\;\;\;
L_{1}|\Delta,\Lambda^2\rangle = \Lambda^2 |\Delta,\Lambda^2\rangle,
\;\;\;
L_{n}|\Delta,\Lambda^2\rangle = 0 \;\;\forall\;n\geq 2.
\end{equation}

The zero flavor $N_f=0$ qunatum irregular block is defined as the inner 
product of the Gaiotto state \cite{Gaiotto:2009,Marshakov:2009}:
\begin{eqnarray}\label{FNf0}
{\cal F}_{c,\Delta}(\Lambda) &=&
\langle\,\Delta, \Lambda^2\,|\,\Delta,\Lambda^2\,\rangle 
=\sum\limits_{n=0}^{\infty}\Lambda^{4n}\,
\Big[ G_{c,\Delta}^{n}\Big]^{(1^n)(1^n)}
\\
&=& 1+
\Lambda^{4}\;\frac{1}{2\Delta} +
\Lambda^{8}\;\frac{c+8 \Delta }{4 \Delta (2 c \Delta +c+2 \Delta (8 \Delta -5))}
\\
&+&
\Lambda^{12}\;\frac{(11 c-26) \Delta +c (c+8)+24 \Delta ^2}
{24 \Delta \left((c-7) \Delta +c+3 \Delta ^2+2\right)
 \left(2 (c-5) \Delta +c+16 \Delta ^2\right)}+\ldots\;.
\end{eqnarray}
In fact, there are much more 
Gaiotto's states and therefore irregular blocks.\footnote{
See for instance \cite{Felinska:2011tn} and refs.~therein.}
In the present paper we confine ourselves to study irregular 
blocks which are built out of (\ref{GaiottoPure}).
Possible extensions of the present work taking into 
account the existence of the other Gaiotto states 
will be discussed soon in a forthcoming publication.\footnote{Cf.~conclusions.}

Let $C_{g,n}$ denotes a Riemann surface with genus $g$ and $n$ punctures.
Let $x$ be the modular parameter of the 4-punctured Riemann sphere $C_{0,4}$.
Then, the $s$-channel conformal block on $C_{0,4}$ is defined as the 
following formal $x$-expansion:
\begin{equation}
\label{four-pointblock}
{\cal F}_{c,\Delta}\!\left[_{\Delta_{1}\;\Delta_{4}}^{\Delta_{2}\;\Delta_{3}}\right]\!(x)=
x^{\Delta-\Delta_{3}-\Delta_{4}}\left( 1 +
\sum_{n=1}^\infty x^{\,n}\,
{\cal F}^{\,n}_{c,\Delta}\!\left[_{\Delta_{1}\;\Delta_{4}}^{\Delta_{2}\;\Delta_{3}}\right] \right),
\end{equation}
where
\begin{eqnarray}\label{4pBC}
{\cal F}^{\,n}_{c,\Delta}\!\left[_{\Delta_{1}\;\Delta_{4}}^{\Delta_{2}\;\Delta_{3}}\right]
&=&
\sum\limits_{|I|=|J|=n}
\rho{^{\Delta_1}_\infty}{^{\Delta_2}_{\:1}}{^{\Delta}_{\,0}}
(\nu_{\Delta_1},\nu_{\Delta_2},\nu_{\Delta, I})
\Big[ G_{c,\Delta}^{n}\Big]^{IJ}
\rho{^{\Delta}_\infty}{^{\Delta_3}_{\:1}}{^{\Delta_4}_{\,0}}
(\nu_{\Delta, J},\nu_{\Delta_3},\nu_{\Delta_4}) \nonumber
\\ &=&
\sum\limits_{|I|=|J|=n}
\gamma_{\Delta}\!\left[\,^{\Delta_2}_{\Delta_1}\right]_I
\Big[ G_{c,\Delta}^{n}\Big]^{IJ}
\gamma_{\Delta}\!\left[\,^{\Delta_3}_{\Delta_4}\right]_J\;.
\end{eqnarray}
Let $q=\textrm{e}^{2\pi i \tau}$  be the elliptic variable on the torus
with modular parameter $\tau$, then
the conformal block on $C_{1,1}$ is given by the following formal $q$-series:
\begin{equation*}
 {\cal F}_{c,\Delta}^{\tilde\Delta}(q)=q^{\Delta-\frac{c}{24}}
\left(1+\sum\limits_{n=1}^{\infty}{\cal F}_{c,\Delta}^{\tilde\Delta,n}q^n \right),
\end{equation*}
where
\begin{equation*}
\label{torusCoeff} 
\mathcal{F}^{\tilde\Delta, n}_{c,\Delta}=
\sum\limits_{|I|=|J|=n}
\rho{^{\Delta\;}_\infty}{^{\tilde\Delta\;}_{\:1}}{^{\Delta}_{\,0}}
(\nu_{\Delta, I},\nu_{\tilde\Delta},\nu_{\Delta, J})
\;\Big[ G_{c,\Delta}^{n}\Big]^{IJ}.
\end{equation*}

The irregular block (\ref{FNf0}) can be recovered from the conformal blocks
on the torus and on the sphere in a properly defined decoupling limit of the
external conformal weights \cite{Marshakov:2009,Alba:2009fp}. 
Indeed, employing the AGT inspired parametrization 
of the external weights $\tilde\Delta$, $\Delta_i$ and the central charge $c$, i.e.:
\begin{eqnarray*}\label{AGTparametrization}
\tilde\Delta=\frac{M \left(\epsilon-M\right)}{\epsilon _1 \epsilon _2},
\;\;\;\;\;\;
\Delta_i = \frac{\alpha_i(\epsilon - \alpha_i)}{\epsilon_1\epsilon_2}\,,
\;&&\;
c = 1 + 6\frac{\epsilon^2}{\epsilon_1\epsilon_2},
\;\;\;\;
\epsilon = \epsilon_1 + \epsilon_2\;,
\\
\alpha_1 = \tfrac{1}{2}\left(\epsilon+\mu_1-\mu_2\right), 
\;\;\;\;
\alpha_2 = \tfrac{1}{2}\left(\mu_1+\mu_2\right),
&&
\alpha_3 = \tfrac{1}{2}\left(\mu_3 + \mu_4\right), 
\;\;\;\;
\alpha_4 = \tfrac{1}{2}\left(\epsilon+\mu_3-\mu_4\right),
\end{eqnarray*}
and introducing the dimensionless expansion parameter
$\Lambda=\hat\Lambda/(-\epsilon_1\epsilon_2)^{\frac{1}{2}}$ 
it is possible to prove the following limits \cite{Marshakov:2009,Alba:2009fp}:
\begin{eqnarray}
q^{\frac{c}{24}-\Delta}\,{\cal F}_{c,\Delta}^{\tilde\Delta}(q)
&\xrightarrow[qM^4=\hat\Lambda^4]{M\,\to\,\infty}&
{\cal F}_{c,\Delta}(\Lambda),\nonumber
\\[8pt]
\label{decoupling}
x^{\Delta_3+\Delta_4-\Delta}\,
{\cal F}_{c,\Delta}\!\left[_{\Delta_{1}\;\Delta_{4}}^{\Delta_{2}\;\Delta_{3}}\right]\!(x)
&\xrightarrow[x\mu_1\mu_2\mu_3\mu_4=\hat\Lambda^4]{\mu_1,\mu_2,\mu_3,\mu_4\,\to\,\infty}&
{\cal F}_{c,\Delta}(\Lambda).
\end{eqnarray}

Due to the `non-conformal' AGT relation, the $N_f\!=\!0$ irregular block can be expressed through 
the SU(2) pure gauge Nekrasov instanton partition function  
\cite{Gaiotto:2009,Hadasz:2010xp,Tan:2013tq,Maulik:2012wi}:
\begin{equation}\label{AGT0}
{\cal F}_{c,\Delta}(\Lambda) = 
{\cal Z}_{\rm inst}^{{\rm SU(2)}, N_f=0}(\hat\Lambda, a, \epsilon_1, \epsilon_2).
\end{equation}
The identity~(\ref{AGT0}), which in particular is understood as term by term 
equality between the coefficients of the
expansions of both sides, holds for
\begin{equation}\label{para1}
\Lambda=\frac{\hat\Lambda}{\sqrt{-\epsilon_1\epsilon_2}},
\;\;\;\;\;\;\;\;\;
\Delta = \frac{\epsilon^2 - 4a^2}{4\epsilon_1\epsilon_2},
\;\;\;\;\;\;\;\;\;
c=1 + 6\frac{\epsilon^2}{\epsilon_1\epsilon_2}
\equiv 1+6{\sf Q}^2
\end{equation}
where
\begin{equation}
\label{para2}
{\sf Q}\;=\;b+\frac{1}{b}\;\equiv\;\sqrt{\frac{\epsilon_2}{\epsilon_1}} + \sqrt{\frac{\epsilon_1}{\epsilon_2}}
\;\;\;\;\Leftrightarrow\;\;\;\; b=\sqrt{\frac{\epsilon_2}{\epsilon_1}}.
\end{equation}

In \cite{NS:2009} it was observed that in the limit $\epsilon_2\to 0$ 
the Nekrasov partition functions ${\cal Z}_{\rm Nekrasov}={\cal Z}_{\rm pert}{\cal Z}_{\rm inst}$
behave exponentially. In particular, for the instantonic sector we have
\begin{equation}\label{asymZ}
\mathcal{Z}_{\rm inst}(\,\cdot\,,\epsilon_1,\epsilon_2)\;\stackrel{\epsilon_2\to 0}{\sim}\;
\exp\left\lbrace \frac{1}{\epsilon_2}\,W_{\rm inst}(\,\cdot\,,\epsilon_1)\right\rbrace.
\end{equation}
Therefore, taking into account the AGT relation~(\ref{AGT0}), the fact that 
$b\!=\!(\frac{\epsilon_2}{\epsilon_1})^{\frac{1}{2}}$ and the Nekrasov--Shatashvili 
limit (\ref{asymZ}) of the instanton function, one can expect that
the irregular block has the following exponential behavior in the limit $b\to 0$:
\begin{equation}\label{ClIrrblock0}\boxed{
{\cal F}_{1+6Q^2,\Delta}(\Lambda)
\;\stackrel{b\to 0}{\sim}\;
\exp\left\lbrace\frac{1}{b^2}f_{\delta}^{\bf 0}\!\left(\hat\Lambda/\epsilon_1\right)\right\rbrace}
\;\;,
\end{equation}
where $\Delta=\frac{1}{b^2}\delta$, $\delta={\cal O}(b^0)$.
The semiclassical asymptotical behavior (\ref{ClIrrblock0}) is a
very nontrivial statement concerning the quantum $N_f=0$ irregular block.\footnote{Cf.~considerations 
in subsection \ref{Towards a proof of classical asymptotic} and conclusions of the present paper.}
First, the existence of the {\it classical zero flavor irregular block} 
$f_{\delta}^{\bf 0}\!\left(\hat\Lambda/\epsilon_1\right)$
can be checked by direct calculation.
Indeed, from the power expansion of the quantum irregular 
block (\ref{FNf0}) and eq.~(\ref{ClIrrblock0}) one finds
\begin{equation}\label{ClIrrexp0}
f_{\delta}^{\bf 0}\!\left(\hat\Lambda/\epsilon_1\right) \;=\;
\lim\limits_{b\to 0} b^2 \log {\cal F}_{1+6Q^2, \frac{1}{b^2}\delta}
\!\left(\hat\Lambda/(\epsilon_1 b)\right)\;=\;
\sum\limits_{n=1}^{\infty}
\left(\hat\Lambda/\epsilon_1\right)^{\!4n}\!\!f_{\delta}^{{\bf 0},n},
\end{equation}
where the coefficients $f_{\delta}^{{\bf 0},n}$ up to $n=6$ take the form:
\begin{eqnarray}\label{coeff0}
&&f_{\delta}^{{\bf 0},1}=\frac{1}{2 \delta},
\;\;\;\;\;\;\;
f_{\delta}^{{\bf 0},2}=\frac{5 \delta -3}{16 \delta ^3 (4 \delta +3)},
\;\;\;\;\;\;\;
f_{\delta}^{{\bf 0},3}=
\frac{9 \delta ^2-19 \delta +6}{48 \delta ^5 \left(4 \delta ^2+11 \delta +6\right)},\nonumber
\\[5pt]
&&f_{\delta}^{{\bf 0},4}=
\frac{5876\, \delta ^5-16489\, \delta ^4-22272\, \delta ^3+17955\, \delta ^2+9045\, \delta -4050}
{512\, \delta ^7 (\delta +2) (4 \delta +3)^3 (4 \delta +15)},\nonumber
\\[5pt]
&&f_{\delta}^{{\bf 0},5}=
\frac{17884\, \delta ^6-96187\, \delta ^5-156432\, \delta ^4+388737\, \delta ^3-7317\, \delta ^2-138348\, \delta +34020}{1280\, \delta ^9 (\delta +2) (\delta +6) (4
   \delta +3)^3 (4 \delta +15)},\nonumber
\\[5pt]
&&f_{\delta}^{{\bf 0},6}=
\left[7756224\, \delta ^{11}-19228160\, \delta ^{10}-456215812\, \delta ^9\right.\nonumber
\\
&&\left.\hspace{80pt}
-971240994\, \delta ^8+1505016987\, \delta ^7+5076827496\, \delta ^6\right.\nonumber
\\
&&\left.\hspace{80pt}
+930371157\, \delta
   ^5-4398704919\, \delta ^4-1494083556\, \delta ^3\right.\nonumber
\\
&&\left.\hspace{80pt}
   +1212636096\, \delta ^2+293932800\, \delta -128595600\,\right]\nonumber
\\
&&\hspace{80pt}\times
   \left[6144\, \delta ^{11} (\delta +2)^3 (\delta +6) (4
   \delta +3)^5 (4 \delta +15) (4 \delta +35)\right]^{-1}.
\end{eqnarray}

As a further consistency check of our approach,
let us observe that combining~(\ref{AGT0})-(\ref{para2}) and 
(\ref{ClIrrblock0}) it is possible to identify 
the classical irregular block with the SU(2) $N_f=0$ effective twisted superpotential:
\begin{equation}\label{ClassAGT}
f_{\delta}^{\bf 0}\!\left(\hat\Lambda/\epsilon_1\right) =
\frac{1}{\epsilon_1}\,W_{\rm inst}^{{\rm SU(2)},\,N_f=0}\!
\left(\hat\Lambda, a,\epsilon_1\right),
\end{equation}
where $\delta=\frac{1}{4}-\frac{a^2}{\epsilon_{1}^2}$.
We stress that the {\it classical conformal weight} 
$\delta$ in eq.~(\ref{ClassAGT}) above is expressed 
in terms of the gauge theory parameters $a$, $\epsilon_1$. Indeed, it
is easy to see that 
$$
\delta = \lim\limits_{b\to 0}b^2\Delta 
= \lim\limits_{\epsilon_2\to 0}\frac{\epsilon_2}{\epsilon_1}\Delta
= \frac{1}{4}-\frac{a^2}{\epsilon_{1}^2}.
$$
By comparison of the expansion (\ref{ClIrrexp0})-(\ref{coeff0})
with that of the twisted superpotential obtained independently from the instanton
partition function, the identity (\ref{ClassAGT}) may be confirmed up to desired order.

\subsection{Towards a proof of the classical limit}
\label{Towards a proof of classical asymptotic}
The equations \eqref{coeff0} constitute a direct premise for the existence of the
classical irregular conformal block.
The rigorous proof of this statement, however, has not yet been performed, 
although there are many convincing arguments in favor 
of its validity.\footnote{See the discussion in the conclusions on this point.} 
In what follows we discuss the leading order of the coefficients of the quantum
irregular block and extend the discussion beyond the leading order to provide yet more arguments for 
the existence of the classical irregular block.

\subsubsection*{Classical irregular block at the leading order}
In order to find the leading contribution to the classical irregular block
we examine the coefficients of the expansion of the quantum irregular block.
Since these are functions of the matrix elements of the
Virasoro algebra, we analyze their dependence on $c$ and $\dl$
to find out how they scale with respect to $b$ within the classical limit.

The quantum irregular block can be rewritten explicitly as
\begin{equation}
\label{irregular_block}
{\mathcal{F}}_{c,\Delta}(\Lambda) = \sum_{n\geq 0} 
\cbr{\frac{\hat\Lambda}{\e_1 b }}^{4n} \cbr{G_{c,\dl}^{(n)}}^{(1^n)\, (1^n)} 
= \sum_{n\geq 0}  \cbr{\frac{\hat\Lambda}{\e_1 b }}^{4n} \frac{M_{p(n),p(n)}(\dl,c)}{\det G_{c,\dl}^{(n)}},
\end{equation}
where $M_{p(n),p(n)}(\dl,c)$ is the greatest principal minor of the Kac determinant at the level $n$ 
$(\ket{\dl}\equiv\ket{\nu_{\dl}})$\footnote{We use 
conventions as in eq. \eqref{Basis}. $I \vdash n$ means that $I$ is a partition of $n$. }
\begin{equation}
\label{KacDet}
\det G_{c,\dl}^{(n)} = \det\cbr{\bra{\dl} L_{I_u}L_{-I_v} \ket{\dl} },
\qquad I_u,I_v \vdash  n,\quad u,v\in\{1,\ldots,p(n)\}.
\end{equation}
As it was mentioned earlier the matrix elements $\bra{\dl} L_{I_u}L_{-I_v} \ket{\dl}$ 
are polynomials in $\dl$, and $c$. Although in the classical limit both parameters are important
we restrict our attention to $\dl$ dependence. The reason is that, due to Virasoro algebra \eqref{Virasoro},
$\dl$ appears as a factor in a matrix element $\bra{\dl} L_{I_u}L_{-I_v} \ket{\dl}$ 
either additively accompanied by $c$ or alone, which takes place when $I_u$ and $I_v$ 
have part one in common with nonzero multiplicity. In order to find the greatest power of $\dl$ 
in general matrix element at level $n$ we take advantage of the argument used by Kac and Raina in ref. \cite{KR}. 
Let us first consider the diagonal matrix element. Making use of the following notation: $\ell_u := \ell(I_u)$,   
$$
I_u = (k_1(I_u),\ldots,k_{\ell_u }(I_u), 0 ,\dots ) 
= (1^{m_{1}(I_u)}\, 2^{m_{2}(I_u)}\ldots ),\ |I_u| 
= \sum\limits_{s\geq 1} k_s(I_u) = \sum\limits_{i\geq 1} i\, m_{i}(I_u)= n,
$$
as well as Virasoro algebra \eqref{Virasoro} we obtain for the arbitrary diagonal matrix element 
\begin{equation}
\label{general_matrix_elem}
\begin{aligned}
\bra{\dl}L_{I_u}L_{-I_u}\ket{\dl} 
&= \bra{\dl}L_{i_l}^{m_{i_{l}}(I_u)}\ldots L_{i_2}^{m_{i_{2}}(I_u)}L_{i_1}^{m_{i_{1}}(I_u)}
L_{-i_1}^{m_{i_{1}}(I_u)}L_{-i_2}^{m_{i_{2}}(I_u)}\ldots L_{-i_l}^{m_{i_{l}}(I_u)}\ket{\dl}
\\
&\overset{\deg}{\sim} \prod_{s=1}^{l} \bra{\dl}L_{i_s}^{m_{i_{s}}(I_u)}L_{-i_s}^{m_{i_{s}}(I_u)}\ket{\dl}
= \prod_{i=1}^{l} \bra{\dl}L_{i}^{m_{i}(I_u)}L_{-i}^{m_{i}(I_u)}\ket{\dl},
\end{aligned}
\end{equation}
which, in accord with the formula
\begin{equation}
\label{commutator_formula1}
\begin{gathered}
\underbrace{[L_{i},\ldots,[L_{i}}_{m},L_{-i}^{m}]\ldots] = \cbr{2L_{0} + \tfrac{c}{12}(i^{2}-1) ; i }_{m},
\\
(a;k)_{n} := \prod_{i=0}^{n-1}(a+ k i),
\qquad (a;1)_{n} = (a)_{n} := \Gamma(a+n)\big/\Gamma(a),
\end{gathered}
\end{equation} 
amounts to
\begin{equation}
\label{matrix_element_uu}
\bra{\dl}L_{I_u}L_{-I_u}\ket{\dl} 
\overset{\deg}{\sim} 
\prod_{i\geq1} i^{m_{i}(I_u)} m_{i}(I_u)! \left( 2\dl + \tfrac{c}{12}(i^{2}-1);i \right)_{m_{i}(I_u)}.
\end{equation} 
The symbol `$\overset{\deg}{\sim}$' indicates that a polynomial on the left hand side
and the one on the right hand side are equal up to the term with the greatest power in variable $\dl$ which, by definition,
determines its degree. Let us now consider arbitrary off-diagonal term $\bra{\dl}L_{I_u}L_{-I_v}\ket{\dl}$. 
Let us assume that both partitions have, say, $N$ parts $\{i_s\}_{s=1}^{N}$ in common with nonzero multiplicities 
$m_{i_s}(I_u)$ and $m_{i_s}(I_{v})$. Then, repeating the above reasoning, one finds that the general off-diagonal 
element takes the form
\begin{displaymath}
\bra{\dl}L_{I_u}L_{-I_v}\ket{\dl} 
= \bra{\dl}L_{i_p}^{m_{i_{p}}(I_u)}\cdots L_{i_N}^{m_{i_{N}}(I_u)}\cdots L_{i_1}^{m_{i_{1}}(I_u)}
L_{-i_1}^{m_{i_{1}}(I_v)}\cdots L_{-i_N}^{m_{i_{N}}(I_v)}\ldots L_{-i_q}^{m_{i_{q}}(I_v)}\ket{\dl}.
\end{displaymath}
Using the generalized formula in eq. \eqref{commutator_formula1} for $m\leq n$
\begin{equation}
\label{commutator_formula2}
\begin{aligned}
\underbrace{[L_{i},\ldots,[L_{i}}_{m},L_{-i}^{n}]\ldots] &= L_{-i}^{n-m} i^{m}
\frac{n!}{(n-m)!}\prod_{s=n-m}^{n-1}\cbr{2L_{0} + \frac{c}{12}(i^{2}-1) +i s }
\\
&=  L_{-i}^{n-m} i^{m} \frac{n!}{(n-m)!} 
\frac{\cbr{2L_{0} + \tfrac{c}{12}(i^{2}-1) ; i }_{n} }{\cbr{2L_{0} + \tfrac{c}{12}(i^{2}-1) ; i }_{n-m}} ,
\end{aligned}
\end{equation}
we get
\begin{equation}
\label{matrix_element_uv}
\bra{\dl}L_{I_u}L_{-I_v}\ket{\dl} \overset{\deg}{\sim}
\prod_{i\geq 1} i^{\min\{m_{i}(I_u),m_{i}(I_{v})\}}
\frac{m_{i}(I_{v})!}{\vartheta\big(m_{i}(I_{v})-m_{i}(I_u)\big)!}
\frac{\cbr{2\dl + \tfrac{c}{12}(i^{2}-1) ; i }_{m_{i}(I_{v})} }
{\cbr{2\dl + \tfrac{c}{12}(i^{2}-1) ; i }_{\vartheta\left( m_{i}(I_{v})-m_{i}(I_{u}) \right)}},
\end{equation} 
where $\vartheta(x) = x\th(x)$ and $\th(x)$ is Heaviside's theta function.
These general results in eqs. \eqref{matrix_element_uu} and \eqref{matrix_element_uv}  
allows us to draw the following conclusions for the matrix elements
as a polynomials in $\dl$ and $c$. For any $I_u,I_v \vdash  n$
\begin{enumerate}
\item $\deg_\dl\bra{\dl}L_{I_u}L_{-I_v}\ket{\dl} \leq \min\{\ell_u,\ell_v\},$

\item $\deg_\dl\bra{\dl}L_{I_u}L_{-I_u}\ket{\dl} = \ell_u,$

\item for $u\neq v$ and $\ell_u = \ell_v$, $\deg_\dl\bra{\dl}L_{I_u}L_{-I_v}\ket{\dl} < \ell_u.$
\end{enumerate}
Since the degree of a matrix element as a polynomial in $\dl$ depends on the length of partition
those of greatest degree yield the leading contribution to both $M_{p(n),p(n)}$ and $\det G^{(n)}_{\dl,c}$ 
within the classical limit. In the following discussion the explicit form of particular matrix elements prove useful 
\begin{equation}
\label{particular_matrix_el}
\begin{gathered}
\bra{\dl}L^n_{1}L^n_{-1}\ket{\dl} = n! (2\dl)_n,
\\
\bra{\dl}L_{I_u}L^n_{-1}\ket{\dl}\! =\! n! \prod_{j\geq1}\Big( (k_j(I_u)\!+\!1)\dl\!+\!\sum_{s>j} k_s(I_u) \Big)
\!=\! n! \prod_{i\geq1}\Big( (i\!+\!1)\dl\! +\! \sum_{s>i} s\, m_{s}(I_u); i \Big)_{m_{i}(I_u)} .
\end{gathered}
\end{equation}
The above results enable to conclude that ($L^n_{-1} = L_{-(1^n)}=L_{-I_{p(n)}}$)
\begin{equation}
\label{deg2deg}
\deg_\dl\bra{\dl}L_{I_u}L_{-I_u}\ket{\dl} = \deg_\dl\bra{\dl}L_{I_u}L_{-I_{p(n)}}\ket{\dl}.
\end{equation}
Moreover, let us note that the minor produced by crossing out
the $i^{\text{th} }$ diagonal element when treated as a polynomial in $\dl$ and $c$ has the
same coefficient in the highest degree term as the product of diagonal elements of the Gram matrix, 
namely 
\begin{equation}
\label{Mii_sim_prodDiag}
M_{u,u}(\dl,c)  \overset{\deg}{\sim} \frac{\prod\limits_{j=1}^{p(n) } \bra{\dl,c}L_{I_j}L_{-I_j}\ket{\dl,c} }
{\bra{\dl,c}L_{I_u}L_{-I_u}\ket{\dl,c}} ,
\end{equation}
With these results at hand we can proceed to estimate the contribution to the classical conformal block
at the leading order within the limit $b\to 0$. In order to do this we expand the Kac determinant along
the $p(n)^{\text{th} }$ row
\begin{equation}
\label{det_expansion}
\frac{\det G_{c,\dl}^{(n)}}{M_{p(n),p(n)}(\dl,c)} = \bra{\dl}L^n_{1}L^n_{-1}\ket{\dl} 
+ \sum_{u=1}^{p(n)-1}(-1)^{p(n)+u}\frac{M_{u,p(n)}(\dl,c) }{M_{p(n),p(n)}(\dl,c)} \bra{\dl}L_{I_u}L^n_{-1}\ket{\dl} .
\end{equation}
Let us observe that in view of our analysis concerning matrix elements as polynomials in $\dl$ 
the leading contribution of the latter to $M_{u,p(n)}$ can be found as follows.
By means of the elementary operations on columns we obtain
\begin{equation}
\label{step1}
\begin{aligned}
M_{u,p(n)} =& \det(c_1,\dots,c_{u-1},c_{u+1},\dots,c_{p(n)-1},c_{p(n)}) 
\\
=& (-1)^{p(n)-u}\det(c_1,\dots,c_{u-1},c_{p(n)},c_{u+1},\dots,c_{p(n)-1})
\\
= & (-1)^{p(n)-u} \tilde M_{u,p(n)},
\end{aligned}
\end{equation}
where $c_{u}\equiv \lbrace\bra{\dl}L_{I_j}L_{-I_u}\ket{\dl}\rbrace_{j=1}^{p(n)-1}$ 
denotes $i^{\rm th}$ column of $G^{(n)}$ with the last entry removed. Using
eq. \eqref{Mii_sim_prodDiag} we find that 
\begin{equation}
\label{step2}
\tilde M_{u,p(n)} 
\overset{\deg}{\sim} \frac{\bra{\dl}L_{I_u}L^n_{-1}\ket{\dl}}{\bra{\dl}L_{I_u}L_{-I_u}\ket{\dl}} M_{p(n),p(n)},
\end{equation}
which, when placed in eq. \eqref{step1}, yields the leading contribution to $M_{u,p(n)}$. 
Combining eqs. \eqref{step1}, \eqref{step2} and \eqref{det_expansion} we obtain 
\begin{equation}
\label{det_expansion_final}
\frac{\det G_{c,\dl}^{(n)}}{M_{p(n),p(n)}(\dl,c)} 
\overset{\deg}{\sim} \bra{\dl}L^n_{1}L^n_{-1}\ket{\dl} 
+ \sum_{u=1}^{p(n)-1}\frac{\bra{\dl}L_{I_u}L^n_{-1}\ket{\dl}}{\bra{\dl}L_{I_u}L_{-I_u}\ket{\dl}} 
\bra{\dl}L_{I_u}L^n_{-1}\ket{\dl} .
\end{equation}
Within the classical limit $\dl\to\infty, \ c\to \infty, \ c/\dl = \textrm{const.}$ for $b\to 0$
the conformal weight and the central charge scale as $\dl\sim\d/b^2$ and $c\sim 6/b^2$. From 
eqs. \eqref{particular_matrix_el} and \eqref{general_matrix_elem} we infer that 
\begin{gather}
\nonumber
\bra{\dl}L^n_{1}L^n_{-1}\ket{\dl}\, \overset{b\to 0}{\sim}\, b^{-2n} n! (2\d)^n , 
\qquad
\bra{\dl}L_{I_u}L^n_{-1}\ket{\dl}\, \overset{b\to 0}{\sim}\, b^{-2\ell_u} n! \prod_{i\geq1}\big( (i+1)\d\big)^{m_{i}(I_u)} ,
\\
\nonumber
\bra{\dl}L_{I_u}L_{-I_u}\ket{\dl}\, \overset{b\to 0}{\sim}\, b^{-2\ell_u} 2^{-\ell_u}
\prod_{i\geq1}m_{i}(I_u)! i^{m_{i}(I_u)} \big(4\d+i^{2}-1\big)^{m_{i}(I_u)},
\end{gather}
and
\begin{equation}
\label{ratio}
\frac{\bra{\dl}L_{I_u}L^n_{-1}\ket{\dl}}{\bra{\dl}L_{I_u}L_{-I_u}\ket{\dl}}
\,\overset{b\to 0}{\sim}\,
\frac{n!}{\prod\limits_{i\geq1}m_{i}(I_u)!} 
\prod_{i\geq1}\left( \frac{2 (i+1)\d}{i \cbr{4\d + i^2-1 }} \right)^{m_{i}(I_u)}.
\end{equation} 
Hence, the formula in eq. \eqref{det_expansion_final} within the classical limit 
amounts to 
\begin{equation*}
\frac{\det G_{c,\dl}^{(n)}}{M_{p(n),p(n)}(\dl,c)} 
\overset{b\to 0}{\sim}
 n! (2\d)^n b^{-2n}
+ n!\! \sum_{u=1}^{p(n)-1} (2\d)^{\ell_u} b^{-2\ell_u}
\frac{n!}{\prod\limits_{i\geq1}m_{i}(I_u)!}
\prod_{i\geq1}\left( \frac{2 (i\!+\!1)^{2}\d}{i \cbr{4\d\! +\! i^2\!-\!1 }} \right)^{m_{i}(I_u)}.
\end{equation*}
Since $n = \ell_{\text{max} }$ then it is seen that the first term dominates over the rest for $b\to0$.
Therefore, the coefficient of the irregular block in eq. \eqref{irregular_block} 
within this limit reads
\begin{equation}
\label{GramElementClassLimit}
\cbr{G_{c,\dl}^{(n)} }^{(1^n)\, (1^n)} = \frac{M_{p(n),p(n)}(\dl,c)}{\det G_{c,\dl}^{(n)}}
\overset{b\to 0}{\sim}\frac{b^{2n}}{n! (2\d)^n},
\end{equation}
This, in accord with eq. \eqref{ClIrrexp0}, yields the first coefficient $f^{\mathbf{0},1}_\d$
of the classical irregular block expansion given in eq. \eqref{coeff0}, namely
$$
{\mathcal{F}}_{c,\Delta}(\Lambda) 
= \sum_{n\geq 0} \cbr{\frac{\hat\Lambda}{\e_1 b }}^{4n} \cbr{G_{c,\dl}^{(n)}}^{(1^n)\, (1^n)} 
\overset{b \to 0}{\sim} \sum_{n\geq 0} \cbr{\frac{\hat\Lambda}{\e_1 b }}^{4n} 
\frac{b^{2n}}{n! (2\d)^n}  = \exp\pbr{\frac{1}{b^2} \cbr{\frac{\hat\Lambda}{\e_1}}^{4} \frac{1}{2\d}}.
$$

\subsubsection*{Classical irregular block beyond the leading order}
The above computations show that in the estimations of the quantum irregular block
coefficients based on the leading order contribution all but the first term in the 
classical irregular block expansion are neglected. 
Therefore a more accurate analysis is required. In general the sought expression takes the form
\begin{equation}
\label{general_expansion}
f_{\delta}^{\bf 0}\!\left(\hat\Lambda/\epsilon_1\right) 
= \lim_{b\to 0} b^2 \log\sbr{1+ \sum_{n\geq 1}\cbr{\frac{\hat\Lambda}{\e_1 b }}^{4n} 
\mathcal{F}^{(n)}(\dl,c) }.
\end{equation}
where for the sake of brevity we have introduced notation 
$\mathcal{F}^{(n)} \equiv \cbr{G^{(n)}}^{(1^n)\, (1^n)}$. 
The logarithm in eq. \eqref{general_expansion} has the following expansion
\begin{multline}
\log\sbr{1+ \sum_{n\geq 1}\cbr{\frac{\hat\Lambda}{\e_1 b }}^{4n} \mathcal{F}^{(n)}(\dl,c) }
\\
= \sum_{n\geq 1} \cbr{\frac{\hat\Lambda}{\e_1 }}^{4n} 
\sum_{\substack{\{m_i\}\geq 0 \\ \sum i m_i = n } }
(-1)^{\sum m_i +1} \cbr{\sum m_i -1}! \prod_{i\geq 1} 
\frac{\sbr{b^{-4 i}\mathcal{F}^{(i)}(\d/b^2,6/b^2)}^{m_i} }{m_i!}.
\end{multline}
In order to find the limit of the above coefficient of $\hat\Lambda/\e_1 $
the knowledge of $\mathcal{F}^{(i)}$ is necessary.
Unfortunately the exact form of $M_{p(n),p(n)}$ as a polynomial in $\dl$ and $c$ is not known and it is 
necessary to compute it term by term which is the major obstacle in finding the limit.
In order for the limit in eq. \eqref{general_expansion} to exist each coefficient should be proportional to $b^2$.
Thus the complete rigorous proof of the mentioned limit is still an open 
problem which in order to be solved must be attacked in
fact by another methods.\footnote{Cf.~conclusions.}

\subsection{The null vector decoupling equations}
In this subsection we shall derive the partial differential equations 
obeyed by the {\it $N_f=0$ degenerate irregular blocks}, cf.~\cite{Maruyoshi:2010}. 
We define the latters as matrix elements of the degenerate chiral vertex operators
$V_{\pm}(z)\!=\!V(|\nu_{\Delta_\pm}\rangle|z)$ between the states (\ref{GaiottoPure}):
\begin{eqnarray}\label{P00}
\Psi^{{\bf 0}}_{\pm}(\Lambda, z) &=&
\langle\,\Delta', \Lambda^2\,|
V_{\pm}(z)|\,\tilde\Delta,\Lambda^2\,\rangle\nonumber
\\[3pt]
&=&
\rho{^{\Delta'}_\infty}{^{\Delta_{\pm}}_{\:z}}{^{\tilde\Delta}_{\;0}}
\left(|\Delta', \Lambda^2\rangle\,,\,|\nu_{\Delta_{\pm}}\rangle\,,\,|\tilde\Delta,\Lambda^2\rangle\right).
\end{eqnarray}
In the above equation:
$$
\Delta_{+} \!\equiv\! \Delta_{21}= -\frac{3}{4}b^2-\frac{1}{2},
\;\;\;\;\;\;\;\;
\Delta_{-} \!\equiv\! \Delta_{12}= -\frac3{4 b^2}-\frac12.
$$
Moreover, in order to apply the null vector decoupling theorem we will assume that the weights 
$\Delta_1\equiv\tilde\Delta$ and $\Delta_3\equiv\Delta'$ of the {\it in} and {\it out} states
are related by the fusion rule:\footnote{In the parameterization   
$\Delta_{\beta_i}=\frac{1}{24}(c-1)+\frac{1}{4}\beta_i$ (see also (\ref{degenerate}))
used in the NVD theorem the fusion rule (\ref{FR}) reads as follows:
$\Delta_{\beta_3}=\Delta_{\beta_1-\beta_{+}}$. In another commonly used
parametrization, in which $\Delta(\alpha_i)=\alpha_i ({\sf Q}-\alpha_i)$, we have 
$\Delta(\alpha_3)=\Delta(\alpha_1+\frac{b}{2})$.}
\begin{equation}\label{FR}
\Delta_1\equiv\tilde\Delta=\Delta\!\left(\sigma-\tfrac{b}{4}\right), 
\;\;\;\;\;\;\;\;
\Delta_3\equiv\Delta'=\Delta\!\left(\sigma+\tfrac{b}{4}\right),
\;\;\;\;{\rm where}\;\;\;\;
\Delta(\sigma)\equiv\frac{{\sf Q}^2}{4}-\sigma^2.
\end{equation}
Let us consider the descendant chiral vertex operator
\begin{equation}\label{nullfield}
\chi_{+}(z)=\left(\widehat{L}_{-2}(z) -
\frac{3}{2(2\Delta_{+} +1)}
\,\widehat{L}_{-1}^{\,2}(z)\right)V_{+}(z)
\equiv
V\!\left(\left(L_{-2}+\tfrac{1}{b^2}L_{-1}^{2}\right)\!|\,\nu_{\Delta_+}\,\rangle\,|\,z\,\right)
\end{equation}
which corresponds to the null vector 
$$
|\,\chi_{+}\,\rangle =
\chi_{+}(0)|\,0\,\rangle =
\left(L_{-2}+\frac{1}{b^2}L_{-1}^{2}\right)\!|\,\nu_{\Delta_+}\,\rangle
$$
appearing at the second
level of the Verma module ${\cal V}_{\Delta_+}$.
Then, by the NVD theorem, we have that
\begin{equation}\label{rho1}
\langle\,\Delta', \Lambda^2\,|\,
\chi_{+}(z)\,|\,\tilde\Delta,\Lambda^2\,\rangle
\;=\;
\rho{^{\Delta'}_\infty}{^{\Delta_+}_{\:z}}{^{\tilde\Delta}_{\;0}}
\left(|\Delta', \Lambda^2\rangle\,,\,
|\,\chi_{+}\,\rangle\,,
\,|\tilde\Delta,\Lambda^2\rangle\right)
\\
\;=\;0.
\end{equation}
In order to convert eq.~(\ref{rho1}) to the PDE obeyed by the degenerate irregular block 
$\Psi^{{\bf 0}}_{+}(\Lambda, z)$,
one needs to employ the following Ward identity:
\begin{eqnarray}\label{CWI0}
\langle\,\Delta', \Lambda^2\,|
\,T(w)V_{+}(z)\,|\,\tilde\Delta,\Lambda^2\,\rangle
&=&
\left[
\frac{z}{w(w-z)}\frac{\partial}{\partial z} + \frac{\Delta_+}{(w-z)^2}
+\left(\frac{\Lambda^{2}}{w}+\frac{\Lambda^2}{w^3}\right)\right.\nonumber
\\
&&\hspace{-50pt}+\left.
\frac{1}{2 w^2}\left(\frac{\Lambda}{2}\frac{\partial}{\partial \Lambda}
+\tilde\Delta+\Delta'-\Delta_{+}-z\frac{\partial}{\partial z}\right)
\right]\Psi^{\bf 0}_{+}(\Lambda, z)\,.
\end{eqnarray} 
Using the formula \cite{Belavin:1984vu}:
$$
\widehat{L}_{-k}(z)\;=\;\frac{1}{2\pi i}\oint\limits_{C_z}dw (w-z)^{1-k}\,T(w)
$$
it is now possible to compute the matrix element 
$\langle\,\Delta', \Lambda^2\,|
\,\widehat{L}_{-2}(z)V_{+}(z)\,|\,\tilde\Delta,\Lambda^2\,\rangle$
with the help of eq.~(\ref{CWI0}). Applying Cauchy's integral formula one 
finds that
\begin{eqnarray}
\label{L20}
\langle\,\Delta', \Lambda^2\,|
\,\widehat{L}_{-2}(z)V_{+}(z)\,|\,\tilde\Delta,\Lambda^2\,\rangle
&=&
\left[
-\frac{1}{z}\frac{\partial}{\partial z}
+\left(\frac{\Lambda^{2}}{z}+\frac{\Lambda^2}{z^3}\right)\right.\nonumber
\\
&&\hspace{-65pt}+\left.
\frac{1}{2 z^2}\left(\frac{\Lambda}{2}\frac{\partial}{\partial \Lambda}
+\tilde\Delta+\Delta'-\Delta_{+}-z\frac{\partial}{\partial z}\right)
\right]\Psi^{\bf 0}_{+}(\Lambda, z)\,.
\end{eqnarray}
Finally, taking into account that the matrix element 
of the descendant operator $\widehat{L}_{-1}^{2}(z)V_{+}(z)$ 
yields $\partial^{2}_{z}\Psi^{\bf 0}_{+}(\Lambda, z)$, we get
from (\ref{nullfield}), (\ref{rho1})   
(\ref{L20}) the desired partial differential equation, determining $\Psi^{\bf 0}_{+}(\Lambda, z)$:
\begin{equation}\label{NVDeq0}
\left[\frac{1}{b^2}\,z^2\frac{\partial^2}{\partial z^2}-
\frac{3z}{2}\frac{\partial}{\partial z}+\Lambda^2 \left(z+\frac{1}{z}\right)
+\frac{\Lambda}{4}\frac{\partial}{\partial\Lambda}+
\frac{\tilde\Delta+\Delta'-\Delta_+}{2}\right]\Psi^{\bf 0}_{+}(\Lambda, z)\;=\;0\,.
\end{equation}
Replacing $\Delta_{+}$ with $\Delta_{-}$ and repeating all the steps described above
one can get an analogous equation for $\Psi^{{\bf 0}}_{-}(\Lambda, z)$.
In the next section we will consider the limit $b\to 0$ of eq.~(\ref{NVDeq0}).
A part of this analysis has been already done in our previous work \cite{Piatek:2014lma}.
The new result here is the derivation from the degenerate zero flavor irregular block
of the formula for the eigenfunction of the Mathieu operator.

\section{The classical irregular block and the spectrum of the Mathieu operator}
\label{sec4}
\subsection{The classical limit of the null vector decoupling equation}
Let us turn for a while to the zero flavor degenerate irregular block 
introduced in eq.~(\ref{P00}). From (\ref{GaiottoPure}) and (\ref{z_dep})
we have that
\begin{eqnarray}
\Psi^{\bf 0}_{+}(\Lambda, z)&=&\langle\,\Delta', \Lambda^2\,|V_{+}(z)|\,\tilde\Delta,\Lambda^2\,\rangle
\;=\;z^{\Delta'-\Delta_+ -\tilde \Delta} \sum_{r,s\geq 0}\Lambda^{2(r+s)}z^{r-s}\nonumber
\\\label{matrix1}
&&\times
\sum_{|I|=r}\sum_{|J|=s} \Big[G^{r}_{c,\Delta'}\Big]^{(1^r) I}
\rho{^{\Delta'}_\infty}{^{\Delta_+}_{\:1}}{^{\tilde\Delta}_{\;0}}
\left(\nu_{\Delta', I}, \nu_{\Delta_+}, \nu_{\tilde\Delta, J}\right)
\Big[G^{s}_{c,\tilde\Delta}\Big]^{J (1^s)}
\\[8pt]\label{matrix2}
&\equiv& z^{\kappa}\,\Phi^{\bf 0}_{+}(\Lambda,z),
\end{eqnarray}
where $\kappa\equiv\Delta'-\Delta_+ -\tilde \Delta$.
Let us observe that $\Phi^{\bf 0}_{+}(\Lambda, z)$ can be split into two parts, i.e. when
$r=s$ and $r\neq s$:
$\label{Phi}
\Phi^{\bf 0}_{+}(\Lambda,z)=\Phi_{r=s}^{\bf 0}(\Lambda)+\Phi_{r\neq s}^{\bf 0}(\Lambda, z),
$
where
\begin{itemize}
\item[$(i)$] for $r=s$,
\begin{equation}\label{Prs}
\Phi_{r=s}^{\bf 0}(\Lambda) =\sum_{r\geq 0}\Lambda^{4r}
\sum_{|I|=|J|=r}\Big[G^{r}_{c,\Delta'}\Big]^{(1^r)\, I}
\rho{^{\Delta'}_\infty}{^{\Delta_+}_{\:1}}{^{\tilde\Delta}_{\;0}}
\left(\nu_{\Delta', I}, \nu_{\Delta_+}, \nu_{\tilde\Delta, J}\right)
\Big[G^{r}_{c,\tilde\Delta}\Big]^{J\, (1^r)},
\end{equation}
\item[$(ii)$] for $r\neq s$,
\begin{equation}\label{Prnots}
\hspace{-25pt}\Phi_{r\neq s}^{\bf 0}(\Lambda, z) =
\sum_{\substack{r\neq s\\r,s\geq 0}}
\Lambda^{2(r+s)}z^{r-s}
\sum_{\substack{|I|=r\\|J|=s} }
\Big[G^{r}_{c,\Delta'}\Big]^{(1^r)\, I}
\rho{^{\Delta'}_\infty}{^{\Delta_+}_{\:1}}{^{\tilde\Delta}_{\;0}}
\left(\nu_{\Delta', I}, \nu_{\Delta_+}, \nu_{\tilde\Delta, J}\right)
\Big[G^{s}_{c,\tilde\Delta}\Big]^{J\, (1^s)}.
\end{equation}
\end{itemize}
Then, one can write
\begin{eqnarray}\label{factors}
\Psi^{\bf 0}_{+}(\Lambda, z)&=& z^\kappa \; \exp\left\lbrace\log\left(
\Phi_{r=s}^{\bf 0}(\Lambda) + \Phi_{r\neq s}^{\bf 0}(\Lambda, z)\right)\right\rbrace\nonumber
\\
&=& z^\kappa \;\exp\left\lbrace\log\Phi_{r=s}^{\bf 0}(\Lambda)+
\log\left(1+\frac{\Phi_{r\neq s}^{\bf 0}(\Lambda, z)}{\Phi_{r=s}^{\bf 0}(\Lambda)}\right)\right\rbrace\nonumber
\\
&=& z^{\kappa}\,{\rm e}^{{\cal Y}^{\bf 0}(\Lambda)}\,{\rm e}^{{\cal X}^{\bf 0}(\Lambda,z)},
\end{eqnarray}
where the following notation has been introduced
\begin{equation}\label{Psi}
{\cal Y}^{\bf 0}(\Lambda)\;=\;\log\Phi_{r=s}^{\bf 0}(\Lambda),
\;\;\;\;\;\;\;
{\cal X}^{\bf 0}(\Lambda,z) \;=\;\log\left(1+\frac{\Phi_{r\neq s}^{\bf 0}(\Lambda, z)}{\Phi_{r=s}^{\bf 0}(\Lambda)}\right).
\end{equation}
Note that the `diagonal' part $\Phi_{r=s}^{\bf 0}$ of the degenerate irregular block and thus
${\cal Y}^{\bf 0}$ do not depend on $z$.

The substitution of (\ref{matrix2}) into eq.~(\ref{NVDeq0}) yields
\begin{eqnarray}\label{NVDeq2}
&&\left[\frac{1}{b^2}\,z^2\frac{\partial^2}{\partial z^2}
+\left(\frac{2\kappa}{b^2}-\frac{3}{2}\right)z\frac{\partial}{\partial z}
+\frac{\Lambda}{4}\,\frac{\partial}{\partial\Lambda} + \frac{\kappa(\kappa -1)}{b^2}
-\frac{3\kappa}{2}\right.
\\
&&\hspace{145pt}\;\left.\,+\,\Lambda^2 \left(z+\frac{1}{z}\right)
+ \frac{\tilde\Delta+\Delta'-\Delta_+}{2}\right] \Phi^{\bf 0}_{+}(\Lambda, z)\;=\;0.\nonumber
\end{eqnarray}

Our aim now is to find the limit $b\to 0$ of eq.~(\ref{NVDeq2}).
To this purpose it is convenient to replace the parameter $\sigma$ in $\tilde\Delta$
and $\Delta'$ (cf.~(\ref{FR})) with $\xi=b\sigma$ and $\Lambda$ with the new parameter
$\hat\Lambda=\Lambda \epsilon_1 b$. After this rescaling, we have
\begin{eqnarray}\label{delta}
&&\Delta',\tilde\Delta\stackrel{b\to 0}{\sim}\frac{1}{b^2}\,\delta,
\;\;\;\;\;\;{\rm where}\;\;\;\;\;\;\delta\;=\;
\lim_{b\to 0}b^2\Delta'\;=\;\lim_{b\to 0}b^2\tilde\Delta\;=\;\tfrac{1}{4}-\xi^2,
\\[8pt]
&&
\tilde\Delta+\Delta'-\Delta_+\stackrel{b\to 0}{\sim}\frac{1}{b^2}
\,2\left(\tfrac{1}{4}-\xi^2\right)\;=\;\frac{1}{b^2}\,2\delta,
\\[8pt]
&&
\kappa\stackrel{b\to 0}{\longrightarrow}\tfrac{1}{2}-\xi,
\;\;\;\;\;\;\;\;\;\;\;
\kappa\left(\kappa-1\right)\stackrel{b\to 0}{\longrightarrow}
-\left(\tfrac{1}{4}-\xi^2\right)\;=\;-\delta.
\end{eqnarray}
Note that $\Delta_+ \stackrel{b\to 0}{\sim} {\cal O}(b^0)$.

The next step needed to complete our task is to determine the behavior 
of the normalized degenerate irregular block $\Phi^{\bf 0}_{+}=z^{-\kappa}\Psi^{\bf 0}_{+}$ 
when $b\to 0$. For $\Lambda=\hat\Lambda/(\epsilon_1 b)$ and
$\Delta',\tilde\Delta\stackrel{b\to 0}{\sim}\frac{1}{b^2}\,\delta$,
it is reasonable to expect that
\begin{equation}\label{ClAsymp}\boxed{
\Phi^{\bf 0}_{+}(\Lambda, z)\;=\;z^{-\kappa}\,\langle\,\Delta', \Lambda^2\,|\,
V_{+}(z)\,|\,\tilde\Delta,\Lambda^2\,\rangle \;\stackrel{b\to 0}{\sim}\;
\varphi^{\bf 0}\!\left(\hat\Lambda/\epsilon_1, z\right)\,
\exp\left\lbrace\frac{1}{b^2}f_{\delta}^{\bf 0}\!\left(\hat\Lambda/\epsilon_1\right)\right\rbrace}
\;\;.
\end{equation}
Moreover, comparing the r.h.s. of eq.~(\ref{ClAsymp}) with eqs.~(\ref{factors})--(\ref{Psi})
one arrives at the following results:
\begin{eqnarray}\label{v}
\varphi^{\bf 0}\!\left(\hat\Lambda/\epsilon_1, z\right) &=&
\lim\limits_{b\to 0}\exp\left\lbrace{\cal X}^{\bf 0}(\Lambda,z)\right\rbrace
=\lim\limits_{b\to 0}
\left(1+\frac{\Phi_{r\neq s}^{\bf 0}(\hat\Lambda/(\epsilon_1 b), z)}
{\Phi_{r=s}^{\bf 0}(\hat\Lambda/(\epsilon_1 b))}\right),
\\[5pt]
\label{f}
f_{\delta}^{\bf 0}\!\left(\hat\Lambda/\epsilon_1\right) &=& 
\lim\limits_{b\to 0}b^2 {\cal Y}^{\bf 0}(\Lambda)
= \lim\limits_{b\to 0}b^2\log\Phi_{r=s}^{\bf 0}\left(\hat\Lambda/(\epsilon_1 b)\right).
\end{eqnarray}
The meaning of eq.~(\ref{ClAsymp}) is that the light
($\Delta_+ \stackrel{b\to 0}{\sim} {\cal O}(b^0)$) degenerate chiral vertex operator 
does not contribute to the classical limit. In other words, its presence in 
the matrix element does not affect the `classical dynamics' (i.e.~the 
`classical action'). Let us note that eq.~(\ref{ClAsymp}) is a `chiral version' of 
Zamolodchikovs' conjecture \cite{Zamolodchikov:1995aa} (see also \cite{Harlow:2011ny})
concerning the semiclassical behavior of the Liouville correlators with heavy and light
vertices on the sphere. Let us stress that there are only a few explicitly 
known tests verifying Zamolodchikovs' hypothesis. For instance, the derivation of 
the large intermediate conformal weight limit $\Delta\to\infty$
of the 4-point block on the sphere as well as its expansion in powers of the 
so-called elliptic variable is based on that assumption in the case of the semiclassical 
behavior of the 5-point function with the light degenerate vertex operator 
\cite{Zam,Z2}. The calculation performed in this section is a new test of the 
semiclassical behavior of the form~(\ref{ClAsymp})
(see also \cite{Litvinov:2013sxa}).
Regardless of the attempts to prove (cf.~subsection \ref{subsect}),
eq.~(\ref{ClAsymp}) can be well confirmed, first, by direct calculations,
secondly, by its consequences. Indeed,
one can check order by order that the limits (\ref{v}) and (\ref{f}) exist.
Moreover, the latter limit reproduces the classical zero flavor irregular block.

Therefore, from (\ref{NVDeq2}) and (\ref{ClAsymp}) for $b\to 0$ one gets 
\begin{equation}\label{M1}
\left[z^2\frac{\partial^2}{\partial z^2}+2(\tfrac{1}{2}-\xi)z\frac{\partial}{\partial z}
+ \frac{\hat\Lambda^2}{\epsilon_1^2 } \left(z+\frac{1}{z}\right)
+\frac{\hat\Lambda}{4}\frac{\partial}{\partial\hat\Lambda}
f_{\delta}^{\bf 0}\!\left(\hat\Lambda/\epsilon_1\right)
\right]\varphi^{\bf 0}\!\left(\hat\Lambda/\epsilon_1, z\right)\;=\;0\,.
\end{equation}
The nontrivial point here is the observation that 
$\lim_{b\to 0}b^2\hat\Lambda\partial_{\hat\Lambda}\varphi^{\bf 0}=0$.
This result has been checked up to high orders of the expansion of (\ref{v}).

At this point, we define the new function $\psi^{\bf 0}(\hat\Lambda/\epsilon_1, z)$ 
related to the old one by
\begin{equation}\label{mpsi}
\varphi^{\bf 0}\!\left(\hat\Lambda/\epsilon_1, z\right)
\;=\; z^{\xi}\,\psi^{\bf 0}\!\left(\hat\Lambda/\epsilon_1, z\right).
\end{equation}
The analogue of eq.~(\ref{M1}) in the case of $\psi^{\bf 0}(\hat\Lambda/\epsilon_1, z)$ is
\begin{equation}\label{NVDeq3}
\left[\,z^2\frac{\partial^2}{\partial z^2}+z\frac{\partial}{\partial z}
+ \frac{\hat\Lambda^2}{\epsilon_1^2 } \left(z+\frac{1}{z}\right)
+\frac{\hat\Lambda}{4}\frac{\partial}{\partial\hat\Lambda}
f_{\delta}^{\bf 0}\!\left(\hat\Lambda/\epsilon_1\right) 
- \xi^2
\right]\psi^{\bf 0}\!\left(\hat\Lambda/\epsilon_1, z\right)\;=\;0\,.
\end{equation}
Since for $z = {\rm e}^w$ the derivatives transform as
$\left(z^2\partial^2_z  + z\partial_z \right)\psi^{\bf 0}(z)
=\partial^2_w \psi^{\bf 0}\!\left({\rm e}^w\right)$,
it turns out that eq.~(\ref{NVDeq3}) becomes
\begin{equation}\label{NVDeq4}
\left[\frac{\mbox{d}^2}{\mbox{d}w^2} + 2\frac{\hat\Lambda^2}{\epsilon_1^2}\cosh(w)
+ \frac{\hat\Lambda}{4}\frac{\partial}{\partial\hat\Lambda} 
f_{\delta}^{\bf 0}\!\left(\hat\Lambda/\epsilon_1\right) 
- \xi^2
\right]\psi^{\bf 0}\!\left(\hat\Lambda/\epsilon_1,{\rm e}^w\right) \;=\; 0 .
\end{equation}
Finally, the substitution $w=-2ix$, $x\in\mathbb{R}$ in (\ref{NVDeq4}) yields
\begin{equation}\label{M2}
\left[-\frac{{\rm d}^2}{{\rm d}x^2} + 8\frac{\hat\Lambda^2}{\epsilon_{1}^{2}}\,\cos 2x
+\hat\Lambda\,\frac{\partial}{\partial\hat\Lambda}
f_{\delta}^{\bf 0}\!\left(\hat\Lambda/\epsilon_1\right)
-4\xi^2\right]\psi^{\bf 0}\!\left(\hat\Lambda/\epsilon_1,{\rm e}^{-2ix}\right)\;=\;0.
\end{equation}
In conclusion, what we have obtained is the following claim: 
{\it\begin{enumerate}
\item 
For the coupling constant $h=2\hat\Lambda/\epsilon_1$ and the
Floquet exponent $\nu=2\xi$ the eigenvalue $\lambda$ of the Mathieu 
operator:
\begin{equation}\label{MO}
\left[-\frac{{\rm d}^2}{{\rm d}x^2}+2h^2\cos2x\right]\psi^{\bf 0} 
\;=\; \lambda\,\psi^{\bf 0}
\end{equation}
is given by the following formula
\begin{equation}\label{eigenvalue}\boxed{
\lambda \;=\; 4\xi^2 - \hat\Lambda\,\frac{\partial}{\partial\hat\Lambda}
f_{\delta}^{\bf 0}\!\left(\hat\Lambda/\epsilon_1\right)}\;\;,
\end{equation}
where $\delta=\frac{1}{4}-\xi^2$.
\item
The corresponding eigenfunction is of the form (cf.~(\ref{v}) and (\ref{mpsi}))
\begin{equation}\label{MathieuF}\boxed{
\psi^{\bf 0}\;\equiv\;\psi^{\bf 0}\!\left(\hat\Lambda/\epsilon_1,{\rm e}^{-2ix}\right)\;=\;
{\rm e}^{2ix\xi}\,
\lim\limits_{b\to 0}
\left(1+\frac{\Phi_{r\neq s}^{\bf 0}(\hat\Lambda/(\epsilon_1 b), {\rm e}^{-2ix})}
{\Phi_{r=s}^{\bf 0}(\hat\Lambda/(\epsilon_1 b))}\right)}\;\;.
\end{equation}
\end{enumerate}}
\noindent
Indeed, using formulae (\ref{coeff0}) for the coefficients of the classical irregular block
$f_{\delta}^{\bf 0}(\hat\Lambda/\epsilon_1)$
with $\delta=\frac{1}{4}-\xi^2$, after postulating
the relation $\xi=\nu/2$ and taking into account that $h^2=4\hat\Lambda^2/\epsilon_{1}^{2}$,
one finds that
\begin{eqnarray}\label{EigenTest}
\lambda & = & 4\xi^2-\hat\Lambda\,\partial_{\hat\Lambda}
\left[\,\sum\limits_{n=1}^{\infty}
\left(\hat\Lambda/\epsilon_1\right)^{\!4n}\!\!f_{\delta}^{{\bf 0},n}\right]\nonumber
\\[10pt]
&=&4\left(\frac{\nu^2}{4}\right)
-\frac{4 h^4}{16}\,f_{\frac{1}{4}-\frac{\nu^2}{4}}^{{\bf 0},1}-
\frac{8 h^8}{256}\,f_{\frac{1}{4}-\frac{\nu^2}{4}}^{{\bf 0},2} -
\frac{12 h^{12}}{4096}\,f_{\frac{1}{4}-\frac{\nu^2}{4}}^{{\bf 0},3} - \ldots\nonumber
\\[10pt]
&&\hspace{-40pt}=\;\nu^2 +
\frac{h^4}{2 \left(\nu ^2-1\right)}+
\frac{\left(5 \nu ^2+7\right)h^8}{32 \left(\nu ^2-4\right) \left(\nu ^2-1\right)^3}+
\frac{\left(9 \nu ^4+58 \nu ^2+29\right)h^{12}}{64 \left(\nu ^2-9\right) \left(\nu ^2-4\right) \left(\nu^2-1\right)^5}
+\ldots\;.
\end{eqnarray}
Hence, the formula (\ref{eigenvalue}) reproduces the well known
weak coupling (small $h^2$) expansion of $\lambda$ for the noninteger Floquet
exponent $\nu\notin\mathbb{Z}$, cf.~\cite{MuellerKirsten:2006}.

\subsection{The classical asymptotic of the degenerate irregular block with the light insertion}
\label{subsect}
In order to understand the factorization phenomenon into `heavy' and `light' factors  
of the degenerate irregular block within the classical limit it suffices to examine 
the behavior of the two factors in eq. \eqref{factors} as $b\to0$. According to eq. \eqref{matrix2}
the main ingredients of the degenerate irregular block expansion are the inverse of the Gram matrix and
the three form rho. Their dependence on $\dl$ is crucial for study of the classical limit. 
As we will see in what follows it is enough to confine oneself to the leading order in $b$.
In section \ref{Towards a proof of classical asymptotic} we found the leading behavior of
the $p(n)\times p(n)$ component of the inverse Gram matrix in the classical limit. However,
from that analysis one can infer also the leading behavior for all the elements of 
$p(n)^{\text{th} }$ column of the inverse Gram matrix. Indeed, from eq. \eqref{step2} 
we find that 
\begin{equation*}
\cbr{G_{c,\dl}^{(n)} }^{(1^n)\, I_u} 
\overset{\deg}{\sim} 
(-1)^{p(n)-u}\frac{\bra{\dl}L_{I_u}L^n_{-1}\ket{\dl}}{\bra{\dl}L_{I_u}L_{-I_u}\ket{\dl}} 
\cbr{G_{c,\dl}^{(n)} }^{(1^n)\, (1^n)},
\end{equation*}
and by virtue of eqs. \eqref{ratio} and \eqref{GramElementClassLimit} we obtain
the leading behavior within the classical limit of the arbitrary matrix element of 
$p(n)^{\text{th} }$ column of the inverse Gram matrix
\begin{equation}
\label{ClassLimitArbitraryElem}
\cbr{G_{c,\dl}^{(n)} }^{(1^n)\, I_u} 
\overset{b\to 0}{\sim} 
\frac{b^{2n}}{(2\d)^n}
\frac{(-1)^{p(n)-u}}{\prod\limits_{i\geq1}m_{i}(I_u)!} 
\prod_{i\geq1}\left( \frac{2 (i+1)\d}{i \cbr{4\d + i^2-1 }} \right)^{m_{i}(I_u)}.
\end{equation}

As for the rho form its analysis is much more involved. By definition, for any two
partitions $I\vdash r,\linebreak\ J\vdash s$ it takes the form
\begin{equation}
\label{rho_form}
\rho\substack{\dl'\,\dl_{+}\,\tilde\dl \\ \!\infty\ \ 1\ \ \ 0}
\left(\nu_{\Delta', I}, \nu_{\Delta_+}, \nu_{\tilde\Delta, J}\right)
= \bra{\dl'}L_{I}V_{\dl_{+}}(z) L_{-J}\ket{\tilde\dl}\big |_{z=1}.
\end{equation}
Making use of the Virasoro algebra \eqref{Virasoro} this can be developed into the form
\begin{multline}
\label{matrix_el_expansion}
\bra{\dl'}L_{I}V_{\dl_{+}}(z) L_{-J}\ket{\tilde\dl}
=\bra{\dl'}V_{\dl_{+}}(z)\ad{I}{L_{-J} } \ket{\tilde\dl}
+ \bra{\dl'}\ad{\ddot I^{(1)}}{V_{\dl_{+}}(z)}\ad{\dot I^{(1)}}{L_{-J}} \ket{\tilde\dl}
\\
+\bra{\dl'}\ad{\ddot I^{(2)}}{V_{\dl_{+}}(z)}\ad{\dot I^{(2)}}{L_{-J}} \ket{\tilde\dl}
+ \ldots
+ \bra{\dl'}\ad{I}{V_{\dl_{+}}(z)}L_{-J}\ket{\tilde\dl} ,
\end{multline}
where for the sake of brevity we have used the following notation
$$
\ad{I}{L_{-J}}\! :=\! [L_{k_{\ell(I)}(I)},\ldots,[L_{k_{1}(I)}, L_{-J}]\ldots],
\quad
\underset{m\in\{1,...,\ell(I)\}}{\forall} \ I = \dot I^{(m)}\! \cup\! \ddot I^{(m)},
\  \dot I^{(m)}\! :=\! \big( k_{1}(I),\dots, k_{\ell(I) - m}(I) \big).
$$
Let us examine a matrix element that contributes to the sum in eq. \eqref{matrix_el_expansion}. 
It takes the form
\begin{multline}
\label{contrib_rho_exp}
\bra{\dl'}\ad{\ddot I^{(m)}}{V_{\dl_{+}}(z)}\ad{\dot I^{(m)}}{L_{-J}} \ket{\tilde\dl}
\\
= \bra{\dl'}[L_{k_{\ell(I)}(I)},\ldots,[L_{k_{\ell(I) - m+1}(I)},V_{\dl_{+}}(z)]\ldots]
[L_{k_{\ell(I) - m}(I)},\ldots,[L_{k_1(I)} , L_{-J}]\ldots]\ket{\tilde\dl}.
\end{multline}
It is nonzero provided $t\equiv|\dot I^{(m)}| = \sum k_{i}(\dot I^{(m)}) \leq|J|= s$.
For definiteness let us assume that $t<s$. From the commutator formula between the Virasoro 
generator and vertex operator \eqref{CVO} we obtain
\begin{displaymath}
\ad{\ddot I^{(m)}}{V_{\dl_{+}}(z)}\!=\!
[L_{k_{\ell(I)}(I)},\ldots,[L_{k_{\ell(I) - m+1}(I)},V_{\dl_{+}}(z)]\ldots]
\!=\! \prod_{i=\ell(I) - m+1}^{\ell(I)} z^{k_i(I)}\Big(z\pt_z\! +\! \big(k_{i}(I)\! +\! 1\big)\dl_{+} \Big)V_{\dl_{+}}(z).
\end{displaymath}
Hence, the matrix element \eqref{contrib_rho_exp} assumes the form
\begin{multline}
\label{matrix_el_structure}
\bra{\dl'}\ad{\ddot I^{(m)}}{V_{\dl_{+}}(z)}\ad{\dot I^{(m)}}{L_{-J}} \ket{\tilde\dl}
\\
=\prod_{i=\ell(I) - m+1}^{\ell(I)} z^{k_i(I)}\Big(z\pt_z + \big(k_{i}(I) +1\big)\dl_{+} \Big)
\bra{\dl'}V_{\dl_{+}}(z) \ad{\dot I^{(m)}}{L_{-J}} \ket{\tilde\dl}.
\end{multline}
The nested commutator encoded in $\ad{\dot I^{(m)}}{L_{-J}}$ on the right hand side of the above formula
provides possible factors containing $\tilde\dl$ and $c$. These factors typically appear
if one or more parts of $\dot I^{(m)}$ coincides with those
in the partition $J$. Let us rewrite the matrix element on the right hand side of eq. \eqref{matrix_el_structure}
in terms of multiplicities
\begin{subequations}
\begin{multline}
\label{matrix_el_mutliplicity}
\bra{\dl'}V_{\dl_{+}}(z) \ad{\dot I^{(m)}}{L_{-J}} \ket{\tilde\dl}
=\bra{\dl'}V_{\dl_{+}}(z) [L_{k_{\ell(I) - m}(I)},\ldots,[L_{k_1(I)} , L_{-J}]\ldots]\ket{\tilde\dl}
\\
= \bra{\dl'}V_{\dl_{+}}(z)
\underbrace{[L_{i_l},\ldots,[L_{i_l}}_{m_{i_l}(I)},\ldots,\underbrace{[L_{i_2},\ldots,[L_{i_2} }_{m_{i_2}(I)},
\underbrace{[L_{i_1},\ldots,[L_{i_1}}_{m_{i_1}(I)} , 
L_{-j_1}^{m_{j_1}(J)}L_{-j_2}^{m_{j_2}(J)}\cdots L_{-j_n}^{m_{j_n}(J)} ]\ldots]\ket{\tilde\dl},
\end{multline}
where 
\begin{equation}
\label{def_of_i_u}
i_1 := k_{1}(I),\, i_{2} := k_{m_{i_1}(I)+1}(I),\,i_{3} := k_{m_{i_1}(I)+m_{i_2}(I)+1}(I),\ldots,\, 
i_l := k_{1+\sum\limits_{u=1}^{l-1} m_{i_u}(I)}(I)=k_{\ell(I) - m}(I),
\end{equation} 
\end{subequations}
and similarly for parts $j_{u}$. Let us assume for definiteness that 
$j_{u} = i_{u}$ for $u\in\{1,\dots,N\}$, $N\leq l$. Then, according 
to the formula in eq. \eqref{commutator_formula2} 
an overall factor that appears in front of the resulting matrix element, 
up to leading term in $\tilde\dl$, reads
\begin{subequations}
\begin{equation}
\label{matrix_el_reduced}
\bra{\dl'}V_{\dl_{+}}(z) \ad{\dot I^{(m)}}{L_{-J}} \ket{\tilde\dl}
\overset{\deg}{\sim} \mathsf{Poly}_{\dot I^{(m)},\dot J}(\tilde\dl,c)\bra{\dl'}V_{\dl_{+}}(z)L_{-\ddot J}\ket{\tilde\dl},
\end{equation} 
where $\dot I^{(m)},\dot J\vdash t$ and $\ddot J \vdash s - t$.
\begin{multline}
\label{Poly_typical}
\mathsf{Poly}_{\dot I^{(m)},\dot J}(\tilde\dl,c)
:= \prod_{u=1}^{N} i_{u}^{\min\{m_{i_u}(I),m_{i_u}(J)\}}
\frac{m_{i_u}(J)!}{\vartheta\big(m_{i_u}(J)-m_{i_u}(I)\big)!}
\\
\times
\frac{\cbr{2\tilde\dl + \tfrac{c}{12}(i_{u}^{2}-1) ; i_{u} }_{m_{i_u}(J)} }
{\cbr{2\tilde\dl + \tfrac{c}{12}(i_{u}^{2}-1) ; i_{u} }_{\vartheta\left( m_{i_u}(J)-m_{i_u}(I) \right)}},
\end{multline}
\end{subequations}
where $\vartheta(x) = x\th(x)$ and $\th(x)$ is Heaviside's theta function.
$\mathsf{Poly}_{\dot I^{(m)},\dot J}(\tilde\dl,c) $ is a polynomial in $\tilde\dl$ and $c$ 
and `$\overset{\deg}{\sim}$' has the same meaning as in section \ref{Towards a proof of classical asymptotic}. 
Its degree in $\tilde\dl$, in general, varies between 
$$
0 \leq \deg_{\tilde\dl} \mathsf{Poly}_{I_{i},I_{j}}(\tilde\dl,c) \leq t,
\qquad I_i,I_j\vdash t,\quad
i,j = \{1,\dots,p(t)\}.
$$ 
Since $\cbr{2\tilde\dl + \tfrac{c}{12}(i^{2}-1) ; i }_{n}\overset{\deg}{\sim} 2^n\tilde\dl^{n}$ as a polynomial in $\tilde\dl$
then from eq. \eqref{Poly_typical} we get 
\begin{equation}
\deg_{\tilde\dl} \mathsf{Poly}_{\dot I^{(m)},\dot J}(\tilde\dl,c) 
= \sum_{u=1}^{N}\Big( m_{i_u}(J) - \vartheta \big( m_{i_u}(J)-m_{i_u}(I) \big) \Big)
\leq \sum_{u = 1}^{l} m_{i_u}(I) = \ell(\dot I^{(m)}).
\end{equation} 
Therefore, the degree of the polynomial is \emph{maximal} if 
$\ell(\dot I^{(m)}) = \sum m_{i_u}(I)= t=|\dot I^{(m)}|=\sum i_{u}\, m_{i_u}(I)$ 
which due to the fact that $i_u>i_v$ for $u<v$ following from eq. \eqref{def_of_i_u} entails that
$$
\sum_{u=1}^{N}(i_{u}-1)m_{i_u}(I) = 0\quad \Rightarrow \quad i_1 := k_1 =1\wedge m_{1}(I) = t,
$$
and $m_{i_u}(I) = 0$ for $u>1$, i.e., $\dot I^{(m)} = (1^{t})$.
Moreover, $\dot I^{(m)}$ by definition consists of first $m$ parts of $I$. Hence, if $I$ is to be a partition
\emph{it must assume the form} $(1^r)$. 

The matrix element \eqref{matrix_el_structure} can now be rewritten as 
\begin{multline}
\bra{\dl'}\ad{\ddot I^{(m)}}{V_{\dl_{+}}(z)}\ad{\dot I^{(m)}}{L_{-J}} \ket{\tilde\dl}
\\
\overset{\deg}{\sim} \mathsf{Poly}_{\dot I^{(m)},\dot J}(\tilde\dl,c)
\prod_{i=\ell(I) - m+1}^{\ell(I)} z^{k_i(I)}\Big(z\pt_z + \big(k_{i}(I) +1\big)\dl_{+} \Big)
\bra{\dl'}V_{\dl_{+}}(z)L_{-\ddot J}\ket{\tilde\dl}.
\end{multline}
Note that the matrix element in the last line of the above equation 
is nothing but the gamma vector given in eq. \eqref{ggg}
with $I\to \ddot J$. Performing necessary computations we find the typical form of the contribution 
to the sum \eqref{matrix_el_expansion}, namely
\begin{multline}
\label{cotrib2sum_typical}
\bra{\dl'}\ad{\ddot I^{(m)}}{V_{\dl_{+}}(z)}\ad{\dot I^{(m)}}{L_{-J}} \ket{\tilde\dl}\big |_{z=1}
\\
\overset{\deg}{\sim} \mathsf{Poly}_{\dot I^{(m)},\dot J}(\tilde\dl,c) 
(-1)^{\ell(\ddot J)}\prod_{i=1}^{\ell(\ddot J) }\cbr{\dl'- k_{i}(\ddot J)\dl_{+} 
-\tilde\dl - \sum_{s>i}^{\ell(\ddot J) }k_{s}(\ddot J)  }
\\
\times
\prod_{j=\ell(I) - m+1}^{\ell(I) }\cbr{\dl'+ k_{j}(I)\dl_{+} 
-\tilde\dl + \sum_{s>j}^{\ell(I)}k_{s}(I) -s +t  }.
\end{multline}
The above formula enables one to estimate the contribution of the corresponding term in the 
sum \eqref{matrix_el_expansion} within the classical limit. Since for $b\to 0$ conformal weights scale as
$\dl',\tilde\dl \sim \d/b^2$ and $\dl_{+}\sim -1/2$ we conclude that the two products in the 
second and third line of eq. \eqref{cotrib2sum_typical} reduce to the $\xi$ dependent numerical factors 
and the only factor that determines the classical behavior of the matrix element is
the polynomial $\mathsf{Poly}_{\dot I^{(m)},\dot J}(\tilde\dl,c)$. The degree of the latter
depends on the number of factors. Therefore, the term with the greatest number of factors
will dominate entire sum within the classical limit. The term in question 
is the first one in eq. \eqref{matrix_el_expansion}. This analysis allows us to conclude that
\begin{align*}
\bra{\dl'}L_{I}V_{\dl_{+}}(1) L_{-J}\ket{\tilde\dl} 
&=\bra{\dl'}V_{\dl_{+}}(z)\ad{I}{L_{-J} } \ket{\tilde\dl}\big|_{z=1}+\ldots
\\
&= (-1)^{\ell(\ddot J)}\mathsf{Poly}_{\dot I^{(m)},\dot J}(\tilde\dl,c) 
\prod_{i=1}^{\ell(\ddot J) }\cbr{\dl'- k_{i}(\ddot J)\dl_{+} 
-\tilde\dl - \sum_{s>i}^{\ell(\ddot J) }k_{s}(\ddot J)  }
 +\ldots
\end{align*}
and in the classical limit and within the leading order approximation the rho form reads
\begin{equation*}
\rho\substack{\dl'\,\dl_{+}\,\tilde\dl \\ \!\infty\ \ 1\ \ \ 0}
\left(\nu_{\Delta', I}, \nu_{\Delta_+}, \nu_{\tilde\Delta, J}\right)
\overset{b\to 0}{\sim}
\mathsf{Poly}_{\dot I^{(m)},\dot J}(\d / b^{2},6 / b^{2}) C_{\ddot J}^{(-)}(\xi),
\end{equation*}
where we introduced the label for the $\xi$ dependent numerical factor
\begin{equation}
\label{C_factor}
C_{\ddot J}^{(-)}(\xi) := \prod_{i=1}^{\ell(\ddot J) }
\bigg(\sum_{s>i}^{\ell(\ddot J) }k_{s}(\ddot J) - \tfrac{1}{2} k_{i}(\ddot J)  + \xi  \bigg).
\end{equation} 
The argument concerning the maximal degree of the polynomial entails that the dominant contribution  
to the sum over all partitions with fixed $|I|=r,\, |J|=s$ comes from the term with the maximal possible
multiplicity, i.e., if $r<s$ then $I=(1^r)$ and $J = \ddot J \cup  (1^r)$. In this case
\begin{equation}
\label{ClassRhoForm}
\begin{aligned}
\rho\substack{\dl'\,\dl_{+}\,\tilde\dl \\ \!\infty\ \ 1\ \ \ 0}
\left(\nu_{\Delta', (1^r)}, \nu_{\Delta_+}, \nu_{\tilde\Delta, J}\right)
=& r! (2\tilde\dl)_{r} (-1)^{\ell(\ddot J)} \prod_{i=1}^{\ell(\ddot J) }\cbr{\dl'- k_{i}(\ddot J)\dl_{+} 
-\tilde\dl - \sum_{s>i}^{\ell(\ddot J) }k_{s}(\ddot J)  } +\ldots
\\
\overset{b\to 0}{\sim}&  r! (2\d)^{r} b^{-2r} C_{\ddot J}^{(-)}(\xi)
= \Big[G^{r}_{c,\d}\Big]_{(1^r)\,(1^r)}C_{\ddot J}^{(-)}(\xi) .
\end{aligned}
\end{equation} 

We are at the point where we have all necessary ingredients to prove the factorization phenomenon
for the three point irregular conformal block stated in eq. \eqref{ClAsymp}. Let us consider the 
case where $r=s$. Then from eqs. \eqref{GramElementClassLimit}
and \eqref{ClassRhoForm} we get that $|\ddot J| = 0$ as well as $C_{\ddot J}^{(-)}(\xi) = 1$ which, 
when applied to eq. \eqref{Prs}, yields
\begin{eqnarray}
\nonumber
\Phi_{r=s}^{\bf 0}(\Lambda) &=&\sum_{r\geq 0}\Lambda^{4r}
\sum_{|I|=|J|=r}\Big[G^{r}_{c,\Delta'}\Big]^{(1^r)\, I}
\rho{^{\Delta'}_\infty}{^{\Delta_+}_{\:1}}{^{\tilde\Delta}_{\;0}}
\left(\nu_{\Delta', I}, \nu_{\Delta_+}, \nu_{\tilde\Delta, J}\right)
\Big[G^{r}_{c,\tilde\Delta}\Big]^{J\, (1^r)}
\\
\nonumber
&\overset{b\to 0}{\sim}&
\sum_{r\geq 0}\cbr{\frac{\hat\Lambda}{\e_1}}^{4r} b^{-4r}
\Big[G^{r}_{c,\d}\Big]^{(1^r)\, (1^r)}
\Big[G^{r}_{c,\d}\Big]_{(1^r)\,(1^r)}
\Big[G^{r}_{c,\d}\Big]^{(1^r)\, (1^r)}
\\
\label{ClassPhi_rr}
&=& \exp\pbr{\frac{1}{b^2 }\,\cbr{\frac{\hat\Lambda}{\e_1}}^{4}\frac{1}{2\d}}.
\end{eqnarray}

Let us now consider the case when $r\neq s$. According to eq. \eqref{Prnots} we have
\begin{equation}
\begin{aligned}
\Phi_{r\neq s}^{\bf 0}(\Lambda, z) &=
\sum_{\substack{r\neq s\\r,s\geq 0}}
\Lambda^{2(r+s)}z^{r-s} 
F^{(r,s)}_{c}\!\left(\Delta', \Delta_{+}, \tilde\Delta\right)
\\
&=\sum_{\substack{s> r\geq 0}}\Lambda^{2(r+s)}
\cbr{F^{(s,r)}_{c}\!\left(\Delta', \Delta_{+}, \tilde\Delta\right) z^{s-r} 
+
F^{(r,s)}_{c}\!\left(\Delta', \Delta_{+}, \tilde\Delta\right)z^{-(s-r) } },
\end{aligned}
\end{equation}
where
\begin{equation}
F^{(r,s)}_{c}\!\left(\Delta', \Delta_{+}, \tilde\Delta\right)
:= \sum_{\substack{|I|=r\\|J|=s} }
\Big[G^{r}_{c,\Delta'}\Big]^{(1^r)\, I}
\rho{^{\Delta'}_\infty}{^{\Delta_+}_{\:1}}{^{\tilde\Delta}_{\;0}}
\left(\nu_{\Delta', I}, \nu_{\Delta_+}, \nu_{\tilde\Delta, J}\right)
\Big[G^{s}_{c,\tilde\Delta}\Big]^{J\, (1^s)}.
\end{equation} 
Thus the function $\Phi_{r\neq s}^{\bf 0}(\Lambda, z)$ splits into two parts with positive and 
negative power of variable $z$
\begin{subequations}
\begin{equation}
\label{Phirnots_splitting}
\Phi_{r\neq s}^{\bf 0}(\Lambda, z) =\f^{r\neq s}_{1}(\Lambda, z) + \f^{r\neq s}_{2}(\Lambda, z),
\end{equation} 
where
\begin{equation}
\label{defs_f1_f2}
\f^{r\neq s}_{1}(\La,z)\!=\!\sum_{\substack{s> r\geq 0}}\Lambda^{2(r+s)}
F^{(s,r)}_{c}\!\left(\Delta',\Delta_{+},\tilde\Delta\right)z^{s-r},
\quad
\f^{r\neq s}_{2}(\La,z) \!=\! \sum_{\substack{s> r\geq 0}}\Lambda^{2(r+s)}
F^{(r,s)}_{c}\!\left(\Delta', \Delta_{+}, \tilde\Delta\right)z^{-(s-r) }.
\end{equation} 
\end{subequations}
Let us consider the second one and compute the classical limit of $F^{(r,s)}$. 
From eqs. \eqref{ClassLimitArbitraryElem} and \eqref{ClassRhoForm} and recalling the notation
$J = \ddot J \cup (1^r)$ we obtain
\begin{align*}
F^{(r,s)}_{c}\!\left(\Delta', \Delta_{+}, \tilde\Delta\right)
&\overset{b\to 0}{\sim} \sum_{|\ddot J| = n} \Big[G^{r}_{c,\d}\Big]^{(1^r)\, (1^r)}
\Big[G^{r}_{c,\d}\Big]_{(1^r)\,(1^r)}\, C_{\ddot J}^{(-)}(\xi) 
\Big[G^{r}_{c,\d}\Big]^{(1^s)\, , \ddot J+ (1^r)}
\\
&= b^{2(r+s)}\frac{1}{b^{2 r}(2\d)^{r} r!}
\frac{1}{(2\d)^{s-r}}\sum_{|\ddot J| = s-r} C_{\ddot J}^{(-)}(\xi)
\prod_{i\geq2}\frac{1}{m_{i}(\ddot J)!} \left( \frac{2 (i+1)\d}{i \cbr{4\d + i^2-1 }} \right)^{m_{i}(\ddot J)}.
\end{align*}
Let us denote 
\begin{equation}
\label{def_varphi}
\z^{(-)}_{n}(\xi) := \frac{1}{(2\d)^{n}}\sum_{|\ddot J| = n} C_{\ddot J}^{(-)}(\xi)
\prod_{i\geq2}\frac{1}{m_{i}(\ddot J)!} \left( \frac{2 (i+1)\d}{i \cbr{4\d + i^2-1 }} \right)^{m_{i}(\ddot J)}.
\end{equation} 
Inserting the result for the classical limit of $F^{(r,s)}$ to $\f^{r\neq s}_{2}$ amounts to
\begin{eqnarray*}
\f^{r\neq s}_{2}(\La,z) &\overset{b\to 0}{\sim}&
\sum_{\substack{s> r\geq 0}} \cbr{\frac{\hat\La}{\e_1}}^{2(r+s)} 
z^{-(s-r) }\frac{1}{b^{2 r}(2\d)^{r} r!} \z^{(-)}_{s-r}(\xi)
\\
&=& \sum_{s\geq 1}\sum_{r=0}^{s-1} \cbr{\frac{\hat\La}{\e_1}}^{4r} 
\frac{1}{b^{2 r}(2\d)^{r} r!}\cbr{\frac{\hat\La}{\e_1}}^{2(s-r)}
z^{-(s-r) } \z^{(-)}_{s-r}(\xi)
\\
&=&\sum_{r\geq 0}\sum_{s\geq r+1}\cbr{\frac{\hat\La}{\e_1}}^{4r} 
\frac{1}{b^{2 r}(2\d)^{r} r!}\cbr{\frac{\hat\La}{\e_1}}^{2(s-r)}
z^{-(s-r) }\z^{(-)}_{s-r}(\xi)
\\
&\stackrel{s-r = n}{=}& \sum_{r\geq 0}
\frac{1}{r!}\cbr{\frac{1}{b^{2}}\cbr{\frac{\hat\La}{\e_1}}^{4}\frac{1}{2\d} }^{r}
\ \sum_{n\geq 1}\cbr{\frac{\hat\La}{\e_1}}^{2n}
\z^{(-)}_{n}(\xi) z^{-n } .
\end{eqnarray*}
The last line of the above formula is noting but the exponent of the first coefficient of the irregular 
classical block as in eq. \eqref{ClassPhi_rr} and the second one is the leading order approximation
to the Mathieu function. The same argument applies to $\f^{r\neq s}_{1}$, 
such that we can conclude with the following formula
\begin{equation}
\label{ClassPhi_rnots}
\Phi_{r\neq s}^{\bf 0}(\Lambda,z)
\overset{b\to 0}{\sim}  
\exp\pbr{\frac{1}{b^2 }\,\cbr{\frac{\hat\Lambda}{\e_1}}^{4}\frac{1}{2\d}}\, 
\sum_{n\geq 1}\cbr{\frac{\hat\La}{\e_1}}^{2n}\cbr{\z^{(-)}_{n}(\xi) z^{-n } + \z^{(+)}_{n}(\xi) z^{n }}.
\end{equation}

Having found the classical limit of $\Phi_{r=s}^{\bf 0}(\Lambda)$ in eq. \eqref{ClassPhi_rr} 
and $\Phi_{r\neq s}^{\bf 0}(\Lambda,z)$ in eq. \eqref{ClassPhi_rnots}
we can combine them to find the limit that defines the factor in eq. \eqref{v} deriving from the light field.
As a result we find
\begin{equation}
\varphi^{\bf 0}\!\left(\hat\Lambda/\epsilon_1, z\right) 
=\lim\limits_{b\to 0}
\left(1+\frac{\Phi_{r\neq s}^{\bf 0}(\hat\Lambda/(\epsilon_1 b), z)}
{\Phi_{r=s}^{\bf 0}(\hat\Lambda/(\epsilon_1 b))}\right)
= \sum_{n\geq 0}\cbr{\frac{\hat\La}{\e_1}}^{2n}\cbr{\z^{(-)}_{n}(\xi) z^{-n } + \z^{(+)}_{n}(\xi) z^{n }}.
\end{equation} 
The above formula does not depend on $b$ and is a finite expression which shows in the leading order approximation
that the factorization of the light operator insertion in the three point irregular conformal block indeed takes 
place.

\subsection{Mathieu functions}
Our next point is to demonstrate that the formula (\ref{MathieuF}) fits the 
{\it noninteger order} Mathieu function which
corresponds to the eigenvalue given by (\ref{eigenvalue}), (\ref{EigenTest}).
As a starting point let us recall that $\Phi_{r=s}^{\bf 0}$ and
$\Phi_{r\neq s}^{\bf 0}$ in eq.~(\ref{MathieuF}) are two parts of 
the normalized $N_f\!=\!0$ degenerate irreagular block
$\Phi^{\bf 0}_{+}(\Lambda,z)=\Phi_{r=s}^{\bf 0}(\Lambda)+\Phi_{r\neq s}^{\bf 0}(\Lambda, z)$
(cf.~eqs.~(\ref{matrix1})--(\ref{NVDeq2})). Explicitely, 
\begin{eqnarray}\label{P0}
\Phi_{r=s}^{\bf 0}(\Lambda) &=& \sum_{r\geq 0}\Lambda^{4r}\,
F^{(r,r)}_{c}\!\left(\Delta', \Delta_{+}, \tilde\Delta\right),
\\
\Phi_{r\neq s}^{\bf 0}(\Lambda, z) &=& \sum\limits_{s\geq 1}\sum\limits_{r=0}^{s-1}
\Lambda^{2(r+s)}\left(F^{(s,r)}_{c}\!\left(\Delta', \Delta_{+}, \tilde\Delta\right)\,z^{s-r} 
+F^{(r,s)}_{c}\!\left(\Delta', \Delta_{+}, \tilde\Delta\right)\,z^{-(s-r)}\right),\nonumber
\end{eqnarray}
where (cf.~eqs.~(\ref{Prs}) and (\ref{Prnots}))
\begin{equation}\label{F}
F^{(r,s)}_{c}\!\left(\Delta', \Delta_{+}, \tilde\Delta\right)\;\equiv\;
\sum_{\substack{|I|=r\\|J|=s} }
\Big[G^{r}_{c,\Delta'}\Big]^{(1^r)\, I}
\rho{^{\Delta'}_\infty}{^{\Delta_+}_{\:1}}{^{\tilde\Delta}_{\;0}}
\left(\nu_{\Delta', I}, \nu_{\Delta_+}, \nu_{\tilde\Delta, J}\right)
\Big[G^{s}_{c,\tilde\Delta}\Big]^{J\, (1^s)}.
\end{equation}
Let us note that $\Phi_{r\neq s}^{\bf 0}(\Lambda, z)$ has two linearly independent 
components to which it can be split, namely 
\begin{equation}\label{PrnotsS}
\Phi_{r\neq s}^{\bf 0}(\Lambda, z)=\phi_{1}^{r\neq s}(\Lambda, z)+\phi_{2}^{r\neq s}(\Lambda, z),
\end{equation}
where
\begin{eqnarray}\label{p1}
\phi_{1}^{r\neq s}(\Lambda, z) &=&
\sum\limits_{s\geq 1}\sum\limits_{r=0}^{s-1}
\Lambda^{2(r+s)} F^{(s,r)}_{c}\!\!\left(\Delta', \Delta_{+}, \tilde\Delta\right)\,z^{s-r}\;,
\\
\label{p2}
\phi_{2}^{r\neq s}(\Lambda, z) &=&
\sum\limits_{s\geq 1}\sum\limits_{r=0}^{s-1}
\Lambda^{2(r+s)} F^{(r,s)}_{c}\!\!\left(\Delta', \Delta_{+}, \tilde\Delta\right)\,z^{-(s-r)}\;.
\end{eqnarray}
Let us consider the ratio $\Phi_{r\neq s}^{\bf 0}/\Phi_{r=s}^{\bf 0}$ in eq.~(\ref{MathieuF}).
From (\ref{P0}), (\ref{PrnotsS}), (\ref{p1}), (\ref{p2}) we get
\begin{equation}\label{decompo}
\frac{\Phi_{r\neq s}^{\bf 0}(\Lambda, z)}{\Phi_{r=s}^{\bf 0}(\Lambda)} = 
\frac{\phi_{1}^{r\neq s}(\Lambda, z)}{\Phi_{r=s}^{\bf 0}(\Lambda)}
+\frac{\phi_{2}^{r\neq s}(\Lambda, z)}{\Phi_{r=s}^{\bf 0}(\Lambda)}\;,
\end{equation}
where
\begin{eqnarray*}
\frac{1}{\Phi_{r=s}^{\bf 0}(\Lambda)}\,\phi_{1}^{r\neq s}(\Lambda, z) &=&
\frac{1}{1+\sum\limits_{s\geq 1}\Lambda^{4s}\,F^{(s,s)}}\, 
\sum\limits_{s\geq 1}\sum\limits_{r=0}^{s-1}
\Lambda^{2(r+s)} F^{(s,r)}\,z^{s-r}\,,
\\
\frac{1}{\Phi_{r=s}^{\bf 0}(\Lambda)}\,\phi_{2}^{r\neq s}(\Lambda, z) &=&
\frac{1}{1+\sum\limits_{s\geq 1}\Lambda^{4s}\,F^{(s,s)}}\, 
\sum\limits_{s\geq 1}\sum\limits_{r=0}^{s-1}
\Lambda^{2(r+s)} F^{(r,s)}\,z^{-(s-r)}.
\end{eqnarray*}
Let us observe that in both equations written above one can expand the factor 
$(1\!+\!\sum_{s\geq 1}\Lambda^{4s}F^{(s,s)})^{-1}$ 
according to the formula for the sum of the geometric series.
Then, collecting the resulting expressions up to $12$ order in $\Lambda$
one can obtain 
\begin{itemize}
\item[---] for the first term in eq.~(\ref{decompo}):
\begin{eqnarray*}
\frac{\phi_{1}^{r\neq s}(\Lambda, z)}{\Phi_{r=s}^{\bf 0}(\Lambda)}&=&
\Lambda^2\, z\, F^{(1,0)} 
+\Lambda^4\, z^2\, F^{(2,0)} 
\\ 
&+&\Lambda^6 \left[z^3\, F^{(3,0)} + z\left(F^{(2,1)}-F^{(1,0)} F^{(1,1)}\right)\right]
\\
&+&
\Lambda^8 \left[z^4\, F^{(4,0)} + z^2\left(F^{(3,1)}-F^{(1,1)} F^{(2,0)}\right)\right]
\\
&+&
\Lambda^{10} \left[z^5\, F^{(5,0)}+ z^3\left(F^{(4,1)}-F^{(1,1)}F^{(3,0)}\right)\right.
\\
&&\hspace{10pt}+\left. z\left(F^{(1,0)} (F^{(1,1)})^2
-F^{(1,1)} F^{(2,1)}-F^{(1,0)} F^{(2,2)}+F^{(3,2)}\right)\right]
\\
&+&
\Lambda^{12}\left[z^6\, F^{(6,0)} + z^4\left(F^{(5,1)}-F^{(1,1)} F^{(4,0)}\right)\right.
\\   
&&\hspace{10pt}+\left. z^2\left((F^{(1,1)})^2 F^{(2,0)}
-F^{(2,0)} F^{(2,2)}-F^{(1,1)} F^{(3,1)}+F^{(4,2)}\right)\right]+\ldots\;,
\end{eqnarray*}
\item[---] for the second term in eq.~(\ref{decompo}):
\begin{eqnarray*}
\frac{\phi_{2}^{r\neq s}(\Lambda, z)}{\Phi_{r=s}^{\bf 0}(\Lambda)}&=&
\Lambda^2\, z^{-1}\, F^{(0,1)}
+\Lambda^4\,z^{-2}\,F^{(0,2)}
\\
&+&
\Lambda^6 \left[z^{-3}\,F^{(0,3)}+z^{-1}\left(F^{(1,2)}-F^{(0,1)} F^{(1,1)}\right)\right]
\\
&+&
\Lambda^8 \left[z^{-4}\,F^{(0,4)} + z^{-2}\left(F^{(1,3)}-F^{(1,1)} F^{(0,2)}\right)\right]
\\
&+&\Lambda^{10}
\left[z^{-5}\,F^{(0,5)}+z^{-3}\left(F^{(1,4)}- F^{(1,1)}F^{(0,3)}\right)\right.
\\
&&\hspace{10pt}+\left. z^{-1}\left(F^{(0,1)}(F^{(1,1)})^2
-F^{(1,1)}F^{(1,2)}-F^{(0,1)} F^{(2,2)}+F^{(2,3)}\right)\right]
\\
&+&\Lambda^{12}
\left[z^{-6}\,F^{(0,6)} + z^{-4}\left(F^{(1,5)}-F^{(1,1)}F^{(0,4)}\right)\right.
\\
&&\hspace{10pt}+\left. z^{-2}\left((F^{(1,1)})^2 F^{(0,2)}-F^{(0,2)}F^{(2,2)}
-F^{(1,1)}F^{(1,3)}+F^{(2,4)}\right)\right]+\ldots\,.
\end{eqnarray*}
\end{itemize}
Thus, eventually we get
\begin{eqnarray}\label{rexp}
\frac{\Phi_{r\neq s}^{\bf 0}(\Lambda, z)}{\Phi_{r=s}^{\bf 0}(\Lambda)}
&=&
\Lambda^2\,{\cal K}_{2} + \Lambda^4\,{\cal K}_{4}+
\Lambda^6\,{\cal K}_{6} + \Lambda^8\,{\cal K}_{8}+ 
\Lambda^{10}\,{\cal K}_{10}+\Lambda^{12}\,{\cal K}_{12}+\ldots\;,
\end{eqnarray}
where
\begin{eqnarray*}
{\cal K}_{2} &=& z^{-1}\,F^{(0,1)} + z\,F^{(1,0)},
\\
{\cal K}_{4} &=& z^2\,F^{(2,0)}+z^{-2}\,F^{(0,2)},
\\
{\cal K}_{6} &=& z^3\,F^{(3,0)}+z^{-3}\,F^{(0,3)}
+z\,\left(F^{(2,1)}-F^{(1,0)}F^{(1,1)}\right)
+z^{-1}\left(F^{(1,2)}-F^{(0,1)}F^{(1,1)}\right),
\\
{\cal K}_{8} &=& z^4\,F^{(4,0)}+z^{-4}\,F^{(0,4)}
+z^2\left(F^{(3,1)}-F^{(1,1)} F^{(2,0)}\right)
+z^{-2}\left(F^{(1,3)}-F^{(1,1)}F^{(0,2)}\right),
\\
{\cal K}_{10} &=& z^5\,F^{(5,0)}+z^{-5}\,F^{(0,5)}
+ z^3\left(F^{(4,1)}-F^{(1,1)}F^{(3,0)}\right)
+z^{-3}\left(F^{(1,4)}-F^{(1,1)}F^{(0,3)}\right)
\\
&&+\; z\left(F^{(1,0)} (F^{(1,1)})^2
-F^{(1,1)}F^{(2,1)}-F^{(1,0)}F^{(2,2)}+F^{(3,2)}\right)
\\
&&+\; z^{-1}\left(F^{(0,1)} (F^{(1,1)})^2-F^{(1,1)}
F^{(1,2)}-F^{(0,1)} F^{(2,2)}+F^{(2,3)}\right),
\end{eqnarray*}
and
\begin{eqnarray*}
{\cal K}_{12} &=& z^6\, F^{(6,0)}+z^{-6}\,F^{(0,6)}
+z^4\left(F^{(5,1)}-F^{(1,1)}F^{(4,0)}\right)
+z^{-4}\left(F^{(1,5)}-F^{(1,1)}F^{(0,4)}\right)
\\
&&+\;z^2\left((F^{(1,1)})^2 F^{(2,0)}
-F^{(2,0)}F^{(2,2)}-F^{(1,1)}F^{(3,1)}+F^{(4,2)}\right)
\\
&&+\;z^{-2}\left((F^{(1,1)})^2 F^{(0,2)}
-F^{(0,2)}F^{(2,2)}-F^{(1,1)} F^{(1,3)}
+F^{(2,4)}\right)\;.
\end{eqnarray*}
Now, having computed the coefficients\footnote{Cf.~eq.~(\ref{F}) and appendix \ref{App1}.}
$F^{(r,s)}_{c}\!\left(\Delta', \Delta_{+}, \tilde\Delta\right)$   
for\footnote{Here, $\Delta(\sigma)\equiv{\sf Q}^2/4-\sigma^2$, cf.~eqs.~(\ref{FR}) and (\ref{delta}).}
$$
\tilde\Delta=\Delta\!\left(\xi/b-\tfrac{b}{4}\right),
\;\;\;\;\;\;
\Delta'=\Delta\!\left(\xi/b+\tfrac{b}{4}\right),
\;\;\;\;\;\;
\Delta_{+}=-\frac{3}{4}b^2-\frac{1}{2},
\;\;\;\;\;\;
c=1+6\left(b+\frac{1}{b}\right)^2,
$$
setting $z={\rm e}^{-2ix}$ and taking into account that $\Lambda=\hat\Lambda/(\epsilon_1 b)$
one can find the limit\footnote{Cf.~eq.~(\ref{MathieuF}).}
$$
\lim\limits_{b\to 0}
\left(1+\frac{\Phi_{r\neq s}^{\bf 0}(\hat\Lambda/(\epsilon_1 b), {\rm e}^{-2ix})}
{\Phi_{r=s}^{\bf 0}(\hat\Lambda/(\epsilon_1 b))}\right) 
$$
order by order, namely 

\medskip\noindent
--- order $\Lambda^2$:
\begin{eqnarray*}
\Lambda^2\,{\cal K}_{2} 
\stackrel{b\to 0}{\longrightarrow}
\left(\frac{\hat\Lambda }{\epsilon _1}\right)^2 
\left(\frac{{\rm e}^{-2ix}}{2\xi-1}-\frac{{\rm e}^{2 i x}}{2 \xi+1}\right)
\equiv
\left(\frac{\hat\Lambda }{\epsilon _1}\right)^2\,{\cal P}_{2}(\xi,x)
\;,
\end{eqnarray*}

\noindent
--- order $\Lambda^4$:
\begin{eqnarray*}
&&\Lambda^4\,{\cal K}_{4} 
\stackrel{b\to 0}{\longrightarrow}
\\
&&\hspace{20pt}
\left(\frac{\hat\Lambda }{\epsilon _1}\right)^4 
\left(\frac{{\rm e}^{4ix}}{4 (\xi +1) (2 \xi +1)}
+\frac{{\rm e}^{-4ix}}{4(\xi -1) (2 \xi -1)}\right)
\equiv\left(\frac{\hat\Lambda }{\epsilon _1}\right)^4{\cal P}_{4}(\xi,x) \;,
\end{eqnarray*}

\noindent
--- order $\Lambda^6$:
\begin{eqnarray*}
&&\Lambda^6\,{\cal K}_{6} 
\stackrel{b\to 0}{\longrightarrow}
\\
&&\hspace{20pt}\left(\frac{\hat\Lambda }{\epsilon _1}\right)^6
\left(\frac{\left(4 \xi ^2-8 \xi +7\right) {\rm e}^{-2 i x}}{4 (\xi -1) (2 \xi -1)^3 (2 \xi
   +1)}-\frac{\left(4 \xi ^2+8 \xi +7\right) {\rm e}^{2 i x}}{4 (\xi +1) (2 \xi -1) (2 \xi
   +1)^3}\right.
\\
&&
+\left.\frac{{\rm e}^{-6 i x}}{12 (\xi -1) (2 \xi -3) (2 \xi -1)}-\frac{{\rm e}^{6 i x}}{12 (\xi +1)
(2 \xi +1) (2 \xi +3)}\right)
\equiv\left(\frac{\hat\Lambda }{\epsilon _1}\right)^6 {\cal P}_{6}(\xi,x) 
\;,
\end{eqnarray*}

\noindent
--- order $\Lambda^8$:
\begin{eqnarray*}
&&\Lambda^8\,{\cal K}_{8} 
\stackrel{b\to 0}{\longrightarrow}
\\[5pt]
&&\hspace{20pt}\left(\frac{\hat\Lambda }{\epsilon _1}\right)^8
\left(\frac{\left(2 \xi ^2-5 \xi +5\right) {\rm e}^{-4 i x}}{3 (\xi -1) (2 \xi -3) (2 \xi -1)^3 (2 \xi
   +1)}+\frac{\left(2 \xi ^2+5 \xi +5\right) {\rm e}^{4 i x}}{3 (\xi +1) (2 \xi -1) (2 \xi +1)^3
   (2 \xi +3)}\right.
\\
&&
   +\left.\frac{{\rm e}^{-8 i x}}{96 (\xi -2) (\xi -1) (2 \xi -3) (2 \xi -1)}+\frac{{\rm e}^{8 i
   x}}{96 (\xi +1) (\xi +2) (2 \xi +1) (2 \xi +3)}\right)
\equiv\left(\frac{\hat\Lambda }{\epsilon _1}\right)^8 {\cal P}_{8}(\xi,x)\;,
\end{eqnarray*}

\noindent
--- order $\Lambda^{10}$:
\begin{eqnarray}
&&\Lambda^{10}\,{\cal K}_{10} 
\stackrel{b\to 0}{\longrightarrow}\nonumber
\\
&&\left(\frac{\hat\Lambda }{\epsilon _1}\right)^{10}
\left(\frac{\left(4 \xi ^2-12 \xi +13\right) {\rm e}^{-6 i x}}{32 (\xi -2) (\xi -1) (2 \xi -3) (2 \xi
   -1)^3 (2 \xi +1)}\right.\nonumber
\\
&&-\left.\frac{\left(4 \xi ^2+12 \xi +13\right) {\rm e}^{6 i x}}{32 (\xi +1) (\xi +2)
   (2 \xi -1) (2 \xi +1)^3 (2 \xi +3)}\right.\nonumber
\\ 
&&+\left.\frac{\left(-32 \xi ^6-80 \xi ^5-64 \xi ^4-32 \xi
   ^3-94 \xi ^2+13 \xi -116\right) {\rm e}^{2 i x}}{6 (\xi -1) (\xi +1) (2 \xi -1)^3 (2 \xi +1)^5
   (2 \xi +3)}\right.\nonumber
\\
&&+\left.\frac{\left(32 \xi ^6-80 \xi ^5+64 \xi ^4-32 \xi ^3+94 \xi ^2+13 \xi
   +116\right) {\rm e}^{-2 i x}}{6 (\xi -1) (\xi +1) (2 \xi -3) (2 \xi -1)^5 (2 \xi
   +1)^3}\right.\nonumber
\\
&&+\left.\frac{{\rm e}^{-10 i x}}{480 (\xi -2) (\xi -1) (2 \xi -5) (2 \xi -3) (2 \xi
   -1)}\right.\nonumber
\\ \label{P10}
&&-\left.\frac{{\rm e}^{10 i x}}{480 (\xi +1) (\xi +2) (2 \xi +1) (2 \xi +3) (2 \xi +5)}\right)
\equiv\left(\frac{\hat\Lambda }{\epsilon _1}\right)^{10} {\cal P}_{10}(\xi,x)\;,
\end{eqnarray}

\noindent
--- order $\Lambda^{12}$:
\begin{eqnarray}
&&\Lambda^{12}\,{\cal K}_{12} 
\stackrel{b\to 0}{\longrightarrow}\nonumber
\\
&&\left(\frac{\hat\Lambda }{\epsilon _1}\right)^{12}
\left(\frac{\left(2 \xi ^2-7 \xi +8\right) {\rm e}^{-8 i x}}{60 (\xi -2) (\xi -1) (2 \xi -5) (2 \xi -3)
   (2 \xi -1)^3 (2 \xi +1)}\right.\nonumber
\\
&+&\left. \frac{\left(2 \xi ^2+7 \xi +8\right) {\rm e}^{8 i x}}{60 (\xi +1)
   (\xi +2) (2 \xi -1) (2 \xi +1)^3 (2 \xi +3) (2 \xi +5)}\right.\nonumber
\\
&&\hspace{-25pt}+\left.\;\frac{\left(704 \xi ^8-3904 \xi
   ^7+8528 \xi ^6-9440 \xi ^5+7780 \xi ^4-5876 \xi ^3+4739 \xi ^2-7672 \xi +4817\right)
   {\rm e}^{-4 i x}}{384 (\xi -2) (\xi -1)^3 (\xi +1) (2 \xi -3) (2 \xi -1)^5 (2 \xi
   +1)^3}\right.\nonumber
\\
&&\left.\hspace{-25pt}+\;\frac{\left(704 \xi ^8+3904 \xi ^7+8528 \xi ^6+9440 \xi ^5+7780 \xi ^4+5876 \xi
   ^3+4739 \xi ^2+7672 \xi +4817\right) {\rm e}^{4 i x}}{384 (\xi -1) (\xi +1)^3 (\xi +2) (2 \xi
   -1)^3 (2 \xi +1)^5 (2 \xi +3)}\right.\nonumber
\\
&&\left.\hspace{-25pt}+\;\frac{{\rm e}^{12 i x}}{5760 (\xi -3) (\xi -2) (\xi -1) (2 \xi
   -5) (2 \xi -3) (2 \xi -1)}\right.\nonumber
\\
\label{P12}
&&\left.\hspace{-25pt}+\;\frac{{\rm e}^{-12 i x}}{5760 (\xi +1) (\xi +2) (\xi +3) (2 \xi +1)
   (2 \xi +3) (2 \xi +5)}\right)
\equiv \left(\frac{\hat\Lambda }{\epsilon _1}\right)^{12}
{\cal P}_{12}(\xi,x).
\end{eqnarray}
The above analysis yields the expansion:
\begin{eqnarray*}
\lim\limits_{b\to 0}
\left(1+\frac{\Phi_{r\neq s}^{\bf 0}(\hat\Lambda/(\epsilon_1 b), {\rm e}^{-2ix})}
{\Phi_{r=s}^{\bf 0}(\hat\Lambda/(\epsilon_1 b))}\right) 
&=& 1 + \left(\frac{\hat\Lambda }{\epsilon _1}\right)^{2}{\cal P}_{2}(\xi,x)
+ \left(\frac{\hat\Lambda }{\epsilon _1}\right)^{4}{\cal P}_{4}(\xi,x)
\\ 
&+&
\left(\frac{\hat\Lambda }{\epsilon _1}\right)^{6}{\cal P}_{6}(\xi,x) +
\ldots + \left(\frac{\hat\Lambda }{\epsilon _1}\right)^{12}{\cal P}_{12}(\xi,x)
+ \ldots
\end{eqnarray*}
which for $\xi=\nu/2$, $h=2\hat\Lambda/\epsilon_1$ and after multiplication
by ${\rm e}^{i\nu x}$ (cf.~eq.~(\ref{MathieuF})) gives the sought eigenfunction:
\begin{eqnarray}\label{MathieuF2}
\psi^{\bf 0}_{\nu}(x) &=& {\rm e}^{i\nu x} + \frac{h^2}{4}\,{\cal R}_{2}(\nu,x)
+\frac{h^4}{32}\,{\cal R}_{4}(\nu,x)+
\frac{h^6}{128}\,{\cal R}_{6}(\nu,x)+
\frac{h^8}{768}\,{\cal R}_{8}(\nu,x)+\ldots\,.\;\;\;\;
\end{eqnarray}
For instance, the coefficients ${\cal R}_{n}(\nu,x)$, $n=2,4,6,8$ explicitly read as follows\footnote{
Coefficients ${\cal R}_{n}$, $n=10, 12$ are 
given by (\ref{P10}), (\ref{P12}) for $\xi=\nu/2$, $h=2\hat\Lambda/\epsilon_1$.}
\begin{eqnarray*}
{\cal R}_{2}&=&\frac{{\rm e}^{i(\nu -2)x}}{\nu -1}-\frac{{\rm e}^{i(\nu +2)x}}{\nu +1}\;,
\\
{\cal R}_{4}&=&
\frac{{\rm e}^{i(\nu +4)x}}{(\nu +1) (\nu +2)}+\frac{{\rm e}^{i(\nu -4)x}}{(\nu -2) (\nu-1)}\;,
\\
{\cal R}_{6}&=&
\frac{\left(\nu ^2-4 \nu +7\right){\rm e}^{i(\nu -2)x}}{(\nu -2) (\nu -1)^3 (\nu
+1)}
-\frac{\left(\nu ^2+4 \nu +7\right){\rm e}^{i(\nu +2)x}}{(\nu -1) (\nu +1)^3 (\nu+2)}
\\
&+&
\frac{{\rm e}^{i(\nu -6)x}}{3 (\nu -3) (\nu -2) (\nu -1)}-\frac{{\rm e}^{i(\nu +6)x}}
{3 (\nu +1) (\nu +2) (\nu +3)}\;,
\\
{\cal R}_{8}&=&
\frac{\left(\nu ^2-5 \nu +10\right) {\rm e}^{i (\nu -4) x}}{(\nu -3) (\nu -2) (\nu -1)^3 (\nu
   +1)}+\frac{\left(\nu ^2+5 \nu +10\right) {\rm e}^{i (\nu +4) x}}{(\nu -1) (\nu +1)^3 (\nu +2)
   (\nu +3)}
\\
&+&\frac{{\rm e}^{i (\nu -8) x}}{8 (\nu -4) (\nu -3) (\nu -2) (\nu -1)}+\frac{{\rm e}^{i (\nu
   +8) x}}{8 (\nu +1) (\nu +2) (\nu +3) (\nu +4)}\;.
\end{eqnarray*}
Finally, let us observe that the combination $\frac{1}{2}(\psi^{\bf 0}_{\nu}(x)+\psi^{\bf 0}_{\nu}(-x))$
yields the {\it noninteger order} ($\nu\notin\mathbb{Z}$) {\it Mathieu cosine function}:\footnote{Cf.~subsection 
$2.16$ in a book by McLachlan \cite{McLachlan:1947}, see also Example $17.1$ in \cite{MuellerKirsten:2006}.}
\begin{eqnarray*}
\frac{1}{2}\left(\psi^{\bf 0}_{\nu}(x)+\psi^{\bf 0}_{\nu}(-x)\right) &=&
\cos(\nu x)-\frac{h^2}{4}
   \left[\frac{\cos ((\nu +2)x)}{\nu +1}-\frac{\cos ((\nu -2)x)}{\nu -1}\right]
   \\
&+&\frac{h^4}{32}\left[\frac{\cos ((\nu -4)x)}{(\nu -2)
   (\nu -1)}+\frac{\cos ((\nu +4)x)}{(\nu +1) (\nu +2)}\right]
   \\
&+&\frac{h^6}{128}\left[\frac{\left(\nu ^2-4 \nu +7\right) \cos ((\nu -2)x)}
{(\nu -2) (\nu -1)^3 (\nu +1)}-\frac{\left(\nu ^2+4 \nu +7\right) \cos ((\nu +2)x)}
{(\nu -1) (\nu +1)^3 (\nu +2)}\right.
\\
&&+\left.\frac{\cos ((\nu -6)x)}
{3(\nu -3) (\nu -2) (\nu -1)}
-\frac{\cos ((\nu +6)x)}{3 (\nu +1) (\nu +2)(\nu +3)}\right]
+\ldots\;
\\
&\equiv& {\rm ce}_{\nu}(x,h^2). 
\end{eqnarray*} 
Therefore, the eigenfunction $\psi^{\bf 0}_{\nu}(x)$ given by our formulae (\ref{MathieuF}), 
(\ref{MathieuF2}) is nothing but the Floquet solution 
${\rm me}_{\nu}(x,h^2)$, $\nu\notin\mathbb{Z}$
known as the {\it Mathieu exponent}
(cf.~Example $17.1$ in \cite{MuellerKirsten:2006}).\footnote{The
coefficient ${\cal R}_4$ in our formula
for $\psi^{\bf 0}_{\nu}(x)$
differs from that presented in 
(\href{http://dlmf.nist.gov/28.15}{{\tt http://dlmf.nist.gov/28.15}}), where
\begin{eqnarray}\label{tab}
{\rm me}_{\nu}(x,h^2) &=& {\rm e}^{i\nu x} - 
\frac{h^2}{4}\left(\frac{1}{\nu+1}{\rm e}^{i(\nu+2)x}-\frac{1}{\nu-1}{\rm e}^{i(\nu-2)x}\right)
\nonumber\\
&+&
\frac{h^4}{32}
\left(\frac{1}{(\nu+1)(\nu+2)}{\rm e}^{i(\nu+4)x} 
+\frac{1}{(\nu-1)(\nu-2)}{\rm e}^{i(\nu-4)x}
\begin{br}-\frac{2(\nu^2+1)}{(\nu^2-1)^2}{\rm e}^{i\nu x}\end{br}\right)
+\ldots\;.
\end{eqnarray}
However, the expression (\ref{tab}) does not satisfy (\ref{cese}).}
The second solution is of the form $\psi^{\bf 0}_{\nu}(-x)\equiv{\rm me}_{\nu}(-x,h^2)$ and it
is the Floquet solution for $-\nu$ and the same eigenvalue (\ref{EigenTest}).
It is known that ${\rm me}_{\nu}(x,h^2)$ and ${\rm me}_{\nu}(-x,h^2)$ obey
(\href{http://dlmf.nist.gov/28.12.iii}{{\tt http://dlmf.nist.gov/28.12.iii}}):
\begin{eqnarray}\label{cese}
{\rm ce}_{\nu}(x,h^2) &=& \tfrac{1}{2}\left({\rm me}_{\nu}(x,h^2)+{\rm me}_{\nu}(-x,h^2)\right),
\nonumber\\
{\rm se}_{\nu}(x,h^2) &=& \tfrac{1}{2}i\left({\rm me}_{\nu}(x,h^2)-{\rm me}_{\nu}(-x,h^2)\right).
\end{eqnarray}
Functions ${\rm ce}_{\nu}$ and ${\rm se}_{\nu}$ constitute another fundamental system of solutions.

So far only the solutions of the noninteger order ($\nu\!\notin\!\mathbb{Z}$) have been discussed.
Hence, the question arises at this point of how to get from the classical limit of the irregular block
the Mathieu eigenvalues and eigenfunctions corresponding to the integer values of the Floquet exponent.
Recall, such solutions are periodic.\footnote{Cf.~appendix \ref{App3}.}
In particular, one can construct the solutions of periods $\pi$ or $2\pi$ ($q=h^2$):
\begin{list}{}{\itemindent=2mm \parsep=0mm \itemsep=0mm \topsep=0mm}
\item[---] the cosine-elliptic ${\rm ce}_{m}(x;q)$, $m=0,1,2,\ldots$, 
that corresponds at $q=0$ with $\cos mx$, for instance:
\begin{eqnarray*}
{\rm ce}_{1}(x;q) &=& \cos x 
- \frac{1}{8}q\cos 3x + \frac{1}{64}q^2\left(\frac{1}{3}\cos 5x -\cos 3x\right)
\\
&-&
\frac{1}{512}q^3\left(\frac{1}{3}\cos 3x -\frac{4}{9}\cos 5x +\frac{1}{18}\cos 7x\right)
+\ldots;
\end{eqnarray*}
\item[---] the sine-elliptic ${\rm se}_{m}(x;q)$, $m=1,2,\ldots$, 
that corresponds at $q=0$ with $\sin mx$, for example:
\begin{eqnarray*}
{\rm se}_{1}(x;q) &=& \sin x - \frac{1}{8}q\sin 3x 
+ \frac{1}{64}q^2\left(\sin 3x + \frac{1}{3}\sin 5x\right)
\\
&-&
\frac{1}{512}q^3\left(\frac{1}{3}\sin 3x -\frac{4}{9}\sin 5x +\frac{1}{18}\sin 7x\right)
+\ldots.
\end{eqnarray*}
\end{list}
The functions ${\rm ce}_{m}$ and ${\rm se}_{m}$ have period $\pi$ if $m$ is even and
period $2\pi$ if $m$ is odd. The corresponding eigenvalues $\lambda$ denoted by $a_{m}(q)$
for ${\rm ce}_{m}$ and $b_{m}(q)$ for ${\rm se}_{m}$ are called {\it characteristic numbers},
e.g.:
\begin{eqnarray*}
a_{1}(q) &=&1+q-\frac{1}{8}q^2-\frac{1}{64}q^3-\frac{1}{1536}q^4+\frac{11}{36864}q^5+\ldots,
\\
b_{1}(q) &=&1-q-\frac{1}{8}q^2+\frac{1}{64}q^3-\frac{1}{1536}q^4-\frac{11}{36864}q^5+\ldots.
\end{eqnarray*}
For any $q>0$ characteristic numbers form the band/gap structure: 
$a_{0}<b_1<a_1<b_2<a_2\ldots$. For large
$m$ the leading terms of the $a_m$ and $b_m$ are 
(\href{http://dlmf.nist.gov/28.6.E14}{{\tt http://dlmf.nist.gov/28.6.E14}}):
\begin{equation*}
  \left.
    \begin{array}{rl}
      a_{m}(q)\\
      b_{m}(q)
    \end{array} \right\}= m^2 +
\frac{q^2}{2 \left(m^2-1\right)}+
\frac{\left(5m^2+7\right)q^4}{32 \left(m^2-4\right) \left(m^2-1\right)^3}+
\frac{\left(9m^4+58m^2+29\right)q^{6}}{64 \left(m^2-9\right) \left(m^2-4\right) \left(m^2-1\right)^5}
+\ldots\;.
\end{equation*}
Notice that for $m\!=\!\nu$ the above expression  matches (\ref{SmallNonIn}) and therefore can be recovered
from the classical irregular block with $\delta\!=\!\frac{1}{4}(1-m^2)$ (cf.~(\ref{eigenvalue}),
(\ref{EigenTest})) or from the gauge theory counterpart of (\ref{eigenvalue}), i.e.: 
$$
\frac{1}{\epsilon_1}\,\hat\Lambda\partial_{\hat\Lambda}
W^{{\rm SU(2)}, N_f=0}\left(\hat\Lambda,a,\epsilon_1\right),
$$
where $a=\frac{1}{2}m\epsilon_1$, cf.~\cite{Piatek:2014lma,Basar:2015xna}.
To conclude, regardless of the coincidence described above, work 
is in progress in order to find a mechanism which allows to derive from the conformal blocks 
the eigenvalues in the case of the finite integer 
values of $m$ and the corresponding integer order solutions.

\section{Concluding remarks and open problems}
\label{sec5}
In the present paper we have shown that the $N_f\!=\!0$ 
classical irregular block solves the eigenvalue problem for the Mathieu operator.
The statement that the Mathieu eigenvalue for  
small $h^2\!=\!4\hat\Lambda^2/\epsilon_{1}^{2}$ (weakly--coupled region)
and noninteger characteristic exponent $\nu\notin\mathbb{Z}$ 
is determined by the classical zero flavor irregular block has been 
already stated in our previous work \cite{Piatek:2014lma}.
The new result of the present work is the derivation of an expression 
of the corresponding eigenfunction. Moreover,
it has been shown that the formula (\ref{MathieuF})
reproduces the known solution of the Mathieu equation with the eigenvalue (\ref{EigenTest}).
Therefore, we have established a link between the Mathieu equation and 
its realization within two-dimensional CFT. 
This result paves the way for a new interesting line of research.
Concretely, one can try to study other regions of 
the spectrum of the Mathieu operator by means of $2d$ CFT tools.
Indeed, two interesting questions arise at this point:
($i$) How is it possible to derive from the
irregular block the solutions with integer values of the Floquet parameter?
($ii$) How within $2d$ CFT one can get the solutions in the other regions of the spectrum?
The answer to the first question needs more studies. It seems that also the second question
is reasonable. Let us remember that the quantum irregular block can be obtained
from the four-point block on the sphere in the so-called decoupling limit of the
external conformal weights (cf.~eq.~(\ref{decoupling})). 
However, to our knowledge, such limit has been discussed  
only for the $s$-channel four-point block \cite{Marshakov:2009}. Therefore, it arises at this point 
the question of what happens if we take the decoupling limit from the four-point blocks
in the other channels, i.e.~$t$ or $u$. The latter case appears to be especially interesting
since in the $u$-channel four-point block the invariant ratio is $\frac{1}{x}$.
Hence, the question is whether we can obtain the irregular block with 
the expansion parameter $\frac{1}{\Lambda}=(\frac{\hat\Lambda}{\epsilon_1 b})^{-1}$.
Secondly, if this is possible, do we get in the classical limit 
the classical irregular block determining the strongly-coupled region of the Mathieu spectrum? 
In addition, let us note that the $s$ and $u$-channel four-point blocks
are related by the {\it braiding relation} which, in general, has the form of an 
integral transform with a complicated kernel 
--- the so-called {\it braiding matrix} (cf.~e.g.~\cite{Hadasz:2004cm}).
However, when one of the four external conformal weights becomes degenerate, then
the integral transform reduces to the known formula for the analytic continuation of the 
hypergeometric function from the vicinity of a point $x$ to that of the point $\frac{1}{x}$.
It seems to be technically possible to take combinations of the decoupling and classical limits
on both sides of the braiding relation (at least) in the degenerate case
and to obtain as a result a `duality relation' for the classical irregular
blocks. As one can expect, in this way it becomes feasible to establish a formula that
continues the Mathieu eigenvalue from the weakly-coupled region to that of strong coupling.
Work is in progress in order to verify this hypothesis.

The derivation of the Mathieu equation within the formalism of two-dimensional
conformal field theory is based on conjectures concerning the asymptotical behavior 
of irregular blocks in the classical limit, cf.~eqs.~(\ref{ClIrrblock0}) and (\ref{ClAsymp}).
We recall that the coefficient of the irregular block expansion is a ratio of polynomials
in $\Delta$ and $c$. Our idea of proving the existence of the classical 
irregular block was to estimate the degree of the irregular block coefficient
as a polynomial in the parameter $b$. To accomplish this task
we used methods which had previously been used to prove the Kac determinant formula, cf.~\cite{KR}.
However, these techniques have proven to be too weak to give a complete answer.
As a result, we have obtained the classical irregular block at the leading order only.
It is therefore necessary to use other methods to try to prove eq.~(\ref{ClIrrblock0}).
It seems that there are two available approaches to solve this problem.
The first way is to use a representation of the Gaiotto states in terms of Jack polynomials.
As S.~Yanagida has shown in ref.~\cite{Yanagida}, it is possible to represent the Gaiotto 
state for the pure gauge theory using the Jack polynomials. The coefficients relating the Jack
polynomials to the Gaiotto state are found explicitly in \cite{Yanagida}. 
The inner product, by means of which one computes the norm of the pure gauge Gaiotto state,
induces by the bosonization map and the parameter dependent isomorphism, 
the inner product in the space spanned by the Jack polynomials. 
However, these Jack polynomials are not orthogonal within this inner product
and one has to expand them in a new basis of Jack polynomials depending on a different parameter 
and orthogonal with respect to the induced inner product.
The coefficients relating the two bases of the Jack polynomials are not known explicitly. 
The lack of the explicit form of the 
coefficients hampers the effort to find the classical limit of the pure gauge conformal block.

The second and in our opinion more promising way to get eq.~(\ref{ClIrrblock0}) in particular
and more in general analogous results for the regular blocks,
is the application of the Fock space free field realization of the conformal blocks. 
In the case of discrete spectrum this approach is even mathematically rigorous \cite{BC,B,B2,ConstanScharf}. 
As a result one gets the Dotsenko--Fateev(-like) integral representations of the conformal blocks and 
hence one can try to use the methods of matrix models (beta--ensembles) 
in the computation of the classical limit of these blocks, cf.~\cite{Rim:2015aha}. 
However, it should be stressed that the link between the integral 
and power series representations of conformal blocks is not completely understood cf.~e.g.~\cite{Mironov:2010zs}. 
In our opinion, also the operator realization of conformal blocks requires further work. 
For instance, the Fock space representation of the chiral vertex operator
with the three independent general conformal weights,
whose compositions in the matrix element lead to integral formulas,
needs to be developed, cf.~\cite{Teschner:2001rv,Teschner:2003en}.

The main claim of this work --- formulas (\ref{eigenvalue}) and (\ref{MathieuF}) --- 
possess fascinating generalizations. The simplest extension is 
to consider irregular blocks with 
$N_f\!=\!1,2$ flavors.\footnote{By the time this research was in progress 
the paper \cite{Rim:2015tsa} appeared which mentioned this problem.}
As a preview of the results which will be reported in our
next papers let us only mention that also in the cases $N_f\!=\!1,2$ the classical limit
of the irregular blocks exists and yields a consistent definition of the classical blocks. 
In these cases we have also found explicit formulas for the eigenvalues and
the eigenfunctions of operators emergent in the classical limit
of the null vector decoupling equations.
For $N_f\!=\!2$ one gets a Schr\"{o}dinger operator containing a generalization
of the Mathieu potential. These further developments will confirm the validities 
of $2d$ CFT technics in the investigation of different regions of the spectra
of the investigated operators.

\appendix
\section{Coefficients \texorpdfstring{$F^{(r,s)}$}{Frs}}
\renewcommand{\theequation}{A.\arabic{equation}}
\setcounter{equation}{0}
\label{App1}
The first few coefficient 
$F^{(r,s)}_{c}(\Delta_1,\Delta_2,\Delta_3)$, e.g.:
\begin{eqnarray*}
F^{(1,0)} &=& \frac{\Delta _{1}+\Delta _{2}-\Delta _{3}}{2 \Delta _{1}},
\\
F^{(0,1)} &=& \frac{-\Delta _{1}+\Delta _{2}+\Delta _{3}}{2 \Delta _{3}},
\\
F^{(2,0)} &=&
\frac{\frac{1}{2} \left(\Delta _1+\Delta _2-\Delta _3\right) \left(\Delta _1+\Delta
   _2-\Delta _3+1\right) \left(c+8 \Delta _1\right)-6 \Delta _1 \left(\Delta _1+2 \Delta
   _2-\Delta _3\right)}{2 \Delta _1 \left(2 \Delta _1 \left(c+8 \Delta
   _1-5\right)+c\right)},
\\
F^{(0,2)} &=&
\frac{\frac{1}{2} \left(\Delta _1-\Delta _2-\Delta _3-1\right) \left(\Delta _1-\Delta
   _2-\Delta _3\right) \left(c+8 \Delta _3\right)-6 \Delta _3 \left(-\Delta _1+2 \Delta
   _2+\Delta _3\right)}{2 \Delta _3 \left(2 \Delta _3 \left(c+8 \Delta
   _3-5\right)+c\right)},
\end{eqnarray*}
and next up to the order $\Lambda^8$ (cf.~(\ref{rexp})) can be easily and quickly computed using computer.
The time of computation dramatically increases for the coefficients appearing in higher orders
of the expansion (\ref{rexp})
and results become very complicated. However, it seems to be possible to find recurrence relations for 
$F^{(r,s)}_{c}(\Delta_1,\Delta_2,\Delta_3)$ which really would improve 
speed of calculations and as a result it would gave an efficient method of calculation of the Mathieu function.

\renewcommand{\theequation}{B.\arabic{equation}}
\setcounter{equation}{0}
\label{App2}

\section{Mathieu equation}
\renewcommand{\theequation}{C.\arabic{equation}}
\setcounter{equation}{0}
\label{App3}
The standard form of the Mathieu equation with parameters $(a,q)$ 
(\href{http://dlmf.nist.gov/28.2}{{\tt http://dlmf.nist.gov/28.2}})
or equivalently \cite{MuellerKirsten:2006} $(\lambda, h^2)$ reads as follows
\begin{equation}\label{M}
\psi'' + \left(a-2q\cos2z\right)\psi\;=\;0
\;\;\;\leftrightarrow\;\;\;
\psi'' + \left(\lambda-2h^2\cos2z\right)\psi\;=\;0.
\end{equation}
A solution $\psi$ with given initial constant values of $\psi$ and $\psi'$
at some point $z_0$ is an entire function of the three variables: $z,a,q$ 
($\Leftrightarrow$ $z,\lambda, h^2$).

The Floquet theorem states that the Mathieu eq.~(\ref{M}) has a nontrivial 
solution $\psi(z)$ such that 
\begin{equation}\label{Fc}
\psi(z+\pi)\;=\;\sigma\,\psi(z)
\end{equation}
with $\sigma$ being a root of the eq.:
$$
\begin{vmatrix}
\,\psi_{1}(\pi)-\sigma  & \psi_{2}(\pi)\, \\
\,\psi_{1}'(\pi) & \psi_{2}'(\pi)-\sigma\,
\end{vmatrix}
\;=\;0\;\;,
$$
where $\psi_{1}(z)$ and $\psi_{2}(z)$ are even and odd, respectively, 
normalized $\left(\psi_{1}\psi_{2}'-\psi_{1}'\psi_{2}\right)\Big|_{z=0}=1$
linearly independent solutions. Equivalently, the coefficients $c_1$ and $c_2$
in the solution $\psi=c_1\psi_1+c_2\psi_2$, which obeys the condition 
(\ref{Fc}), are given as an eigenvector of eq.:
$$
\begin{pmatrix}
\,\psi_{1}(\pi)  & \psi_{2}(\pi)\, \\
\,\psi_{1}'(\pi) & \psi_{2}'(\pi)\,
\end{pmatrix}
\begin{pmatrix}
\,c_1 \\
\,c_2 \,
\end{pmatrix}
\;=\; \sigma
\begin{pmatrix}
\,c_1 \\
\,c_2 \,
\end{pmatrix}\;\;.
$$

A solution of the Mathieu eq.~in the form given by the Floquet theorem
is called a Floquet solution. In order to gain more understanding of the Floquet solution
let us define the quantities $\nu$ and $y$ such that 
$$
\sigma\;=\;{\rm e}^{i\pi\nu}, \;\;\;\;\;\;\;\;
y(z)\;=\;{\rm e}^{-i\nu z}\,\psi(z),
$$
where $\psi(z)$ fulfills (\ref{Fc}). The definition has the effect that
$$
y(z+\pi)={\rm e}^{-i\nu (z+\pi)}\,\psi(z+\pi)={\rm e}^{-i\nu z}\psi(z) = y(z)
$$
which shows that $y(z)$ is a periodic function of $z$ with period $\pi$.
Moreover, 
\begin{equation}\label{Fl}
\psi(z)\;=\;{\rm e}^{i\nu z}\,y(z)
\end{equation}
showing that a Floquet solution $\psi(z)$ consists of a periodic function of $z$
multiplied by a complex exponential in $z$.
The quantity $\nu$ which controls the exponential behavior is known as the 
characteristic or Floquet exponent of $\psi$.

The Floquet exponent $\nu$ is determined by the eq.:
\begin{equation}\label{qc}
\cos\pi\nu\;=\; \psi_{1}(\pi;a,q).
\end{equation}
Eq.~(\ref{qc}) allows to express the eigenvalue $a$ (or $\lambda$) in terms of $q$ (or $h^2$) 
and Floquet parameter $\nu$. 
However, usefulness of eq.~(\ref{qc}) is restricted by an ability to calculate the normalized 
Mathieu function $\psi_1$. The eigenfunction $\psi_1$ can be found as an expansion in terms 
of other functions, in particular, in terms of trigonometric functions. Indeed, the meaning of 
eq.~(\ref{qc}) is that the Floquet exponent is determined by the value at $z=\pi$ of the solution
which is even around $z=0$. To the lowest order, i.e.~for $q=h^2=0$, the even solution
around $z=0$ is $\psi_{1}^{(0)}\!(z)=\cos\!\sqrt{\lambda}z$. Therefore, from~(\ref{qc})
for $q=h^2=0$ one gets $\nu=\!\!\sqrt{\lambda}$, and more in general 
$\nu^2=\lambda+O\!\left(h^2\right)$.
One can derive various terms of this expansion perturbatively. Indeed, for small $q=h^2$
the eigenvalue $\lambda$ as a function of $\nu$ and $h^2$ explicitly reads as follows 
(\href{http://dlmf.nist.gov/28.15}{{\tt http://dlmf.nist.gov/28.15}} and
cf.~Example $17.1$ in \cite{MuellerKirsten:2006})
\begin{equation}\label{SmallNonIn}
\lambda_{\nu}(h^2)\;=\;\nu^2 +
\frac{h^4}{2 \left(\nu ^2-1\right)}+
\frac{\left(5 \nu ^2+7\right)h^8}{32 \left(\nu ^2-4\right) \left(\nu ^2-1\right)^3}+
\frac{\left(9 \nu ^4+58 \nu ^2+29\right)h^{12}}{64 \left(\nu ^2-9\right) \left(\nu ^2-4\right) \left(\nu^2-1\right)^5}
+\ldots\;.
\end{equation}
The expansion (\ref{SmallNonIn}) holds for noninteger values of $\nu\notin\mathbb{Z}$. 
The corresponding eigenfunction for small $q=h^2$ and $\nu\notin\mathbb{Z}$ is of the form
\begin{eqnarray*}
{\rm me}_{\nu}(z,h^2) &=& {\rm e}^{i\nu z} - 
\frac{h^2}{4}\left(\frac{1}{\nu+1}{\rm e}^{i(\nu+2)z}-\frac{1}{\nu-1}{\rm e}^{i(\nu-2)z}\right)
+\ldots\;.
\end{eqnarray*}

The Mathieu equation admits periodic solutions. Indeed, the Floquet solution will be 
periodic for special values of $\sigma$. A necessary condition for periodicity 
is that $|\sigma|=1$. Since the Floquet solution (\ref{Fl}) contains the factor $y$ 
that is periodic with period $\pi$, $\psi$ will be periodic with period
\begin{list}{}{\itemindent=2mm \parsep=0mm \itemsep=0mm \topsep=0mm}
\item[a)] $\pi$ if $\nu=0,2,\ldots$ $\Leftrightarrow$ $\sigma=1$,
\item[b)] $2\pi$ if $\nu=1,3,\ldots$ $\Leftrightarrow$ $\sigma=-1$,
\item[c)] $s\pi$ if $\nu=2r/s$, where $r,s>2$ are integers with no common divisors.
\end{list}
For physical reasons the solutions of most importance are those 
with periods $\pi$ or $2\pi$, and these are the ${\rm ce}_{m}$ and
${\rm se}_{m}$ introduced in the main text.

\section*{Acknowledgments}
The authors are grateful to Franco Ferrari for useful discussions, very valuable advices and careful reading of the manuscript. M.P.~is also grateful to Franco Ferrari for his kind hospitality during stays in Szczecin.


\begin{thebibliography}{10}

\bibitem{Piatek:2014lma}
M.~Piatek and A.R.~Pietrykowski, {\it Classical irregular block, ${\cal N} = 2$ pure gauge theory and Mathieu equation}, JHEP {\bf 12} (2014) 032, [\href{http://xxx.lanl.gov/abs/1407.0305}{{\tt arXiv:1407.0305}}].


\bibitem{Nekrasov:2010ka}
N.~Nekrasov, E.~Witten, {\it The Omega Deformation, Branes, Integrability, and Liouville Theory},
JHEP {\bf 09} (2010) 092, [\href{http://xxx.lanl.gov/abs/1002.0888}{{\tt arXiv:1002.0888}}]. 


\bibitem{Teschner:2010je}
J.~Teschner, {\it Quantization of the Hitchin moduli spaces, Liouville theory, and the geometric Langlands correspondence I}, Adv.
Theor. Math. Phys. {\bf 15} (2011) 471--564, [\href{http://xxx.lanl.gov/abs/1005.2846}{{\tt arXiv:1005.2846}}]. 


\bibitem{Kozlowski:2010tv}
K.K.~Kozlowski and J.~Teschner, {\it TBA for the Toda chain},  [\href{http://xxx.lanl.gov/abs/1006.2906}{{\tt arXiv:1006.2906}}].


\bibitem{Meneghelli:2013tia}
C.~Meneghelli and G.~Yang, {\it Mayer-Cluster Expansion of Instanton Partition Functions and Thermodynamic Bethe Ansatz}, 
JHEP {\bf 05} (2014) 112, [\href{http://xxx.lanl.gov/abs/1312.4537}{{\tt arXiv:1312.4537}}].


\bibitem{Vartanov:2013ima}
J.~Teschner and G.S.~Vartanov, {\it Supersymmetric gauge theories, quantization of $\mathcal{M}\_{\mathrm{flat}}$, and
conformal field theory}, Adv. Theor. Math. Phys. {\bf 19} (2015) 1--135, 
[\href{http://xxx.lanl.gov/abs/1302.3778}{{\tt arXiv:1302.3778}}].


\bibitem{Belavin:2011js}
A. Belavin, V. Belavin, {\it AGT conjecture and Integrable structure of Conformal field theory for $c=1$},
Nucl. Phys. {\bf B 850} (2011) 199--213, [\href{http://xxx.lanl.gov/abs/1102.0343}{{\tt arXiv:1102.0343}}].


\bibitem{Fateev:2011hq}
V.A. Fateev, A.V. Litvinov, {\it Integrable structure, W-symmetry and AGT relation}, 
JHEP {\bf 01} (2012) 051, [\href{http://xxx.lanl.gov/abs/1109.4042}{{\tt arXiv:1109.4042}}].


\bibitem{Alba:2010qc}
V.A. Alba, V.A. Fateev, A.V. Litvinov, G.M. Tarnopolsky,
{\it On combinatorial expansion of the conformal blocks arising from AGT conjecture},
Lett. Math. Phys. {\bf 98} (2011) 33-64, [\href{http://xxx.lanl.gov/abs/1012.1312}{{\tt arXiv:1012.1312}}].


\bibitem{Tai:2010ps}
Ta-Sheng Tai, {\it Uniformization, Calogero-Moser/Heun duality and Sutherland/bubbling pants},
JHEP {\bf 10} (2010) 107, [\href{http://xxx.lanl.gov/abs/1008.4332}{{\tt arXiv:1008.4332}}].


\bibitem{Muneyuki:2011qu}
K.~Muneyuki, T.~Tai, N.~Yonezawa, R.~Yoshioka, {\it Baxter's T-Q equation,
$SU(N)/SU(2)^{N-3}$ correspondence and $\Omega$-deformed Seiberg-Witten prepotential},
JHEP {\bf 09} (2011) 125, [\href{http://xxx.lanl.gov/abs/1107.3756}{{\tt arXiv:1107.3756}}].


\bibitem{Itoyama:2015xia}
H.~Itoyama and R.~Yoshioka, {\it Developments of theory of effective prepotential from extended Seiberg-Witten system and
matrix models}, [\href{http://xxx.lanl.gov/abs/1507.00260}{{\tt arXiv:1507.00260}}].


\bibitem{Poghossian:2010pn}
R.~Poghossian, {\it Deforming SW curve}, JHEP {\bf 04} (2011) 033, 
[\href{http://xxx.lanl.gov/abs/1006.4822}{{\tt arXiv:1006.4822}}].


\bibitem{Fucito:2011pn}
F.~Fucito, J.F.~Morales, D.~Ricci Pacifici, and R.~Poghossian, {\it Gauge theories on $\Omega$-backgrounds from non commutative
Seiberg-Witten curves}, JHEP {\bf 05} (2011) 098, [\href{http://xxx.lanl.gov/abs/1103.4495}{{\tt arXiv:1103.4495}}].


\bibitem{Maruyoshi:2010}
K.~Maruyoshi and M.~Taki, {\it {Deformed Prepotential, Quantum Integrable
  System and Liouville Field Theory}},  {\em Nucl. Phys.} {\bf B841} (2010)
  388--425, [\href{http://xxx.lanl.gov/abs/1006.4505}{{\tt arXiv:1006.4505}}].


\bibitem{Bonelli:2009zp}
G. Bonelli, A. Tanzini, {\it Hitchin systems, N=2 gauge theories and W-gravity},
Phys. Lett. {\bf B 691} (2010) 111--115, 
[\href{http://xxx.lanl.gov/abs/0909.4031}{{\tt arXiv:0909.4031}}].


\bibitem{Bonelli:2011na}
G.~Bonelli, K.~Maruyoshi and A.~Tanzini, {\it Quantum Hitchin Systems via beta-deformed Matrix Models},
[\href{http://xxx.lanl.gov/abs/1104.4016}{{\tt arXiv:1104.4016}}].


\bibitem{Piatek:2011tp}
M. Piatek, {\it Classical conformal blocks from TBA for the elliptic Calogero-Moser system},
JHEP {\bf 06} (2011) 050, [\href{http://xxx.lanl.gov/abs/1102.5403}{{\tt arXiv:1102.5403}}].


\bibitem{Ferrari:2012gc}
F.~Ferrari and M.~Piatek, {\it Liouville theory, N=2 gauge theories and accessory parameters}, 
JHEP {\bf 05} (2012) 025, [\href{http://xxx.lanl.gov/abs/1202.2149}{{\tt arXiv:1202.2149}}].


\bibitem{Ferrari:2012qj}
F.~Ferrari and M.~Piatek, {\it On a singular Fredholm-type integral equation arising in N=2 super Yang-Mills theories},
Phys. Lett. {\bf B718} (2013) 1142–1147, [\href{http://xxx.lanl.gov/abs/1202.5135}{{\tt arXiv:1202.5135}}].


\bibitem{Ferrari:2014nba}
F.~Ferrari and M.~Piatek, {\it On a path integral representation of the Nekrasov instanton partition function and its Nekrasov--Shatashvili limit}, Can. J. Phys. {\bf 92} (2014) 267--270.


\bibitem{Tan:2013tq}
M.-C. Tan, {\it {M-Theoretic Derivations of 4d-2d Dualities: From a Geometric
  Langlands Duality for Surfaces, to the AGT Correspondence, to Integrable
  Systems}},  JHEP {\bf 1307} (2013) 171,
  [\href{http://xxx.lanl.gov/abs/1301.1977}{{\tt arXiv:1301.1977}}].
  

\bibitem{Teschner:2014oja}
J.~Teschner, {\it Exact results on N=2 supersymmetric gauge theories},  
[\href{http://xxx.lanl.gov/abs/1412.7145}{{\tt arXiv:1412.7145}}].


\bibitem{Alday:2009aq}
L.F.~Alday, D.~Gaiotto, and Y.~Tachikawa, {\it Liouville Correlation Functions from Four-dimensional Gauge Theories}, Lett. Math.
Phys. {\bf 91} (2010) 167--197, [\href{http://xxx.lanl.gov/abs/0906.3219}{{\tt arXiv:0906.3219}}].
  

\bibitem{NS:2009}  
N.A.~Nekrasov and S.L.~Shatashvili, {\it Quantization of Integrable Systems 
and Four Dimensional Gauge Theories}, in {\it Proceedings of XVIth International 
Congress on Mathematical Physics}, Prague Czech Republic (2009), 
[\href{http://xxx.lanl.gov/abs/0908.4052}{{\tt arXiv:0908.4052}}].


\bibitem{Nekrasov:2009uh}
N. Nekrasov, S. Shatashvili, {\it Supersymmetric vacua and Bethe
ansatz}, Nucl. Phys. Proc. Suppl. {\bf 192-193} (2009) 91--112, 
[\href{http://xxx.lanl.gov/abs/0901.4744}{{\tt arXiv:0901.4744}}].


\bibitem{Nekrasov:2009ui}
N. Nekrasov, S. Shatashvili, {\it Quantum integrability and
supersymmetric vacua}, Prog. Theor. Phys. Suppl. {\bf 177} (2009) 105--119, 
[\href{http://xxx.lanl.gov/abs/0901.4748}{{\tt arXiv:0901.4748}}].


\bibitem{Nekrasov:2011bc}
N. Nekrasov, A. Rosly, S. Shatashvili, {\it Darboux coordinates, Yang-Yang functional, and gauge theory},
Nucl. Phys. Proc. Suppl. {\bf 216} (2011) 69--93,
[\href{http://xxx.lanl.gov/abs/1103.3919}{{\tt arXiv:1103.3919}}].


\bibitem{Nekrasov:2012xe}
N.~Nekrasov and V.~Pestun, {\it Seiberg-Witten geometry of four dimensional N=2 quiver gauge theories},
[\href{http://xxx.lanl.gov/abs/1211.2240}{{\tt arXiv:1211.2240}}].


\bibitem{Nekrasov:2013xda}
N.~Nekrasov, V.~Pestun, and S.~Shatashvili, {\it Quantum geometry and quiver gauge theories},
[\href{http://xxx.lanl.gov/abs/1312.6689}{{\tt arXiv:1312.6689}}].


\bibitem{Dorn:1994xn}
H.~Dorn, H.J.~Otto, {\it Two  and  three  point  functions  in  Liouville  theory},  Nucl.  Phys.
{\bf B 429}  (1994) 375--388, [\href{http://xxx.lanl.gov/abs/hep-th/9403141}{{\tt hep-th/9403141}}].


\bibitem{Zamolodchikov:1995aa}
A.B.~Zamolodchikov and A.B.~Zamolodchikov, {\it {Structure constants and
  conformal bootstrap in Liouville field theory}},  Nucl. Phys. {\bf B 477}
  (1996) 577--605, [\href{http://xxx.lanl.gov/abs/hep-th/9506136}{{\tt
  hep-th/9506136}}].


\bibitem{Nekrasov:2002qd}
N.~Nekrasov, {\it Seiberg-Witten  prepotential  from instanton counting}, 
Adv. Theor. Math. Phys. {\bf 7} (2004) 831--864, [\href{http://xxx.lanl.gov/abs/hep-th/0206161}{{\tt hep-th/0206161}}].


\bibitem{Nekrasov:Okounkov:2003}
N.~Nekrasov, A.~Okounkov, {\it Seiberg-Witten theory and random partitions},
[\href{http://xxx.lanl.gov/abs/hep-th/0306238}{{\tt hep-th/0306238}}].


\bibitem{Wyllard:2009hg}
N.~Wyllard, {\it $A_{N-1}$ conformal Toda field theory correlation
functions from conformal N = 2 SU(N) quiver gauge theories}, JHEP {\bf 11} (2009) 002,
[\href{http://xxx.lanl.gov/abs/0907.2189}{{\tt arXiv:0907.2189}}].


\bibitem{Mironov:2009by}
A.~Mironov, A.~Morozov, {\it On AGT relation in the case of U(3)},
Nucl. Phys. {\bf B 825} (2010) 1--37, [\href{http://xxx.lanl.gov/abs/0908.2569}{{\tt arXiv:0908.2569}}].


\bibitem{Gaiotto:2009}
D.~Gaiotto, {\it {Asymptotically free N=2 theories and irregular conformal
  blocks}},  [\href{http://xxx.lanl.gov/abs/0908.0307}{{\tt arXiv:0908.0307}}].


\bibitem{Marshakov:2009}
A.~Marshakov, A.~Mironov, and A.~Morozov, {\it {On non-conformal limit of the
  AGT relations}},  {\em Phys. Lett.} {\bf B682} (2009) 125--129,
  [\href{http://xxx.lanl.gov/abs/0909.2052}{{\tt arXiv:0909.2052}}].
  
  
\bibitem{Alba:2009fp}
V.~Alba and A.~Morozov, {\it {Non-conformal limit of AGT relation from the
  1-point torus conformal block}},  {\em JETP Lett.} {\bf 90} (2009) 708--712,
  [\href{http://xxx.lanl.gov/abs/0911.0363}{{\tt arXiv:0911.0363}}].  


\bibitem{Hadasz:2010xp}
L.~Hadasz, Z.~Jaskolski, and P.~Suchanek, {\it {Proving the AGT relation for
  $N_f = 0,1,2$ antifundamentals}},  JHEP {\bf 1006} (2010) 046,
  [\href{http://xxx.lanl.gov/abs/1004.1841}{{\tt arXiv:1004.1841}}].


\bibitem{Gaiotto:2012sf}
D.~Gaiotto and J.~Teschner, {\it Irregular singularities in Liouville theory and Argyres-Douglas type gauge theories, I},
JHEP {\bf 1212} (2012) 050, [\href{http://xxx.lanl.gov/abs/1203.1052}{{\tt arXiv:1203.1052}}].


\bibitem{Maulik:2012wi}
D.~Maulik and A.~Okounkov, {\it {Quantum Groups and Quantum Cohomology}},
  [\href{http://xxx.lanl.gov/abs/1211.1287}{{\tt arXiv:1211.1287}}].  


\bibitem{YY}
C.N.~Yang,  C.P.~Yang, {\it Thermodinamics  of  a
one-dimensional  system  of  bosons  with  repulsive delta-function
interaction}, J. Math. Phys. {\bf 10} (1969) 1115.


\bibitem{Basar:2015xna}
G.~Basar and G.V.~Dunne, {\it Resurgence and the Nekrasov-Shatashvili limit: 
connecting weak and strong coupling in the Mathieu and Lam\'{e} systems}, JHEP {\bf 02} (2015) 160, 
[\href{http://xxx.lanl.gov/abs/1501.05671}{{\tt arXiv:1501.05671}}].


\bibitem{Piatek:2013ifa}
M.~Piatek, {\it Classical torus conformal block, ${\cal N} = 2^*$ twisted superpotential and the accessory parameter of Lam\'{e} 
equation}, JHEP {\bf 1403} (2014) 124,
[\href{http://xxx.lanl.gov/abs/1309.7672}{{\tt arXiv:1309.7672}}].


\bibitem{Belavin:1984vu}
A.~Belavin, A.~M. Polyakov, and A.~Zamolodchikov, {\it {Infinite Conformal
Symmetry in Two-Dimensional Quantum Field Theory}},  {\em Nucl. Phys.} {\bf
B241} (1984) 333--380.


\bibitem{Moore:1988qv}
G.~W.~Moore and N.~Seiberg,
{\it Polynomial Equations For Rational Conformal Field Theories},
Phys.\ Lett.\ B {\bf 212} (1988) 451.
{\it Classical And Quantum Conformal Field Theory},
Commun.\ Math.\ Phys.\  {\bf 123} (1989) 177.


\bibitem{FFK}
G.~Felder, J.~Fr\"{o}hlich, G. Keller, 
{\it On the structure of unitary conformal field theory}, 
I, Commun. Math. Phys. {\bf 124} (1989) 417-463; II, Commun. Math. Phys. {\bf 130} (1990) 1-49.


\bibitem{Teschner:2001rv}
J.~Teschner, {\it {Liouville theory revisited}},  {\em Class. Quant. Grav.}
{\bf 18} (2001) R153--R222,
[\href{http://xxx.lanl.gov/abs/hep-th/0104158}{{\tt hep-th/0104158}}].


\bibitem{Teschner:2003en}
J.~Teschner, {\it A lecture on the Liouville vertex operators}, hep-th/0303150.


\bibitem{K1}
V.~G. Kac, {\it Contravariant form for infinite-dimensional Lie algebras and superalgebras}, 
Lect. Notes in Phys. {\bf 94} (1979) 441-445.


\bibitem{FF1}
B.~L. Feigin, D.~B. Fuchs, {\it Invariant skew-symmetric differential operators 
on the line and Verma modules over the Virasoro algebra}, Funkt. Anal. i ego Prilozh. 16 (1982) No. 2, 47-63; Funct. Anal. Appl. 16 (1982) 114-126.


\bibitem{FF2}
B.~L. Feigin, D.~B. Fuchs, {\it Representations of the Virasoro algebra, in: Representations of Lie groups and related topics}, eds. A.M. Vershik, D.P. Zhelobenko (Gordon and Breach, London 1990).


\bibitem{Th}
C.~B. Thorn, {\it Computing The Kac determinant using dual model techniques and more about the no-ghost theorem}, Nucl. Phys. {\bf B248} (1984) 551-569.


\bibitem{KW}
V.~G. Kac, M. Wakimoto, {\it Unitarizable highest weight representations of the Virasoro, Neveu-Schwarz and Ramond algebras}, Lect. Notes in Phys. {\bf 261} (1986) 345-371.


\bibitem{KR}
V.~G. Kac, A.~K. Raina, {\it Bombay lectures on highest weight representations of infinite dimensional Lie algebras}, Advanced Series in Mathematical Physics Vol.2, World Scientific Publishing Co.~Pte.~Ltd.


\bibitem{FF3}
B.~Feigin and D.~Fuchs, {\it Representations of the Virasoro algebra}, 
Adv. Stud. Contemp. Math. {\bf 7} (1990) 465.
  
  
\bibitem{Felinska:2011tn}
E.~Felinska, Z.~Jaskolski and M.~Kosztolowicz,
{\it Whittaker pairs for the Virasoro algebra and the Gaiotto-BMT states},
J.\ Math.\ Phys.\  {\bf 53} (2012) 033504 [Erratum-ibid.\  {\bf 53} (2012) 129902],
[arXiv:1112.4453 [math-ph]].
  

\bibitem{Zam}
A.~B.~Zamolodchikov, {\it Conformal symmetry in two-dimensional
space: recursion representation of conformal block},
Theor. Math. Phys. {\bf 73} (1987) 1088.


\bibitem{Zamolodchikov:ie}
A.~B.~Zamolodchikov, {\it Conformal Symmetry In Two-Dimensions: An
Explicit Recurrence Formula For The Conformal Partial Wave
Amplitude}, Commun.\ Math.\ Phys.\  {\bf 96} (1984) 419.


\bibitem{ZZ}
A.~B.~Zamolodchikov, A.~B.~Zamolodchikov, {\it Conformal Field
Theory And Critical Phenomena In Two-Dimensional Systems}, Sov.\
Sci.\ Rev.\ A.\ Phys.\ Vol.\ 10 (1989) 269-433.
    
  
\bibitem{Harlow:2011ny}
D.~Harlow, J.~Maltz, and E.~Witten, {\it {Analytic Continuation of Liouville
  Theory}},  {\em JHEP} {\bf 1112} (2011) 071,
  [\href{http://xxx.lanl.gov/abs/1108.4417}{{\tt arXiv:1108.4417}}].
  
  
\bibitem{Z2}
A.~Zamolodchikov, {\it {Two-dimensional conformal symmetry and critical
  four-spin correlation functions in the Ashkin-Teller model}},  {\em Sov.
  Phys. JEPT} {\bf 63} (1986) 1061.
  

\bibitem{Litvinov:2013sxa}
A.~Litvinov, S.~Lukyanov, N.~Nekrasov, and A.~Zamolodchikov, {\it {Classical
  Conformal Blocks and Painleve VI}},
  [\href{http://xxx.lanl.gov/abs/1309.4700}{{\tt arXiv:1309.4700}}].


\bibitem{MuellerKirsten:2006}
H.~M{\"u}ller-Kirsten, {\em {Introduction to Quantum Mechanics: Schr{\"o}dinger
  Equation and Path Integral}}.
\newblock World Scientific, 2006.  


\bibitem{McLachlan:1947}
N.~W. McLachlan, {\em {Theory and application of Mathieu functions}}.
\newblock Oxford: Clarendon Press, 1947.


\bibitem{Yanagida}
S.~Yanagida,
{\it Whittaker vectors of the Virasoro algebra in terms of Jack symmetric polynomial}, 
Journal of Algebra \textbf{333} (2011) 273.


\bibitem{Hadasz:2004cm}
L.~Hadasz, Z.~Jaskolski and M.~Piatek, {\it Analytic continuation formulae for the BPZ conformal block}, Acta Phys. Polon. {\bf B36} (2005) 845--864,
[\href{http://xxx.lanl.gov/abs/hep-th/0409258}{{\tt hep-th/0409258}}].


\bibitem{BC}
W.~Boenkost, F.~Constantinescu, {\it Vertex operators in Hilbert space}, J. Math. Phys. 34 (1993).


\bibitem{B}
W.~Boenkost, {\it Vertex-Operatoren, Darstellungen der Virasoro-Algebra und konforme Quantenfeldtheorie}, Dissertation, Frankfurt am Main, 1994.


\bibitem{B2}
W.~Boenkost, {\it Vertex operators are not closeable}, Rev. Math. Phys. 7 (1995) 51--56, hep-th/9401004.


\bibitem{ConstanScharf}
F.~Constantinescu, G.~Scharf, {\it Smeared and unsmeared chiral vertex operators},
Comm. Math. Phys. 200, 275--296, (1999), hep-th/9712174.


\bibitem{Rim:2015tsa}
C.~Rim and H.~Zhang, {\it Classical Virasoro irregular conformal block}, JHEP {\bf 07} (2015) 163,
[\href{http://xxx.lanl.gov/abs/1504.07910}{{\tt arXiv:1504.07910}}].


\bibitem{Rim:2015aha}
C.~Rim and H.~Zhang, {\it Classical Virasoro irregular conformal block II}, 
[\href{http://xxx.lanl.gov/abs/1506.03561}{{\tt arXiv:1506.03561}}].


\bibitem{Mironov:2010zs}
A.~Mironov, A.~Morozov, Sh.~Shakirov, {\it Conformal blocks as Dotsenko-Fateev Integral Discriminants}, 
Int. J. Mod. Phys. {\bf A25} (2010), 3173--3207,
[\href{http://xxx.lanl.gov/abs/1001.0563}{{\tt arXiv:1001.0563}}].
\end{thebibliography}

\providecommand{\href}[2]{#2}\begingroup\raggedright
\end{document}